\documentclass{emulateapj}
\usepackage{subfigure}
\makeatletter

\usepackage{times}
\usepackage{amssymb}
\usepackage{subfigure}
\usepackage{graphicx}

\newcommand\ionpat[2]{#1$\;${\scshape{#2}}}

\def\laundry{redshift $z$, stellar mass $M$, and ellipticity $\epsilon$}

\def\vecspace{x = (log$_{10}(M/0.1)$, $z/0.01$, $\epsilon/0.05$)}
\def\lgrbmedian{0.55}
\def\ccmedian{0.07}
\def\darkgrbfrac{2/15}

\def\KSnonomasspegsigmahalphasnCCohfullperUniformIb/Ic{0.0\%}
\def\KSnonomasspegsigmahalphasnCCohfullperUniformIc-BL{0.0\%}

\def\KSnonomasspegsigmahalphasnCCohfullperUniformPTF12gzk{0.0\%}

\def\KSnonomasspegsigmahalphasnCCohfullpercPTF12gzkIbIc{49\%}
\def\KSnonomasspegsigmahalphasnCCohfullpercPTF12gzkSLSN{-10000.0\%}
\def\KSnonomasspegsigmahalphasnCCohfullpercPTF12gzkII{49\%}
\def\KSnonomasspegsigmahalphasnCCohfullpercPTF12gzkIcBL{49\%}
\def\KSnonomasspegsigmahalphasnCCohfullpercPTF12gzkLGRB{-10000.0\%}

\def\KSnonomasspegsigmahalphasnCCohfullresidualRelIIPTF12gzk{1.7\%}

\def\KSnonomasspegsigmahalphasnCCohfullresidualRelIbIcPTF12gzk{2.1\%}

\def\KSnonomasspegsigmahalphasnCCohfullresidualRelPTF12gzkIcBL{2.9\%}

\def\KSnonomasspegsigmahalphasnCCohfullresidualGloIIPTF12gzk{8.8\%}
\def\KSnonomasspegsigmahalphasnCCohfullresidualGloIIIcBL{0.02\%}
\def\KSnonomasspegsigmahalphasnCCohfullresidualGloIbIcPTF12gzk{26\%}
\def\KSnonomasspegsigmahalphasnCCohfullresidualGloIbIcIcBL{0.6\%}
\def\KSnonomasspegsigmahalphasnCCohfullresidualGloPTF12gzkIcBL{2.9\%}

\def\KSnonomasspegmasspegdensityrsnCCohfullperUniformIc-BL{0.0\%}

\def\KSnonosfrPhothlkpcrsnCCohfullperUniformIb/Ic{0.0\%}
\def\KSnonosfrPhothlkpcrsnCCohfullperUniformIc-BL{0.0\%}

\def\KSnonosfrPhothlkpcrsnCCohfullperUniformPTF12gzk{9.8\%}

\def\KSnonosfrPhothlkpcrsnCCohfullpercPTF12gzkIbIc{49\%}
\def\KSnonosfrPhothlkpcrsnCCohfullpercPTF12gzkSLSN{-10000.0\%}
\def\KSnonosfrPhothlkpcrsnCCohfullpercPTF12gzkII{49\%}
\def\KSnonosfrPhothlkpcrsnCCohfullpercPTF12gzkIcBL{49\%}
\def\KSnonosfrPhothlkpcrsnCCohfullpercPTF12gzkLGRB{49\%}

\def\KSnonosfrPhothlkpcrsnCCohfullresidualGloIIPTF12gzk{2.3\%}

\def\KSnonosfrPhothlkpcrsnCCohfullresidualGloIbIcPTF12gzk{2.7\%}

\def\KSnonosfrPhothlkpcrsnCCohfullresidualGloPTF12gzkIcBL{8.0\%}
\def\KSnonosfrPhothlkpcrsnCCohfullresidualGloPTF12gzkLGRB{16\%}

\def\KSnonomasspegtohfoursnCCohfullperUniformIb/Ic{0.0\%}
\def\KSnonomasspegtohfoursnCCohfullperUniformIc-BL{0.0\%}

\def\KSnonomasspegtohfoursnCCohfullperUniformPTF12gzk{0.0\%}

\def\KSnonomasspegtohfoursnCCohfullpercPTF12gzkIbIc{49\%}
\def\KSnonomasspegtohfoursnCCohfullpercPTF12gzkSLSN{-10000.0\%}
\def\KSnonomasspegtohfoursnCCohfullpercPTF12gzkII{49\%}
\def\KSnonomasspegtohfoursnCCohfullpercPTF12gzkIcBL{49\%}
\def\KSnonomasspegtohfoursnCCohfullpercPTF12gzkLGRB{-10000.0\%}

\def\KSnonomasspegtohfoursnCCohfullresidualRelIIPTF12gzk{2.9\%}

\def\KSnonomasspegtohfoursnCCohfullresidualRelIbIcPTF12gzk{15\%}

\def\KSnonomasspegtohfoursnCCohfullresidualRelPTF12gzkIcBL{22\%}

\def\KSnonomasspegtohfoursnCCohfullresidualGloIIPTF12gzk{11\%}
\def\KSnonomasspegtohfoursnCCohfullresidualGloIIIcBL{19\%}
\def\KSnonomasspegtohfoursnCCohfullresidualGloIbIcPTF12gzk{61\%}
\def\KSnonomasspegtohfoursnCCohfullresidualGloIbIcIcBL{65\%}
\def\KSnonomasspegtohfoursnCCohfullresidualGloPTF12gzkIcBL{49\%}

\def\KSnonomasspegsfrsnCCohfullperUniformIb/Ic{0.0\%}
\def\KSnonomasspegsfrsnCCohfullperUniformIc-BL{0.0\%}

\def\KSnonomasspegsfrsnCCohfullperUniformPTF12gzk{0.0\%}

\def\KSnonomasspegsfrsnCCohfullpercPTF12gzkIbIc{49\%}
\def\KSnonomasspegsfrsnCCohfullpercPTF12gzkSLSN{-10000.0\%}
\def\KSnonomasspegsfrsnCCohfullpercPTF12gzkII{49\%}
\def\KSnonomasspegsfrsnCCohfullpercPTF12gzkIcBL{49\%}
\def\KSnonomasspegsfrsnCCohfullpercPTF12gzkLGRB{-10000.0\%}

\def\KSnonomasspegsfrsnCCohfullresidualRelIIPTF12gzk{4.2\%}

\def\KSnonomasspegsfrsnCCohfullresidualRelIbIcPTF12gzk{8.3\%}

\def\KSnonomasspegsfrsnCCohfullresidualRelPTF12gzkIcBL{2.9\%}

\def\KSnonomasspegsfrsnCCohfullresidualGloIIPTF12gzk{1.9\%}

\def\KSnonomasspegsfrsnCCohfullresidualGloIbIcPTF12gzk{8.3\%}

\def\KSnonomasspegsfrsnCCohfullresidualGloPTF12gzkIcBL{2.9\%}

\def\KSnonomasspeghlkpcrsnCCohfullresidualRelIIPTF12gzk{16\%}

\def\KSnonomasspeghlkpcrsnCCohfullresidualRelIbIcPTF12gzk{41\%}

\def\KSnonomasspeghlkpcrsnCCohfullresidualRelPTF12gzkIcBL{2.9\%}
\def\KSnonomasspegmasspegdensityrsnCCohfullperUniformIb/Ic{0.0\%}

\def\KSnonosfrhlkpcrsnCCohfullperUniformIb/Ic{0.0\%}
\def\KSnonosfrhlkpcrsnCCohfullperUniformIc-BL{0.0\%}

\def\KSnonosfrhlkpcrsnCCohfullperUniformPTF12gzk{0.0\%}

\def\KSnonosfrhlkpcrsnCCohfullpercPTF12gzkIbIc{49\%}
\def\KSnonosfrhlkpcrsnCCohfullpercPTF12gzkSLSN{-10000.0\%}
\def\KSnonosfrhlkpcrsnCCohfullpercPTF12gzkII{49\%}
\def\KSnonosfrhlkpcrsnCCohfullpercPTF12gzkIcBL{49\%}
\def\KSnonosfrhlkpcrsnCCohfullpercPTF12gzkLGRB{-10000.0\%}

\def\KSnonosfrhlkpcrsnCCohfullresidualRelIIPTF12gzk{17\%}

\def\KSnonosfrhlkpcrsnCCohfullresidualRelIbIcPTF12gzk{41\%}

\def\KSnonosfrhlkpcrsnCCohfullresidualRelPTF12gzkIcBL{2.9\%}

\def\KSnonosfrhlkpcrsnCCohfullresidualGloIIPTF12gzk{53\%}

\def\KSnonosfrhlkpcrsnCCohfullresidualGloIbIcPTF12gzk{41\%}

\def\KSnonosfrhlkpcrsnCCohfullresidualGloPTF12gzkIcBL{2.9\%}

\def\KSnonomasspegppohfoursnCCohfullperUniformIb/Ic{0.0\%}
\def\KSnonomasspegppohfoursnCCohfullperUniformIc-BL{0.0\%}

\def\KSnonomasspegppohfoursnCCohfullperUniformPTF12gzk{0.0\%}

\def\KSnonomasspegppohfoursnCCohfullpercPTF12gzkIbIc{49\%}
\def\KSnonomasspegppohfoursnCCohfullpercPTF12gzkSLSN{-10000.0\%}
\def\KSnonomasspegppohfoursnCCohfullpercPTF12gzkII{49\%}
\def\KSnonomasspegppohfoursnCCohfullpercPTF12gzkIcBL{49\%}
\def\KSnonomasspegppohfoursnCCohfullpercPTF12gzkLGRB{-10000.0\%}

\def\KSnonomasspegppohfoursnCCohfullresidualRelIIPTF12gzk{21\%}

\def\KSnonomasspegppohfoursnCCohfullresidualRelIbIcPTF12gzk{61\%}

\def\KSnonomasspegppohfoursnCCohfullresidualRelPTF12gzkIcBL{49\%}

\def\KSnonomasspegppohfoursnCCohfullresidualGloIIPTF12gzk{1.2\%}

\def\KSnonomasspegppohfoursnCCohfullresidualGloIbIcPTF12gzk{2.1\%}

\def\KSnonomasspegppohfoursnCCohfullresidualGloPTF12gzkIcBL{2.9\%}
\def\LGRBnumPTF12gzk{1}

\def\KSnonomasspegtohfoursnCCohfullnumPTF12gzk{1}
\def\KSnonomasspegtohfoursnCCohfullnumIbIc{11}

\def\KSnonomasspegtohfoursnCCohfullnumII{59}
\def\KSnonomasspegtohfoursnCCohfullnumIcBL{6}

\def\KSnonosfrPhothlkpcrsnCCohfullnumPTF12gzk{1}

\def\KSnonomasspegsfrpegdensityrsnCCohfullperUniformIb/Ic{0.0\%}
\def\KSnonomasspegsfrpegdensityrsnCCohfullperUniformIc-BL{0.0\%}

\def\KSnonomasspegsfrpegdensityrsnCCohfullperUniformPTF12gzk{9.9\%}

\def\KSnonomasspegsfrpegdensityrsnCCohfullnumPTF12gzk{1}
\def\KSnonomasspegsfrpegdensityrsnCCohfullpercPTF12gzkIbIc{49\%}
\def\KSnonomasspegsfrpegdensityrsnCCohfullpercPTF12gzkSLSN{-10000.0\%}
\def\KSnonomasspegsfrpegdensityrsnCCohfullpercPTF12gzkII{49\%}
\def\KSnonomasspegsfrpegdensityrsnCCohfullpercPTF12gzkIcBL{49\%}
\def\KSnonomasspegsfrpegdensityrsnCCohfullpercPTF12gzkLGRB{49\%}

\def\KSnonomasspegsfrpegdensityrsnCCohfullresidualGloIIPTF12gzk{2.8\%}
\def\KSnonomasspegsfrpegdensityrsnCCohfullresidualGloIIIcBL{1.7\%}

\def\KSnonomasspegsfrpegdensityrsnCCohfullresidualGloIbIcPTF12gzk{2.3\%}
\def\KSnonomasspegsfrpegdensityrsnCCohfullresidualGloIbIcIcBL{5.6\%}
\def\KSnonomasspegsfrpegdensityrsnCCohfullresidualGloIbIcLGRB{0.0006\%}
\def\KSnonomasspegsfrpegdensityrsnCCohfullresidualGloPTF12gzkIcBL{18\%}
\def\KSnonomasspegsfrpegdensityrsnCCohfullresidualGloPTF12gzkLGRB{24\%}
\def\KSnonomasspegsfrpegdensityrsnCCohfullresidualGloIcBLLGRB{13\%}

\def\KSnonomasspegsfrmpadensityrsnCCohfullperUniformIb/Ic{0.0\%}
\def\KSnonomasspegsfrmpadensityrsnCCohfullperUniformIc-BL{0.0\%}

\def\KSnonomasspegsfrmpadensityrsnCCohfullperUniformPTF12gzk{0.0\%}

\def\KSnonomasspegsfrmpadensityrsnCCohfullnumPTF12gzk{1}
\def\KSnonomasspegsfrmpadensityrsnCCohfullpercPTF12gzkIbIc{49\%}
\def\KSnonomasspegsfrmpadensityrsnCCohfullpercPTF12gzkSLSN{-10000.0\%}
\def\KSnonomasspegsfrmpadensityrsnCCohfullpercPTF12gzkII{49\%}
\def\KSnonomasspegsfrmpadensityrsnCCohfullpercPTF12gzkIcBL{49\%}
\def\KSnonomasspegsfrmpadensityrsnCCohfullpercPTF12gzkLGRB{-10000.0\%}

\def\KSnonomasspegsfrmpadensityrsnCCohfullnumII{70}

\def\KSnonomasspegsfrmpadensityrsnCCohfullnumIcBL{6}

\def\KSnonomasspegsfrmpadensityrsnCCohfullresidualRelIIPTF12gzk{10\%}
\def\KSnonomasspegsfrmpadensityrsnCCohfullresidualRelIIIcBL{3.2\%}
\def\KSnonomasspegsfrmpadensityrsnCCohfullresidualRelIbIcPTF12gzk{15\%}

\def\KSnonomasspegsfrmpadensityrsnCCohfullresidualRelPTF12gzkIcBL{2.9\%}

\def\KSnonomasspegsfrmpadensityrsnCCohfullresidualGloIIPTF12gzk{8.5\%}

\def\KSnonomasspegsfrmpadensityrsnCCohfullresidualGloIbIcPTF12gzk{15\%}

\def\KSnonomasspegsfrmpadensityrsnCCohfullresidualGloPTF12gzkIcBL{2.9\%}
\def\KSnonomasspegmasspegdensityrsnCCohfullperUniformPTF12gzk{10\%}

\def\KSnonomasspegmasspegdensityrsnCCohfullnumPTF12gzk{1}
\def\KSnonomasspegmasspegdensityrsnCCohfullpercPTF12gzkIbIc{49\%}
\def\KSnonomasspegmasspegdensityrsnCCohfullpercPTF12gzkSLSN{-10000.0\%}
\def\KSnonomasspegmasspegdensityrsnCCohfullpercPTF12gzkII{49\%}
\def\KSnonomasspegmasspegdensityrsnCCohfullpercPTF12gzkIcBL{49\%}
\def\KSnonomasspegmasspegdensityrsnCCohfullpercPTF12gzkLGRB{49\%}
\def\KSnonomasspegmasspegdensityrsnCCohfullnumIbIc{52}

\def\KSnonomasspegmasspegdensityrsnCCohfullnumII{251}

\def\KSnonomasspegmasspegdensityrsnCCohfullnumIcBL{16}

\def\KSnonomasspegmasspegdensityrsnCCohfullnumLGRB{14}

\def\KSnonomasspegmasspegdensityrsnCCohfullresidualGloIIPTF12gzk{3.9\%}
\def\KSnonomasspegmasspegdensityrsnCCohfullresidualGloIIIcBL{0.03\%}
\def\KSnonomasspegmasspegdensityrsnCCohfullresidualGloIILGRB{0.07\%}
\def\KSnonomasspegmasspegdensityrsnCCohfullresidualGloIbIcPTF12gzk{3.8\%}
\def\KSnonomasspegmasspegdensityrsnCCohfullresidualGloIbIcIcBL{0.03\%}
\def\KSnonomasspegmasspegdensityrsnCCohfullresidualGloIbIcLGRB{0.1\%}
\def\KSnonomasspegmasspegdensityrsnCCohfullresidualGloPTF12gzkIcBL{51\%}
\def\KSnonomasspegmasspegdensityrsnCCohfullresidualGloPTF12gzkLGRB{36\%}
\def\KSnonomasspegmasspegdensityrsnCCohfullresidualGloIcBLLGRB{88\%}

\def\StellarMassnum12gzk{2}

\def\StellarMassvalsII12gzk{1.3\%}

\def\StellarMassvalsIbIc12gzk{1.4\%}

\def\StellarMassvals12gzkIcBL{1.7\%}
\def\StellarMassvals12gzkLGRB{1.8\%}

\def\SFRkpcminussqnum12gzk{2}

\def\SFRkpcminussqvalsII12gzk{13\%}

\def\SFRkpcminussqvalsIbIc12gzk{21\%}

\def\SFRkpcminussqvals12gzkIcBL{51\%}
\def\SFRkpcminussqvals12gzkLGRB{9.9\%}

\submitted{Submitted to The Astrophysical Journal}

\begin{document} 

\title{The Host Galaxies of Fast-Ejecta Core-Collapse Supernovae}
\shorttitle{Hosts of fast-ejecta core-collapse supernovae}

\author{Patrick L. Kelly\altaffilmark{1}}
\author{Alexei V. Filippenko\altaffilmark{1}}
\author{Maryam Modjaz\altaffilmark{2}}
\author{Daniel Kocevski\altaffilmark{3}}
 
\altaffiltext{1}{Department of Astronomy, University of California, Berkeley, CA 94720-3411, USA}
\altaffiltext{2}{CCPP, New York University, 4 Washington Place, New York, NY 10003, USA}
\altaffiltext{3}{NASA/Goddard Space Flight Center, Code 662, Greenbelt, MD 20771, USA}


\begin{abstract}
Spectra of broad-lined Type Ic supernovae (SN~Ic-BL),
the only kind of SN observed at the locations of 
long-duration gamma-ray bursts (LGRBs),
exhibit wide features indicative of high ejecta velocities ($\sim0.1c$).
We study the host galaxies of a sample of 245 low-redshift ($z<0.2$) core-collapse SN, including 17 SN~Ic-BL, discovered by galaxy-untargeted searches, 
and 15 optically luminous and dust-obscured $z<1.2$ LGRBs. We show that, in comparison with SDSS galaxies having similar stellar masses, the hosts of low-redshift SN~Ic-BL and $z<1.2$ LGRBs
have high stellar-mass and star-formation-rate densities. 
Core-collapse SN having typical ejecta velocities, in contrast, show no preference for 
such galaxies. Moreover, we find that the hosts of SN~Ic-BL, unlike those of 
SN~Ib/Ic and SN~II, exhibit high gas velocity dispersions for their stellar masses. 
The patterns likely reflect variations among star-forming environments, and suggest that LGRBs can be used as probes of conditions in high-redshift galaxies.
They may be caused by efficient formation of massive binary progenitors systems in densely star-forming regions, or, less probably, a higher fraction of stars created with the initial masses required for a SN~Ic-BL or LGRB.
Finally, we show that the preference of SN~Ic-BL and LGRBs for galaxies with high stellar-mass and
star-formation-rate densities cannot be attributed to a preference for low metal abundances but must reflect the influence of a separate environmental factor.
\end{abstract}

\maketitle 

\section{Introduction}
In the cases of at least ten nearby (redshift $z\lesssim0.5$) LGRBs, observations have revealed a SN~Ic-BL spectrum superimposed on the power-law continuum of the fading optical afterglow (\citealt{ga98,ma03,st03,hj03}; see \citealt{woosleybloom06} and \citealt{modjaz11rev} for reviews). 
The spectra are characterized by wide features consistent with high ejecta velocities ($\sim$\,20,000--30,000\,km\,s$^{-1}$), and an absence of hydrogen and helium.
The other principal classes of core-collapse SN, in contrast, exhibit spectroscopic features consistent with more slowly moving ejecta.
The most common core-collapse SN are Type~II ($\sim 60$\%; \citealt{lileaman11}; \citealt{smartteldridge09}), which exhibit hydrogen (H) in their spectra; they are the final eruptions of stars that have retained their outer H shell. 
When the progenitor sheds, transfers to a companion, or internally mixes its outer H envelope during pre-SN evolution, 
the explosion will produce an H-deficient SN~Ib, or an H- and He-deficient SN~Ic \citep[e.g.,][and references therein]{fil97}. 

While the absence of H and He in the spectra of SN~Ic-BL indicates that their progenitors have lost their envelopes prior to core collapse, 
simulations additionally suggest that the progenitors of SN~Ic-BL with associated LGRBs may also have rapid speeds of rotation.
In models, only quickly rotating stars without an H envelope produce the outflowing jets that yield $\gamma$-ray emission \citep{hirschimeynet05,yoonlanger05,langernorman06} 
after core collapse to a black hole \citep{wo93,macfadyenwoosley99}.

Close massive binary systems that experience mass transfer, common-envelope evolution, or a merger \citep{podsialowskiivanova10,langer12} 
are possible progenitors of LGRBs, because they can likely produce the required rapidly rotating massive stars without
H or He envelopes. 
Recent observations of Galactic O-type stars show that, in fact, $\gtrsim 70$\% of massive stars experience mass transfer with a companion and $\sim30$\% undergo a merger \citep{sanademink12}. 
Alternative progenitor candidates include quickly rotating, metal-poor stars that internally mix their outer envelopes \citep{yoonlanger05, woosleyheger06}.
Single stars with high abundances, however, are considered improbable progenitors, because their comparatively strong winds \citep{vinkdekoter01} are expected to reduce their angular momentum. 

Spectroscopy of $z \lesssim 0.3$ host galaxies shows, in fact, that nearby
SN~Ic-BL with an associated GRB prefer more metal-poor environments than nearby Type~Ic-BL SN having no obvious $\gamma$-ray emission \citep{mod08,kocevski09,grahamfruchter13}.  
The latter, in turn, prefer more metal-poor \citep{kelkir12, sanderssoderberg12} and blue \citep{kelkir12} environments than do SN~Ic without broad features. 

In several cases, however, the positions of low- and moderate-redshift LGRBs are spatially coincident or closely associated with massive, metal-rich galaxies (e.g., \citealt{lev10highmet,perleylevan13, elliottkruhler13}). 
Absorption spectroscopy also finds evidence for high-redshift $z \gtrsim 2$  systems along the line of sight to LGRBs
with metallicities exceeding the $z \gtrsim 2$ cosmic average \citep{prochaska07}.
\citet{savagliorau12} has inferred a supersolar abundance from $z = 3.57$ absorption features consistent with a pair galaxies 
in the GRB 090323 afterglow spectrum.
GRB 130702A \citep{singercenko13} occurred, however, in a metal-poor faint satellite of a $z=0.145$ massive galaxy \citep{kelfil13}, raising the possibilty of  superpositions or associations for some LGRBs that would be difficult to resolve at high redshift. 

Here we measure the host-galaxy properties of nearby \mbox{($z<0.2$)} core-collapse SN explosions using the imaging and photometry of the Sloan Digital Sky Survey (SDSS). For $z<1.2$ LGRB hosts, we estimate these host properties from published photometry and archival {\it HST} images.  
We show that SN Ic-BL and LGRBs 
exhibit a strong preference for galaxies that have high stellar-mass density and star-formation-rate density 
for their stellar mass.
We also use SDSS spectra to show that the gas kinematics of SN~Ic-BL hosts are exceptional.  
In \S \ref{sec:data}, we describe the core-collapse SN and LGRB samples as well as the SDSS and {\it HST} galaxy data that we use in this analysis.
Section~\ref{sec:methods} presents our techniques to analyze the galaxy imaging and spectroscopy, and the statistical methods we employ.
In \S \ref{sec:results}, we describe the results of our analysis, while
\S \ref{sec:discussion} discusses the interpretation of the observed
patterns. Our conclusions are presented in \S \ref{sec:conclusions}.

\section{Data}
\label{sec:data}
We study the host galaxies of both nearby ($z<0.2$) core-collapse SN discovered by ``galaxy-untargeted'' transient searches (e.g., the Palomar Transient Factory; PTF; \citealt{raukulkarni09,lawkulkarni09}) which do not target specific potential hosts, and $z<1.2$ LGRBs  detected by $\gamma$-ray satellites.
We use the SDSS spectroscopic sample to build a control sample of low-redshift star-forming galaxies, and SDSS photometry and spectroscopy to measure properties of both the sample of low-redshift star-forming galaxies and the host galaxies of the nearby SN.
For the host galaxies of $z < 1.2$ LGRBs, we estimate host properties using published 
photometry and {\it HST} imaging.

\subsection{SDSS DR10 Photometry and Spectroscopy}
The SDSS galaxy photometry and fiber spectra are from 
Data Release 10 \citep[DR10;][]{sdssdrten13}, and they were collected with the 2.5\,m telescope at Apache Point, New Mexico. The imaging survey, which spans 14,555 square degrees, consists of 53.9\,s integrations through the SDSS $ugriz$ filters, and the typical limiting $r$-band AB magnitude is 22.2.
The typical sensitivity of available SDSS imaging makes possible detection of $z=0.1$ galaxies with absolute Vega magnitudes $M_B$ or $M_V$ brighter than about $-15.2$. 
Each Sloan $2048 \times 1498$ pixel CCD array records a $13.5' \times 9.9'$ field of view.

The SDSS spectroscopic survey acquired approximately 45\,min of total integration in clear conditions, 
split into a series of three successive exposures. The spectrograph comprises $3''$ ($2''$ for BOSS) diameter fiber-optic cables placed at the positions of targets on the focal plane. 
Adjacent fibers can be no closer than $55''$ in a single fiber mask because of 
engineering constraints \citep{stra02}, and the SDSS spectrographs record light with wavelengths 3800--9200\,\AA. 

The targets selected for SDSS fiber spectroscopy consist of three primary ``Legacy'' categories of objects, 
and were also taken from several dozen ancillary programs, some of which were limited to specific parts of the survey (e.g., Stripe 82). Objects detected with $5\sigma$ significance in the imaging survey, with an extended light distribution and having $r$-band magnitude brighter than 17.77, as well as QSO candidates and luminous red galaxies (LRGs), formed the Legacy samples \citep{baldry05}. More limited special-program science targets included, for example, a $u$-band galaxy sample selected to investigate the properties of blue, faint galaxies.  

Table~\ref{tab:selection} shows the construction of our sample of SDSS galaxy spectra.

\subsection{Transient and Host-Galaxy Samples}
\subsubsection{Nearby SN Sample}

The core-collapse SN sample is constructed from $z < 0.2$ discoveries by SN surveys that do not target specific 
galaxies. The ``galaxy-untargeted'' SN search technique is akin to that of wide-field $\gamma$-ray satellite (e.g., {\it Swift}) searches for LGRBs. The SN in our sample consist of discoveries by galaxy-untargeted searches reported to the International Astronomical Union (IAU), as well as those published \citep{arcavi10} or reported via Astronomical Telegrams by the PTF, from 1990 January 1 through 2012 May 10.
The SN searches that we considered to be galaxy-untargeted are identical those listed by \citet{kelkir12}.
Our sample consists of only those nearby SN whose host galaxies have SDSS $ugriz$ imaging and, when appropriate, a fiber spectrum. 

The probability of detecting a SN in an image depends on the limiting magnitude of the transient search, as well as on 
the SN distance, luminosity, light-curve shape, and dust attenuation along the line of sight. 
Comparative analysis of the redshifts of core-collapse SN discovered by galaxy-untargeted searches
found no significant evidence that variation among the principal species strongly affects their discovery rate \citep{kelkir12}. 

\citet{kel08} showed that SN~Ic are more closely associated with 
the highest surface brightness regions of their host galaxies than SN~Ib, 
but here we combine these two SN types to assemble a larger sample of stripped-envelope
SN that do not show high ejecta velocities. 
We also show, for comparison, the host of PTF12gzk, a peculiar SN~Ic that did not exhibit broad spectroscopic features \citep{benamigalyam12} 
but in which radio observations found evidence for $\sim0.3c$ ejecta speeds \citep{horeshkulkarni13}. 
Lists of the nearby SN and the LGRBs are presented in Tables~2 and 3 (full tables are available in the 
electronic version).

\subsubsection{LGRB Sample}
Our LGRB sample includes both objects with 
a detected optical afterglow, and ``dark'' bursts without a luminous optical afterglow \citep{taylorfrail98} whose position was determined using their X-ray flux \citep{cenkokeleman09,perleycenko09}. 
Analysis of the host galaxies of dark GRBs shows that they are both more massive \citep{perleylevan13} and dust-obscured \citep{djorgovskifrail01,klosehenden03,perleylevan13} than the hosts of GRBs with detected optical afterglows.
The fraction of dark bursts in our sample (\darkgrbfrac) is approximately representative of the fraction of dust-obscured GRBs below the $z = 1.2$ redshift upper limit of our LGRB sample reported by \citet{perleylevan13}. 

We assemble $z<1.2$ LGRB host galaxies having archival {\it HST} images from the 
unobscured LGRBs assembled by \citet{savaglio09} and dust-obscured LGRBs assembled by
\citet{perleylevan13}, 
after rejecting several datasets that showed evidence of residual LGRB light. 
In many cases, {\it HST} images were acquired at least a year
after the explosion; these are the data we use when available to measure 
galaxy sizes. When only images taken closer to the time of the GRB are 
available, we inspect the host galaxy to determine whether evidence for a 
point source at the explosion site exists. LGRBs exhibit a strong association with the 
brightest pixels of their hosts \citep{fru06}, so a potential concern is that 
we could reject data where the LGRB coincided with a bright star-forming 
region. In practice, possible confusion was minimal for the data taken within 
$\sim4$ months after the GRB trigger. While we do not expect any significant
contamination, here we are interested in the integrated properties of the host galaxy
and not the specific region where the LGRB occurred. 

Even when an optical afterglow can be detected, identifying a coincident SN 
in the fading afterglow light curve, or by identifying a SN spectrum superimposed on the
power-law afterglow continuum, requires high signal-to-noise ratio (S/N) data. 
Sufficient follow-up observations have only been possible for LGRBs with $z \lesssim 0.6$. 
Our GRB sample includes events with and 
without a detected optical afterglow or coincident SN. 
Since a representative fraction of the LGRBs in our sample have no optical counterpart, 
our findings should be robust to the effects of dust obscuration on LGRB detection.

\begin{deluxetable}{lc}
\tablecaption{Construction of SDSS Galaxy Spectra Sample}
\tablecolumns{2}
\tablehead{\colhead{Criterion}&\colhead{SDSS Fibers}}
\startdata
(1) Full Catalog & 948,205 \\
(2) Reliable Line Measurements & 910,532 \\
(3) Star Forming (Low or High S/N) & 377,763 \\
(4) H$\alpha$ S/N $>$ 20 &302,865\\
(5) Fiber Offset $< 0.2\,R_{\rm P}$ & 289,474
\enddata
\tablecomments{SDSS spectroscopic fiber sample construction. SDSS galaxies remaining of each spectroscopic type after  applying each inclusion criterion. (1) Spectra in full SDSS catalog; (2) deemed reliable line measurement with no redshift fit warning; (3) spectrum classified as star forming, with low or high S/N, according to its position on the Baldwin, Phillips, \& Terlevich \citep[BPT;][]{baldwin81} diagram of the [\ionpat{O}{iii}] $\lambda$5007/H$\beta$ and [\ionpat{N}{ii}] $\lambda$6584/H$\alpha$ line ratios and line-flux S/N, respectively; (4) galaxies whose H$\alpha$ line-flux measurements have S/N $> 5$; and (5) center of the SDSS fiber is within 20\% of the Petrosian radius $R_{\rm P}$. }
\label{tab:selection}
\end{deluxetable}

\section{Methods}
\label{sec:methods}

\subsection{Measurements of Galaxy Properties}
For both the $z<0.2$ core-collapse SN and $z<1.2$ LGRB samples, as well as the SDSS star-forming population, we estimate host-galaxy stellar masses $M$ and photometric star-formation rates (SFRs) by fitting PEGASE2 \citep{fi99} stellar population synthesis models to broadband photometry; 
see \citet{kel10} for detailed information 
on the star-formation histories and
initial-mass functions (IMFs) used. 
For nearby core-collapse SN host galaxies and the SDSS star-forming population, we fit Sloan $ugriz$ magnitudes. 
The multi-band photometry of the host galaxies of LGRBs was assembled from 
the GHostS database\footnote{http://www.grbhosts.org}.

We describe a second, complementary set of SFR estimates in the following section that is
available only for the SDSS star-forming spectroscopic sample and which uses both the fiber spectrum and 
broadband $ugriz$ photometry. When comparing among samples, however, we only 
compare SFR values estimated using the same method.

\subsubsection{Analyses of SDSS Galaxy Spectra}
Several teams have performed and made available detailed measurements of SDSS galaxy properties from the Sloan
photometry and spectroscopy.
For the $z < 0.2$ samples of core-collapse SN host galaxies and the SDSS star-forming population, we use SFRs estimated 
from fitting both spectra and photometry, and gas velocity dispersions $\sigma_{\rm gas}$ measured from the H$\alpha$ emission-line profile. 

We use the star-forming classifications for SDSS galaxies and hybrid spectroscopic and photometric SFR estimates made available by a collaboration that was both at the Max Planck Institute for Astronomy (MPA) and Johns Hopkins University (JHU)
(S. Charlot, G. Kauffmann, S. White, T. Heckman, C. Tremonti, and J. Brinchmann; MPA-JHU).
SDSS fiber apertures generally do not cover the entire light distribution of each target galaxy.
The MPA-JHU team therefore estimates the total SFR of each galaxy as the sum of the SFR
within the fiber aperture determined from fitting the spectrum, and from a fit to the $ugriz$
light outside the fiber opening. 

We adopt the gas velocity dispersions estimated from SDSS emission-line profiles 
by the Portsmouth group \citep{thomassteele13}, which apply the public Penalized PiXel Fitting \citep{cappellariemsellem04} (pPXF) and the Gas and Absorption Line Fitting \citep{sarzifalconbarroso06} (GANDALF v1.5) codes. 
The velocity dispersion of the gas is estimated from the widths of emission lines (e.g., H$\alpha$, [\ionpat{O}{iii}]),
and here we take the dispersion for H$\alpha$.

\subsubsection{Host-Galaxy Sizes}
The SDSS Photo pipeline  performs separate fits of a de Vaucouleurs $r^{1/4}$ law and an exponential profile to the light distribution of each extended object. 
The pipeline next finds the linear combination of the two models (holding all parameters except flux fixed)
that minimizes the $\chi^2$ statistic. 
To obtain an estimate of the half-light radius $r_{50}$, we compute the weighted average of the two components' $r_{50}$ parameters 
and weight each by its fractional contribution to the total model flux.

We use the GALFIT \citep{pe02} program to perform the same surface-brightness fitting analysis on archival {\it HST} images of LGRB host galaxies.
We apply Source Extractor \citep[SExtractor;][]{bert96} to drizzled and cosmic-ray-rejected images to estimate 
object positions, ellipticities, and magnitudes, and these are used as input GALFIT parameters. 
Extended sources except the LGRB hosts are modeled with Sersic profiles, and the instrument point-spread function (PSF) is used to model stars.  
The host-galaxy $r_{50}$ estimates show good agreement with existing estimates of $r_{50}$ from profile fitting \citep{conselicevreeswijk05}, and with published $r_{80}$ SExtractor measurements \citep{svensson10}.

The host galaxy of GRB 020903 is part of a complex association of interacting clumps, 
so the galaxy is not well approximated by a simple surface-brightness model. 
An additional complication is that much of the host-galaxy photometry was taken from the ground,
where the galaxy components cannot be resolved. We therefore have excluded this host
galaxy from the analysis. 

The angular diameter of ESO184-G82, the host galaxy of the nearby GRB 980425 ($z=0.0087$), 
approximately spans the available {\it HST} images. The physical resolution of the {\it HST} data
is also substantially higher than that of the images of the SDSS SN host galaxies, or other LGRB hosts.
Measurement of the host effective radius from model fitting requires significant 
imaging area without galaxy light to be able to fit robustly for the background level. 
We therefore do not include GRB 980425, which does not have SDSS exposures, in the LGRB sample.

The host galaxy of GRB 051022 is also a system with two peaks that 
may possibly correspond to two strongly interacting galaxies, or else may
instead be a galaxy with irregular morphology. We expect that the 
SDSS Photo pipeline would be most likely to model the host as a single system,
so we construct a GALFIT model that consists of a single galaxy.

\subsubsection{Stellar-Mass and Star-Formation Densities Estimates}

We calculate the projected stellar-mass density $\Sigma_{M}$ and the projected star-formation density $\Sigma_{\rm SFR}$ from the galaxy stellar mass $M$ and
star-formation rate SFR, respectively, and the parameters of the model $r$-band isophotal
ellipse that encloses half of the galaxy light. 
We compute, for example, $\Sigma_{M} = {\rm log}_{10}(M \mathbin{/} 2 \mathbin{/} \pi A B)$, where 
$M$ is the stellar mass (in $M_{\odot}$), while $A$ and $B$ are the semimajor and semiminor axes (in kpc) of the 
isophotal ellipse that contains half of the galaxy $r$-band flux, determined from fitting its surface-brightness distribution. 

\subsection{Comparison of Host-Galaxy Properties}

\subsubsection{Median Relationship for Photometric Host Properties}
Photometric magnitudes measured through an appropriate aperture designed to enclose 
a specific percentage of the host light (e.g., a Petrosian aperture; \citeyear{petrosian76}) can be expected to sample an approximately consistent fraction of galaxy light with increasing distance to sources. 
Physical properties estimated from broadband fluxes (e.g., stellar mass $M$) should therefore not exhibit strong aperture biases with increasing redshift in our sample, although the effects of surface-brightness dimming may become important at high redshifts.

To study the properties of galaxies derived from photometric measurements, 
we therefore find the best-fitting $M$--$\Sigma_{M}$, $M$--$\Sigma_{\rm SFR}$, $M$--$r_{50}$,  and SFR--$r_{50}$
relations for the complete $0.03 < z < 0.1$ SDSS star-forming catalog. 
We fit a second-degree polynomial to the median ordinate value (e.g., $\Sigma_{\rm SFR}$) across at least five bins in galaxy stellar mass.
We include an additional point at (log $M$, $r_{50}$) = (0,0) and (log SFR, $r_{50}$) = ($-10$,0), respectively, when fitting for the SDSS $M$--SFR relation so that $r_{50}$ approaches zero for galaxies with negligible stellar mass $M$ or SFR. 
We use the Main SDSS spectroscopic sample Legacy
``GALAXY'' targets (i.e., where the 64 bit of the primTarget bitmask was set)
to measure median relations of the properties of the low-redshift galaxy population.

\subsubsection{Spectroscopic Host Properties}
The fixed angular size of SDSS $3''$ fibers (or $2''$ for the Baryon Acoustic Oscillation Survey; BOSS), in contrast to (for example) a Petrosian aperture, samples 
a percentage of the light of each extended target that depends on the redshift and intrinsic size of the galaxy. 
To reach conclusions about spectroscopic measurements that are not sensitive to aperture effects, we apply a separate predictive algorithm that we have 
developed that yields robust comparisons. 

To predict the expected value of an observable (e.g., $\sigma_{\rm gas}$) given other host-galaxy properties (e.g., mass $M$),
we perform, for each $z < 0.2$ SN host galaxy, a locally weighted multiple linear least-squares fit to the SDSS catalog. 
We model the observable of interest $O_j$ as
\begin{equation}
O_j = \sum\limits_{i = 0}^{n} A_i \, x_{ij}, 
\end{equation}
where $j$ indexes the SDSS galaxies, and each $x_{ij}$ is a galaxy property, a combination (e.g., multiplicative product) of galaxy properties, or unity.
For example, we model the gas velocity dispersion $\sigma_{\rm gas}$ as
\begin{equation}
\sigma_{{\rm gas},j} =  A_0 + A_1{\rm log}_{10}M_j + A_2({\rm log}_{10}M_j)^2 + A_3\epsilon_j,
\end{equation}
where $M$ is stellar mass and $\epsilon$ is galaxy ellipticity.
Each row $j$ of the design matrix $A_i^j$ and $O^j$ is multiplied by the weight $W_j = e^{-u_j}$, where $u_j$ is
\begin{equation}
u_j = \frac{(z_j - z^{\rm SN})^2}{(0.01)^2} + {\rm log}_{10}\left(\frac{M_j}{M^{\rm SN}}\right)^2 + \frac{(F_{{\rm aper},j} - F_{\rm aper}^{\rm SN})^2}{(0.1)^2}
\end{equation}
and $z$ is galaxy redshift.
To be able to predict a specific property $O_{\rm pred}$ of each host galaxy (e.g., $\sigma_{\rm gas}$)
independent of its observed value, $O_{\rm obs}$, we 
exclude the host galaxy itself when fitting the predictive model to the properties of SDSS galaxies. 
This approach assigns greater weight to galaxies with similar observables, and mitigates any 
possible selection effect with galaxy redshift or fiber fraction. 

The best-fit parameters $A_i$ for each fit are used to compute a prediction for the value of the
galaxy property, $O_{\rm pred}$, which can then be compared to the 
observed value,  $O_{\rm obs}$, as a residual,

\begin{equation}
\Delta O \equiv O_{\rm obs} - O_{\rm pred}.
\label{eqn:residual}
\end{equation}
\noindent
Here the fitting analysis requires only modest numbers of nearby datapoints from the SDSS spectroscopic catalog.

\begin{figure*}
\centering
\subfigure{\includegraphics[angle=0,width=5in]{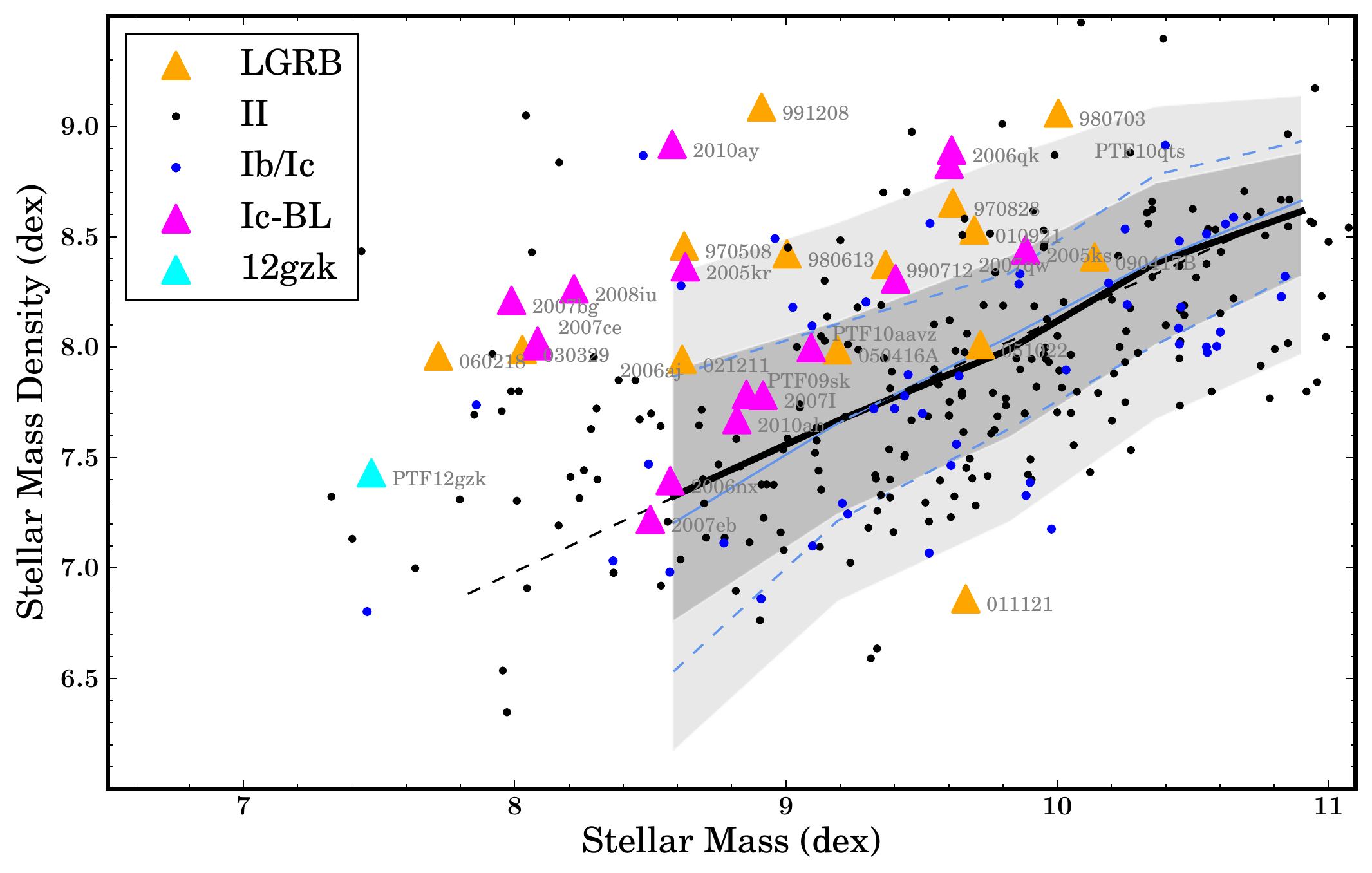}}
\subfigure{\includegraphics[angle=0,width=5in]{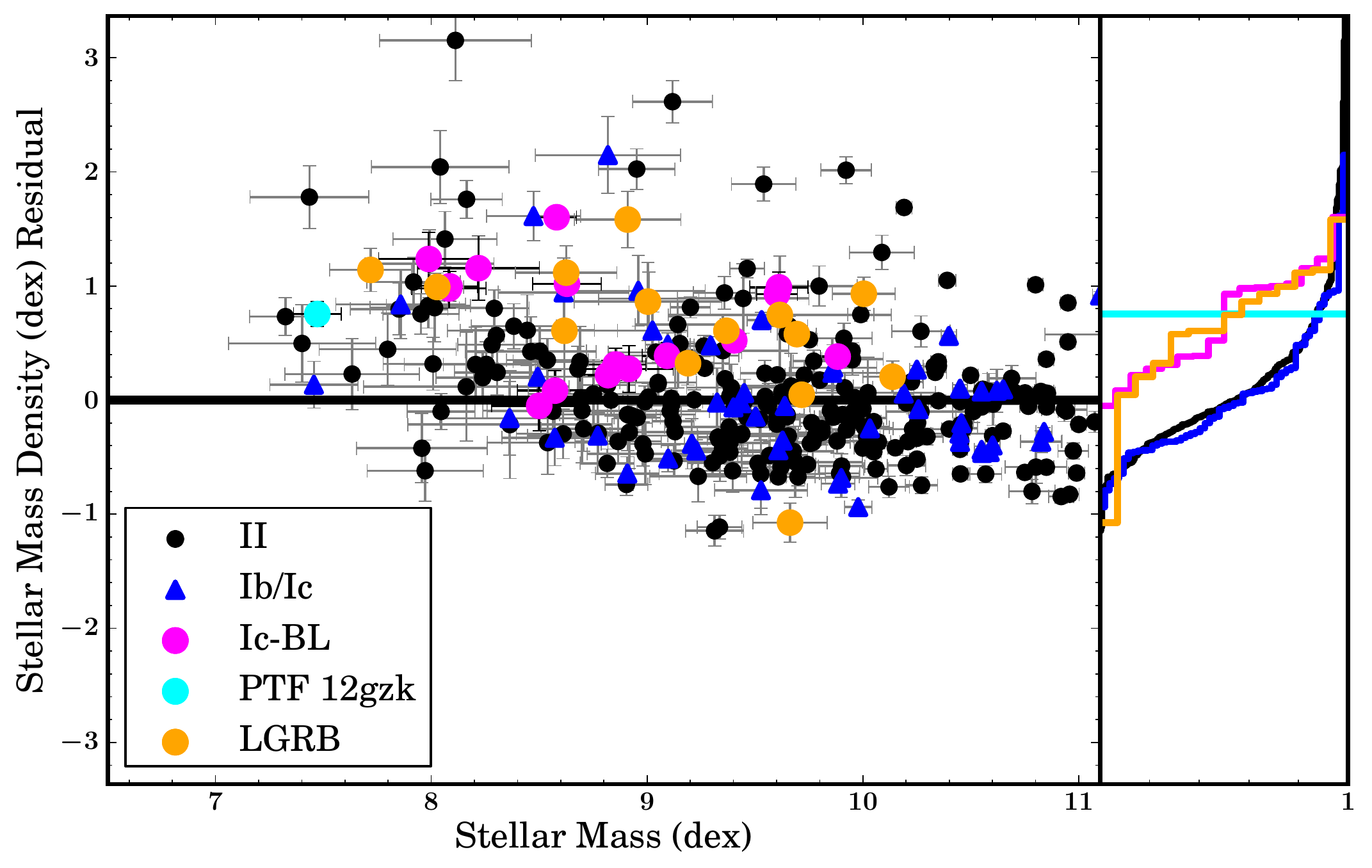}}
\caption{Stellar mass density ($\Sigma_{M}$) against host stellar mass for $z<0.2$ core-collapse SN as well as $z<1.2$ LGRBs. The host galaxies of broad-lined SN~Ic and LGRBs show higher $\Sigma_{M}$ for their stellar 
masses. In the upper panel, the solid black line (median), dark-grey region (68\%), and light-grey region (95\%) show the distribution of SDSS star-forming galaxies.
The blue solid (median) and dashed (68\%) lines show the distribution expected for a transient population that follows the SFR. 
The black dashed line shows the second-order polynomial fit to the median SDSS $M$--$\Sigma_{M}$ relation used
to compute a $\Sigma_{M}$ residual for each galaxy.
Broad features were not evident in the optical spectrum of PTF12gzk, a peculiar SN~Ic, but radio observations show evidence for ejecta speeds that reach $\sim0.3c$ \citep{horeshkulkarni13}.   
In the lower panel, we plot the residuals of host-galaxy $\Sigma_{M}$
from the SDSS $M$--$\Sigma_{M}$ relation, and the cumulative distribution for each host-galaxy sample.
 }
\label{fig:massmassdensity}
\end{figure*}

\begin{figure*}
\centering
\subfigure{\includegraphics[angle=0,width=5in]{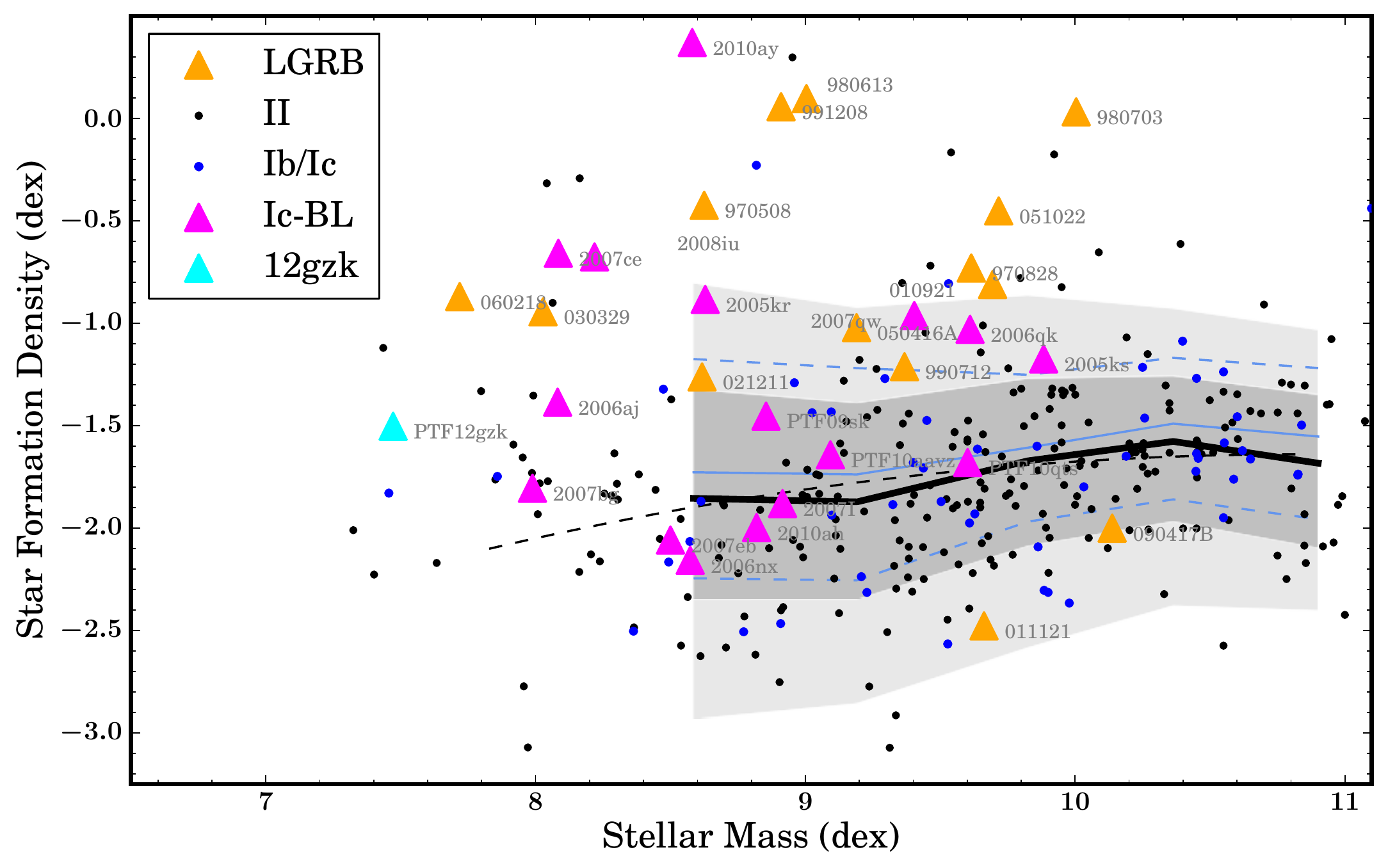}}
\subfigure{\includegraphics[angle=0,width=5in]{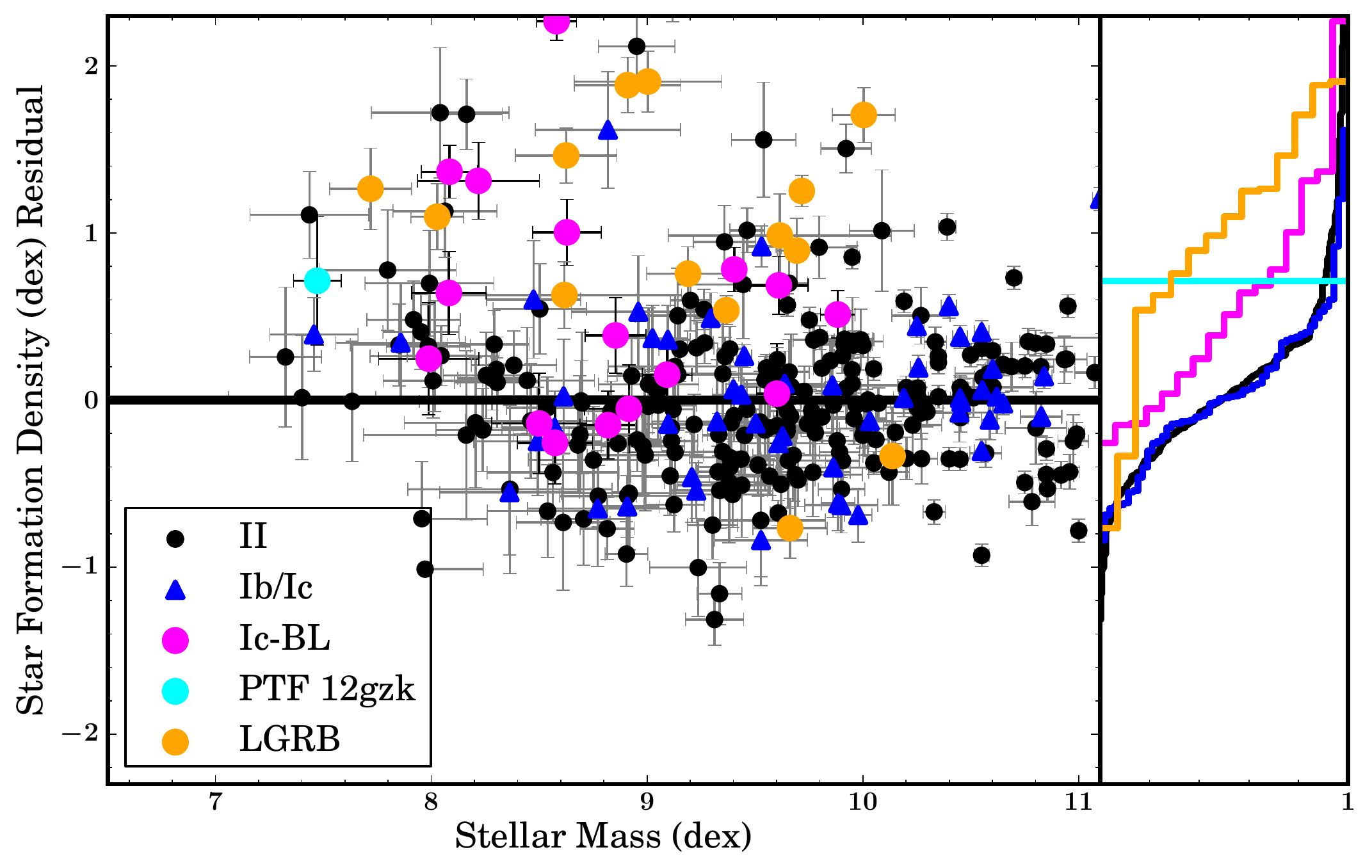}}
\caption{As in Figure~\ref{fig:massmassdensity}, but showing the star-formation density ($\Sigma_{\rm SFR}$) against 
host stellar mass for $z<0.2$ core-collapse SN as well as $z<1.2$ LGRBs. Host galaxies of broad-lined SN~Ic and 
LGRBs show high star-formation densities for their stellar masses.
 }
\label{fig:masssfrdensity}
\end{figure*}

\begin{figure*}
\centering
\subfigure{\includegraphics[angle=0,width=5in]{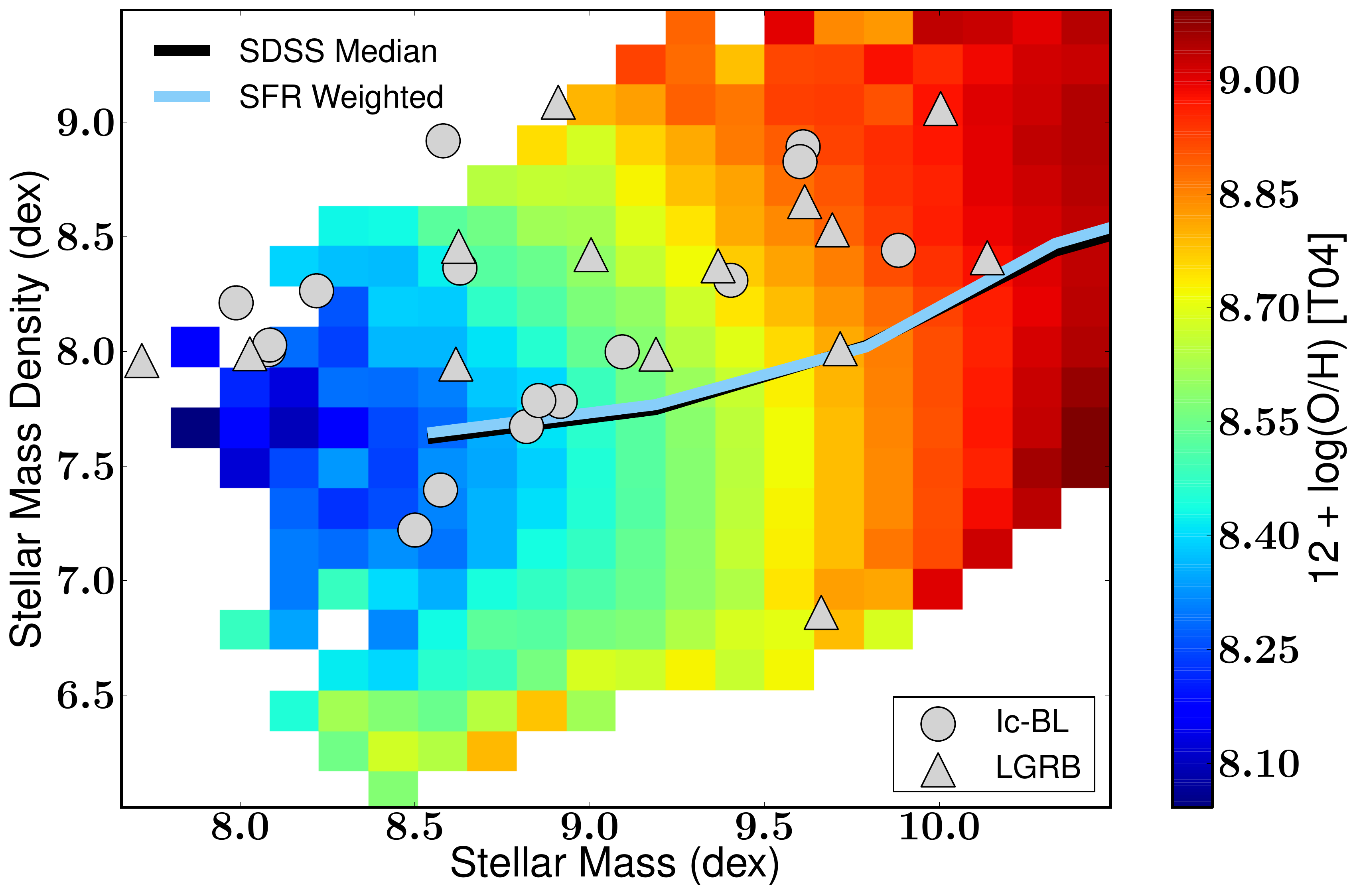}}
\subfigure{\includegraphics[angle=0,width=5in]{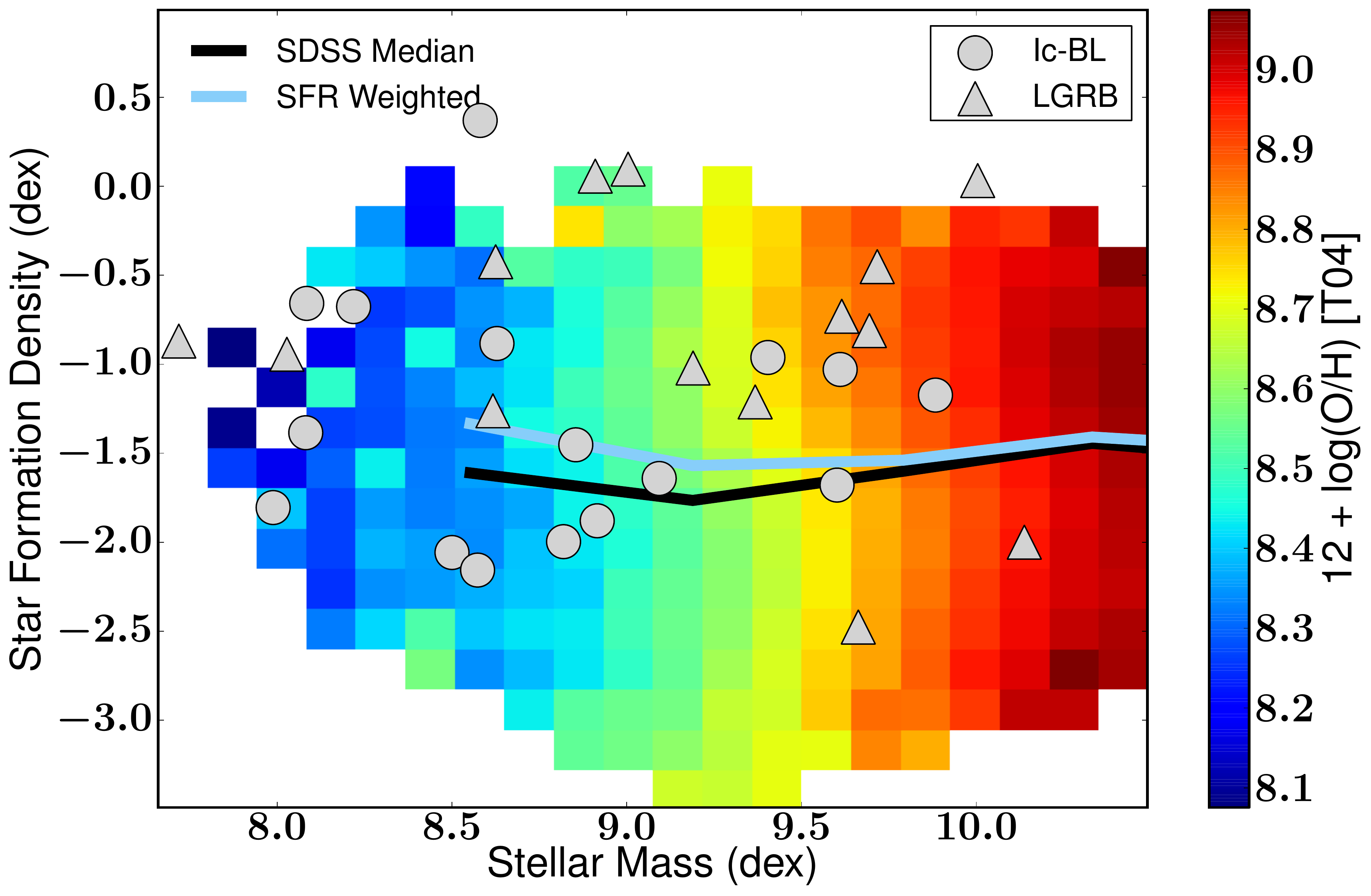}}
\caption{Variation in mean SDSS oxygen abundance as a function of 
galaxy stellar mass ($M$) and stellar-mass density ($\Sigma_{M}$) as well as 
star-formation density ($\Sigma_{\rm SFR}$). 
The host galaxies of nearby SN~Ic-BL (grey circles) and $z < 1.2$ LGRBs (grey triangles) are overplotted.
Among galaxies with masses typical of those
of SN~Ic-BL and LGRB hosts ($\lesssim 10^{10}$\,M$_{\odot}$), galaxies with high $\Sigma_{M}$ have (on average)
modestly higher abundances, while those with high $\Sigma_{\rm SFR}$ are at least as metal-rich as those with low $\Sigma_{\rm SFR}$. 
Solid black and blue lines are as in Figure~\ref{fig:massmassdensity}.
If the production SN Ic-BL and LGRB progenitors depended only on metallicity (and increased at low abundance), 
we would expect them to occur more frequently among the metal-poor galaxies below the blue SFR-weighted $M-\Sigma_{M}$ relation, and
approximately equally above and below the blue SFR-weighted $M-\Sigma_{\rm SFR}$ relation.
Since their actual distribution differs strongly from this prediction, 
this suggests that the SN Ic-BL and LGRB rate is enhanced by a factor other than low metal abundance that
varies with $\Sigma_{M}$ and $\Sigma_{\rm SFR}$.
An average number of 520 SDSS galaxies populate each plotted bin, and the average uncertainty of the mean 
abundance is 0.022 dex. 
Average uncertainties of plotted SN~Ic-BL and LGRB host galaxy stellar-mass and star-formation-rate densities are 0.18 and 0.19 dex, respectively (see Figures~\ref{fig:massmassdensity} and \ref{fig:masssfrdensity}). 
}
\label{fig:sdsspop}
\end{figure*}

\begin{figure*}
\centering
\subfigure{\includegraphics[angle=0,width=5in]{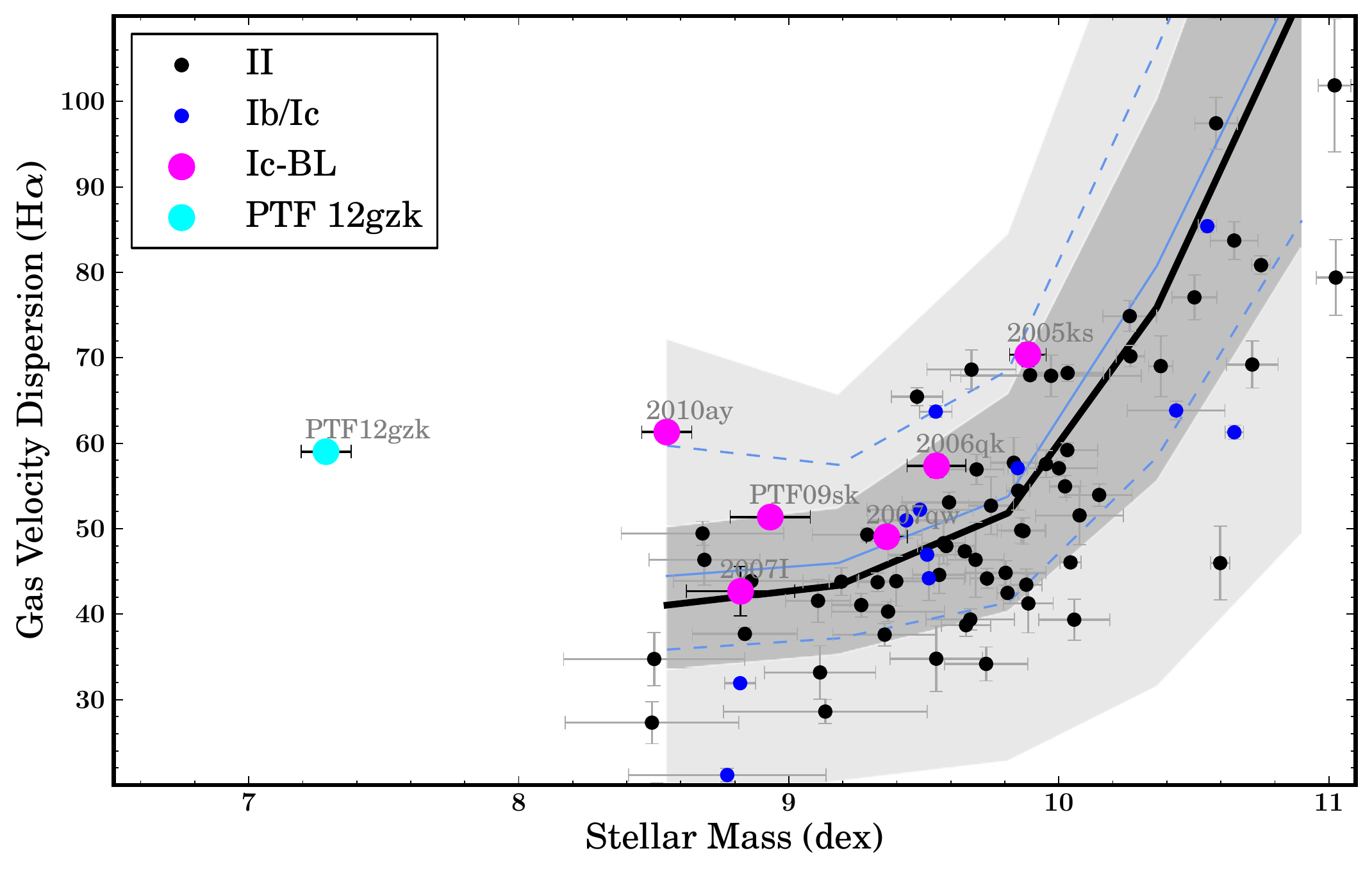}}
\subfigure{\includegraphics[angle=0,width=5in]{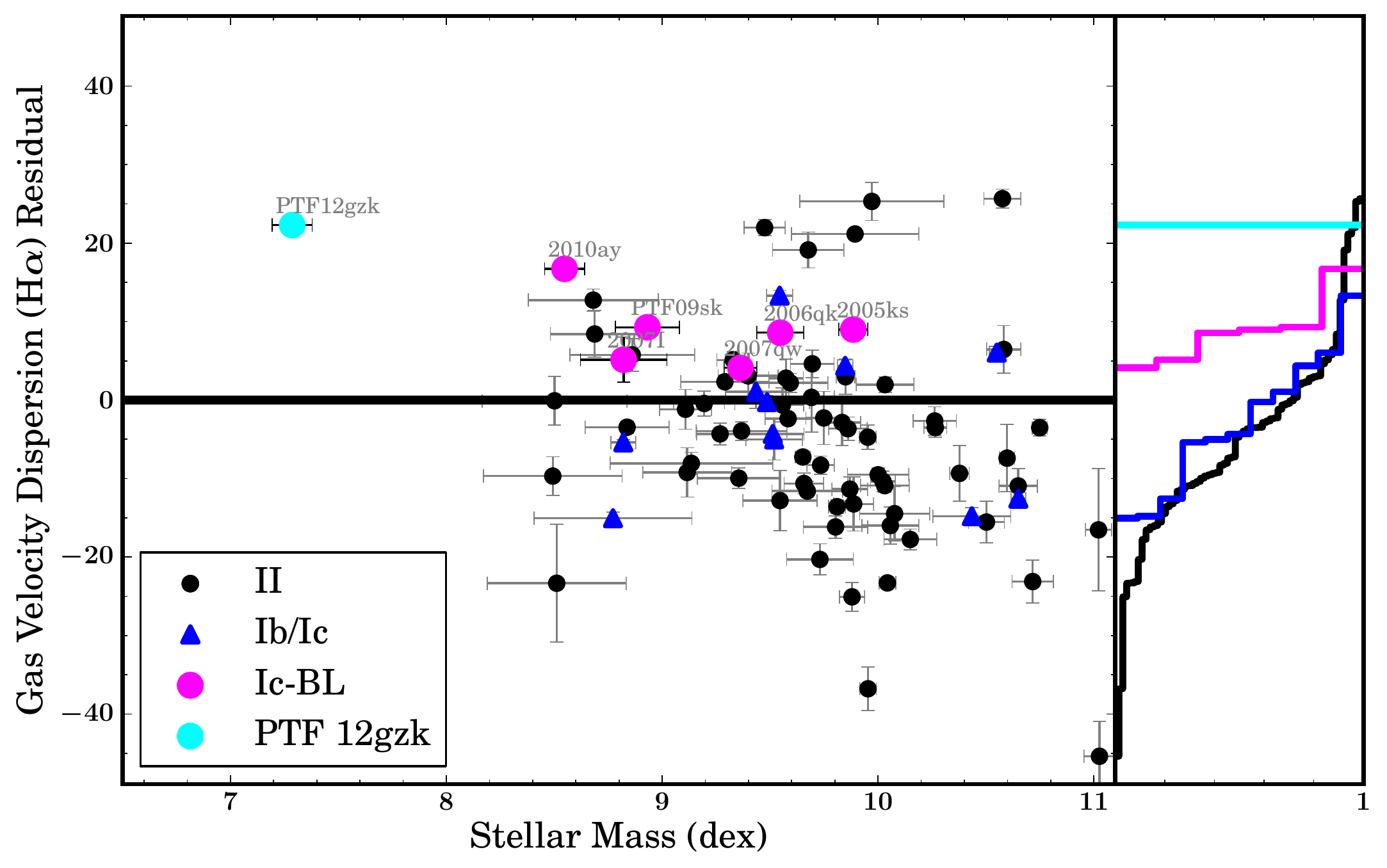}}
\caption{Host-galaxy gas velocity dispersion inferred from the H$\alpha$ emission-line profile against host stellar mass.
Host galaxies of broad-lined SN~Ic (magenta) as well as that of PTF12gzk show high gas velocity dispersion.
Residuals plotted below are from the values predicted from locally weighted linear least-squares fits performed on the SDSS catalog. 
At bottom right, we show the cumulative distribution for each SN
sample.
}
\label{fig:halphaveldisp}
\end{figure*}

\begin{figure*}[htp!]
\centering
\subfigure{\includegraphics[angle=0,width=6in]{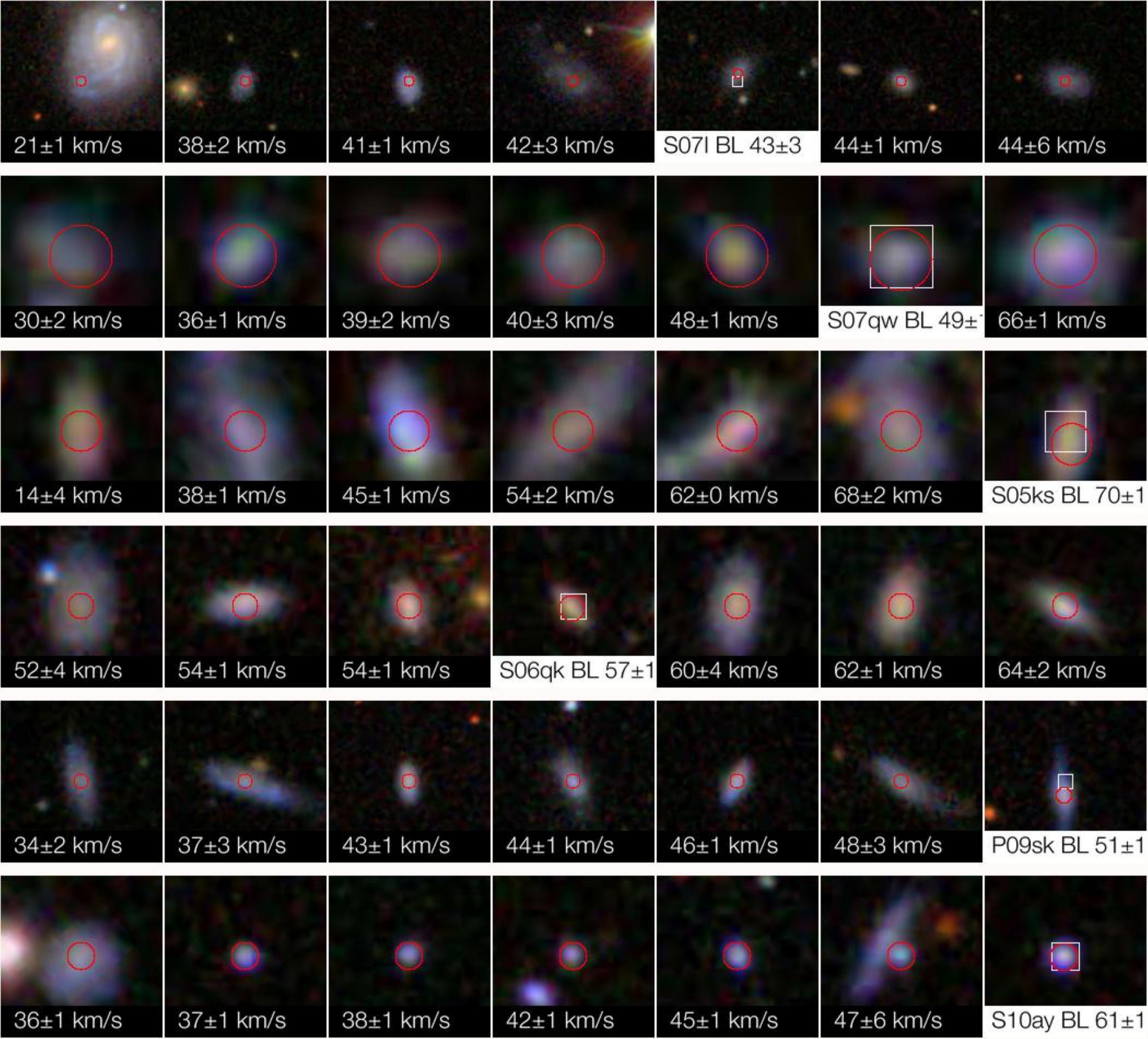}}
\caption{Comparison between the host galaxies of broad-lined SN~Ic and SDSS galaxies with similar
\laundry. 
Each row, ordered according the gaseous velocity dispersion estimated from the H$\alpha$ emission line, 
shows the host galaxy of a broad-lined SN~Ic and the six nearest neighbors in the vector space \vecspace.
These comparisons show that the host galaxies of broad-lined SN~Ic have comparatively high velocity 
dispersions for their host-galaxy stellar masses. 
Red circles represent the $3''$ SDSS fiber aperture, while the white squares are positioned at the location of the
SN explosion.
From the top to the bottom row, host galaxies are those of SN 2007I, SN 2007qw, SN 2005ks, SN 2006qk, PTF 09sk, and SN 2010ay.
}
\label{fig:icblmosaic}
\end{figure*}

\begin{figure*}[htp!]
\centering
\subfigure{\includegraphics[angle=0,width=6in]{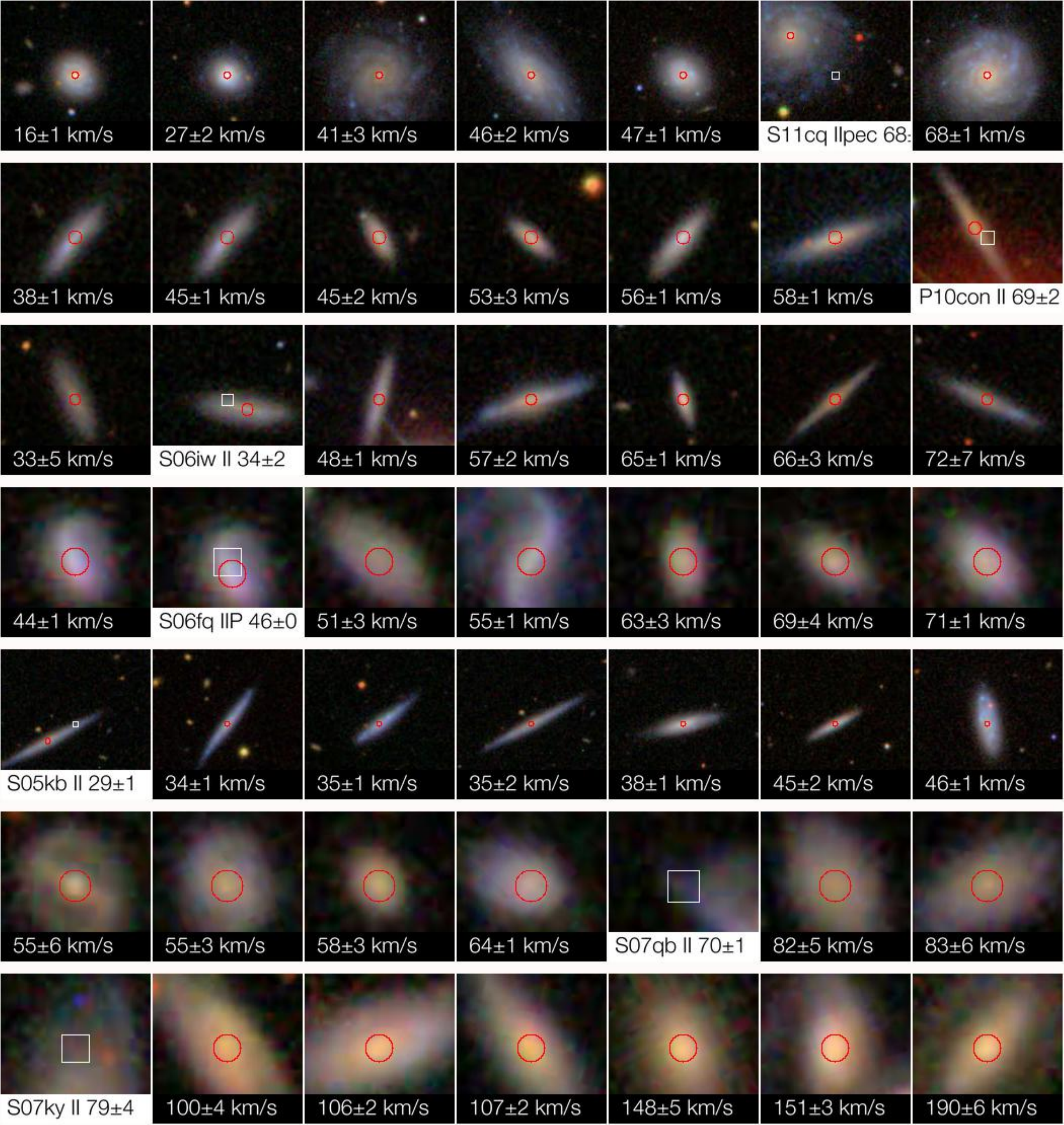}}
\caption{Same as  Figure~\ref{fig:icblmosaic}, but for SN~II host galaxies.
In contrast to the hosts of SN~Ic-BL, the host galaxies of SN~II have $\sigma_{\rm vel}$ 
measurements that are both low and high in comparison to galaxies having similar \laundry. 
From the top to the bottom row, host galaxies are those of SN 2011cq, PTF 10con, SN 2006iw, SN 2006fq, SN 2005kb, SN 2007qb, and SN 2007ky.
}
\label{fig:othermosaic}
\end{figure*}

\begin{figure*}
\centering
\subfigure{\includegraphics[angle=0,width=5in]{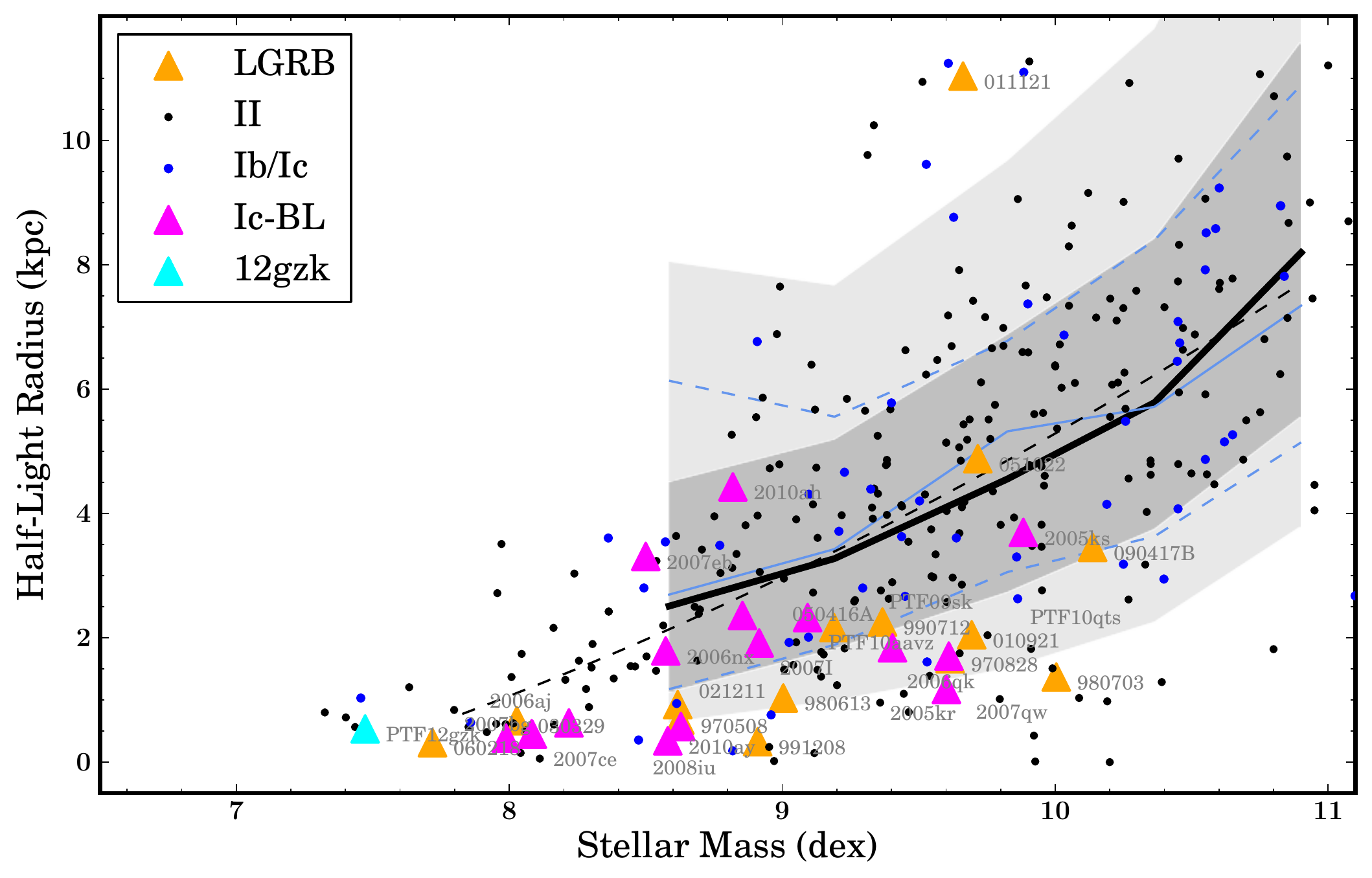}}
\subfigure{\includegraphics[angle=0,width=5in]{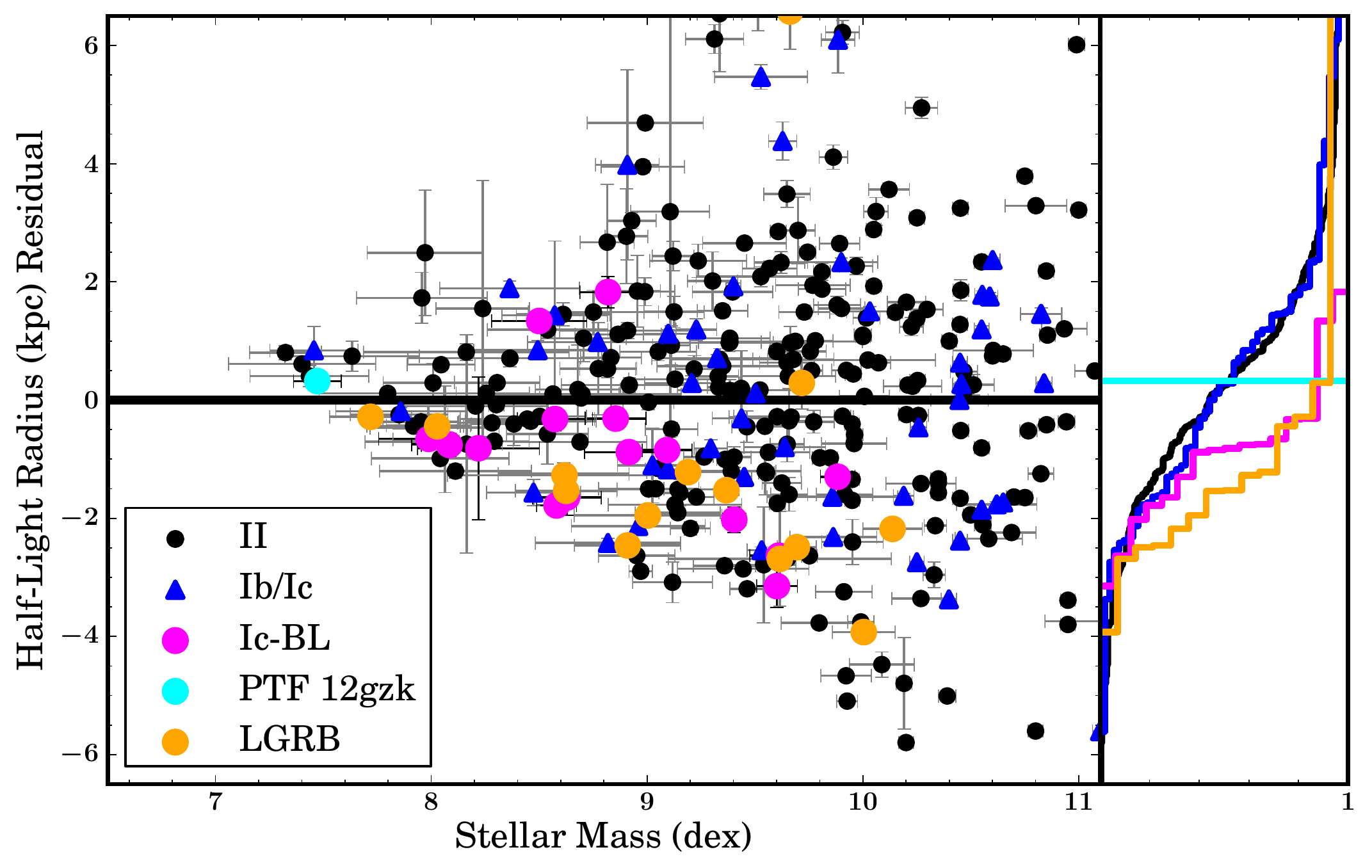}}
\caption{As in Figure~\ref{fig:massmassdensity}, but showing the effective host-galaxy radius ($r_{50}$) against host SFR for $z < 0.2$ core-collapse SN as well as $z<1.2$ LGRBs. Host galaxies of broad-lined SN~Ic and LGRBs are compact for their stellar masses.
 }
\label{fig:masshl}
\end{figure*}

\begin{figure*}
\centering
\subfigure{\includegraphics[angle=0,width=5in]{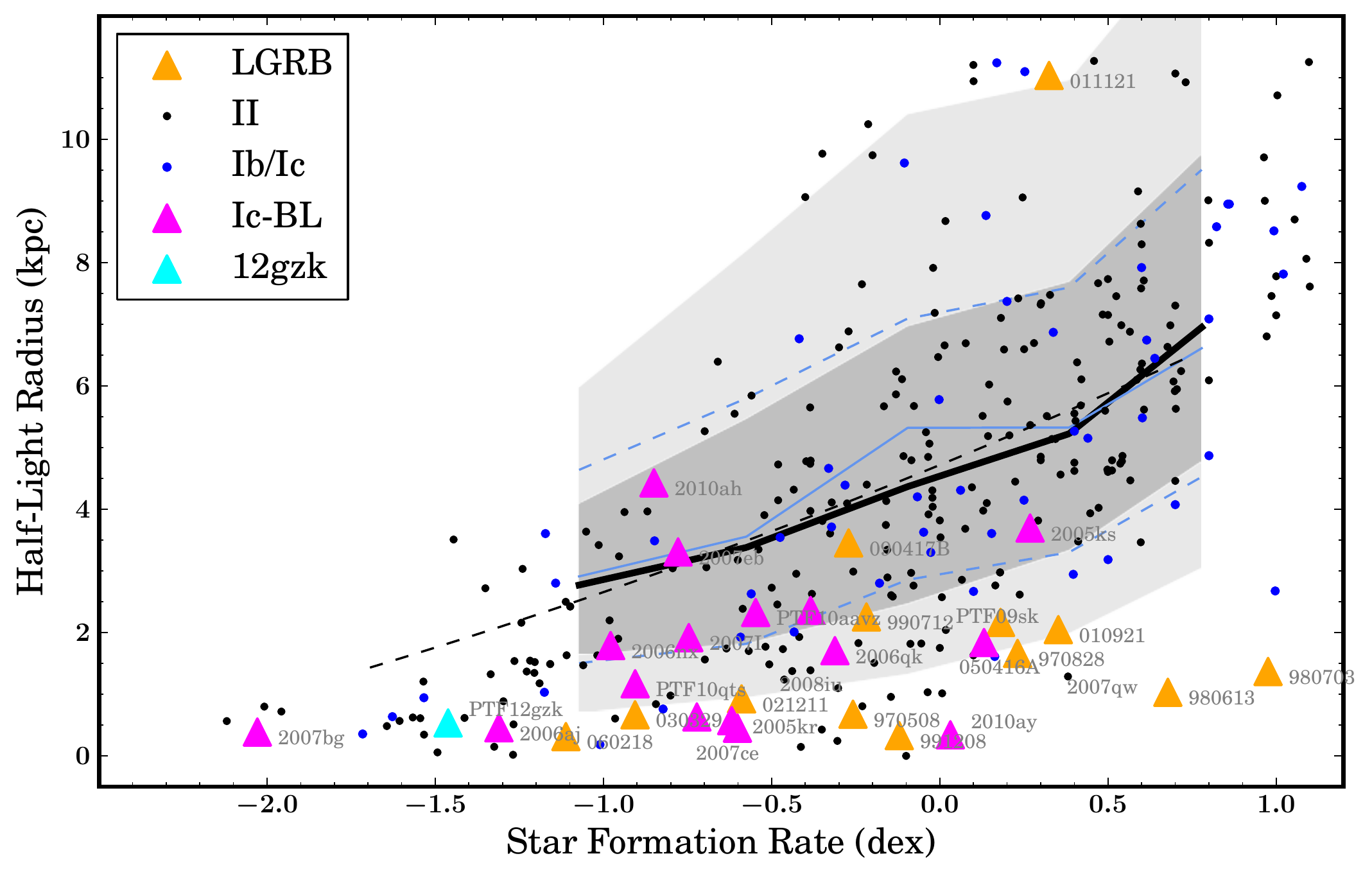}}
\subfigure{\includegraphics[angle=0,width=5in]{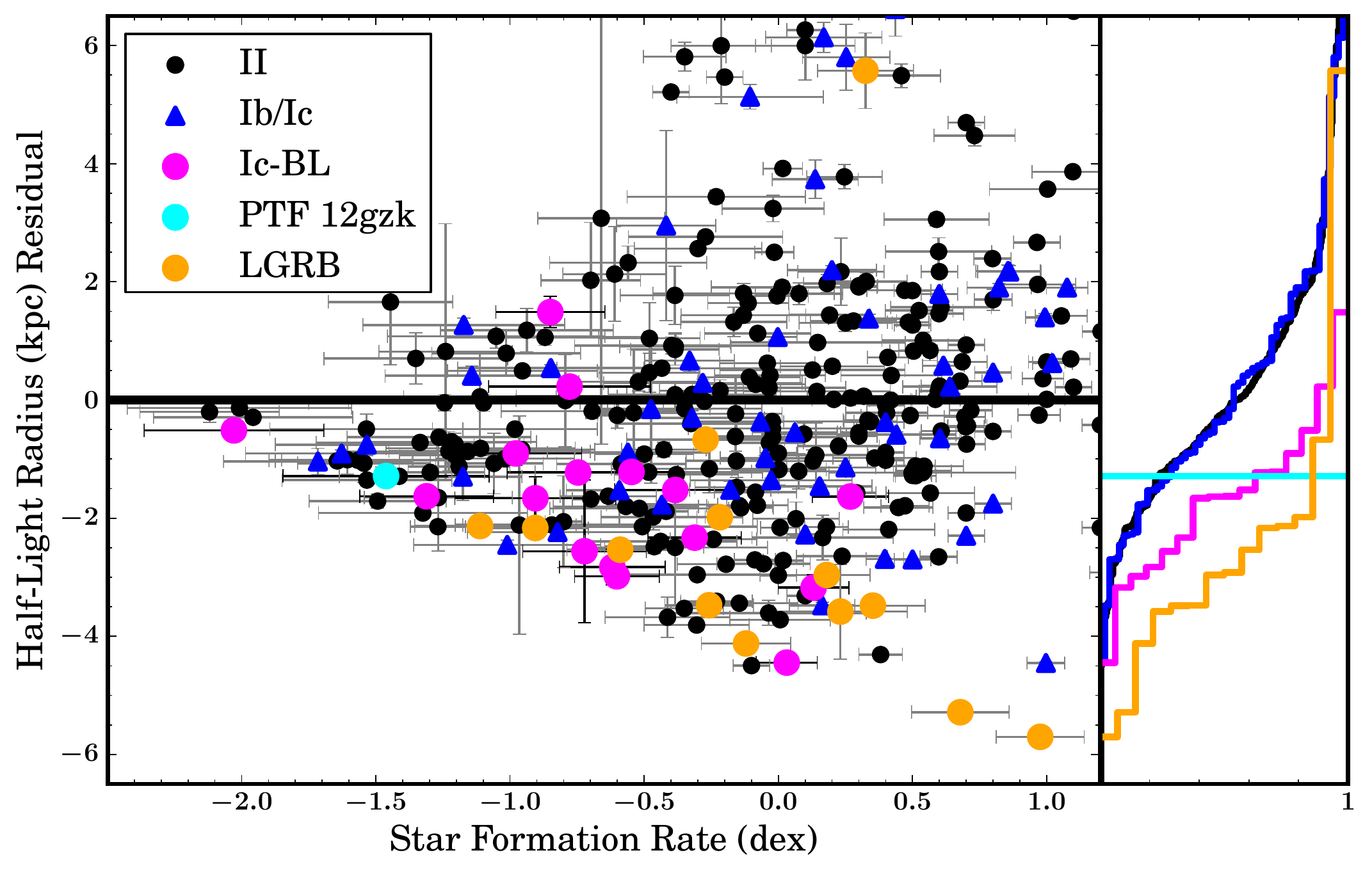}}
\caption{As in Figure~\ref{fig:massmassdensity}, but showing the effective host-galaxy radius ($r_{50}$) against host SFR for $z<0.2$ core-collapse SN as well as $z<1.2$ LGRBs. Host galaxies of broad-lined SN~Ic and LGRBs are compact for their SFR.
 }
\label{fig:sfrhl}
\end{figure*}

\section{Results}
\label{sec:results}

\subsection{Host-Galaxy $\Sigma_{M}$ and $\Sigma_{\rm SFR}$}
In Figures~\ref{fig:massmassdensity} and~\ref{fig:masssfrdensity}, we plot galaxy stellar-mass density $\Sigma_{M}$ and star-formation density $\Sigma_{\rm SFR}$, estimated from broadband magnitudes, against stellar mass $M$.
These show that $z<0.2$ SN~Ic-BL and $z < 1.2$~LGRB host galaxies have high stellar-mass density and star-formation density for their stellar masses, compared with the low-redshift ($z<0.2$) SDSS star-forming galaxy population.
In contrast, SN~Ib/Ic (with slower ejecta speeds) and SN~II show no preference for galaxies that have high stellar-mass density or high star-formation density for their stellar masses. 

We compute the residuals of each host galaxy from the SDSS $M$--$\Sigma_{M}$ and $M$--$\Sigma_{\rm SFR}$ relations.
Our statistical method is to determine the probability that each pair of residual distributions is identical using the Kolmogorov-Smirnov (KS) two-sample test.
We find significant evidence for differences between the SN~Ic-BL ($n=$ \KSnonomasspegmasspegdensityrsnCCohfullnumIcBL) and the SN~II ($n=$ \KSnonomasspegmasspegdensityrsnCCohfullnumII; \KSnonomasspegmasspegdensityrsnCCohfullresidualGloIIIcBL~and \KSnonomasspegsfrpegdensityrsnCCohfullresidualGloIIIcBL~for 
$\Sigma_{M}$ and $\Sigma_{\rm SFR}$, respectively) host distributions.  
Comparison between the SN~Ic-BL and SN~Ib/Ic
($n=$ \KSnonomasspegmasspegdensityrsnCCohfullnumIbIc) host distributions likewise finds evidence for distinct underlying distributions  (\KSnonomasspegmasspegdensityrsnCCohfullresidualGloIbIcIcBL~and 
\KSnonomasspegsfrpegdensityrsnCCohfullresidualGloIbIcIcBL).

The residuals of LGRB ($n=$ \KSnonomasspegmasspegdensityrsnCCohfullnumLGRB) host galaxies from the SDSS $M$--$\Sigma_{M}$ and $M$--$\Sigma_{\rm SFR}$ relations are more positive than (i.e., $\Sigma_{M}$ and $\Sigma_{\rm SFR}$ are greater than) those of SN~Ib/Ic (\KSnonomasspegmasspegdensityrsnCCohfullresidualGloIbIcLGRB~and \KSnonomasspegsfrpegdensityrsnCCohfullresidualGloIbIcLGRB, respectively) and SN~II (\KSnonomasspegmasspegdensityrsnCCohfullresidualGloIILGRB~and \KSnonomasspegmasspegdensityrsnCCohfullresidualGloIILGRB) hosts,
while we find no statistically significant difference with the SN~Ic-BL host residual distribution (\KSnonomasspegmasspegdensityrsnCCohfullresidualGloIcBLLGRB~and \KSnonomasspegsfrpegdensityrsnCCohfullresidualGloIcBLLGRB).
The galaxy $M$--$\Sigma_{M}$ relation shows no significant
change with increasing redshift to $z\approx1.1$  (e.g., \citealt{bardenrix05}), so 
comparisons involving the LGRB host  $M$--$\Sigma_{M}$ relation should not be strongly affected by evolution in the galaxy population with redshift. 

\subsection{Preference for Overdense Galaxies}

We study the $z<0.2$ SDSS star-forming population to investigate whether
galaxies with relatively high stellar-mass and star-formation densities have comparatively low chemical abundances.
In Figure~\ref{fig:sdsspop}, we show the average \citet{tre04} oxygen abundance of $z<0.2$ SDSS star-forming galaxies as a function of $M$ and $\Sigma_{M}$ as well as $\Sigma_{\rm SFR}$.  
The star-forming SDSS galaxies with the highest stellar-mass or star-formation-rate densities (in each stellar-mass bin) are not, on average,  
comparatively metal poor. The galaxies with the highest stellar-mass densities 
are, on average, more metal-rich by $\sim$0.2 dex.

An important question is whether the high stellar-mass and
star-formation-rate densities of SN~Ic-BL and $z < 1.2$ LGRB host galaxies can be attributed entirely to 
the preference for low metal abundance observed among low-redshift SN~Ic-BL \citep{kelkir12,sanderssoderberg12} and LGRBs \citep{mod08,grahamfruchter13}. 
We can represent such a preference as a
progenitor formation efficiency $\eta$ per unit stellar mass created that depends solely on metallicity $Z$
and diminishes with increasing $Z$.
If $\eta$ is a function of only metallicity $Z$, then 
the rate $r$ of fast-ejecta transients in a galaxy with a given SFR is
\begin{equation}
r \propto \eta(Z) \times {\rm SFR}.
\end{equation}

Half of the star formation in each galaxy mass $M$ bin plotted in Figure~\ref{fig:sdsspop} 
occurs in galaxies below the $M-\Sigma_{M}$ and $M-\Sigma_{\rm SFR}$
relations marked by thick blue lines.
A transient population whose $\eta$ has no environmental dependence would be expected to 
explode equally in galaxies above and below these SFR-weighted relations. 
If $\eta$ instead increases at low abundance (and depends only on $Z$), we would 
expect greater numbers of SN Ic-BL and $z < 1.2$ LGRBs on the side of the 
relation, either above or below, that is comparatively metal poor.

Inspection of Figure~\ref{fig:sdsspop} shows that, on average, higher stellar-mass density corresponds to higher metallicity, 
while metallicity does not vary strongly with star-formation density among galaxies in each bin in stellar mass. 
Consequently, if $\eta$ increases at low abundance (and depends only on $Z$), we would expect a plurality of SN Ic-BL and $z < 1.2$ LGRBs to be below
the SFR-weighted $M-\Sigma_{M}$ relation, and approximately equal numbers above and below 
the SFR-weighted $M-\Sigma_{\rm SFR}$ relation.
Since this prediction is very different from the pattern we observe, the production of SN Ic-BL and LGRB progenitors much be enhanced 
by an additional factor other than low metallicity that varies with $\Sigma_{M}$ as well as $\Sigma_{\rm SFR}$.

\subsection{Host-Galaxy Gaseous Velocity Dispersions}
High gas dispersion velocities provide additional, independent evidence for distinct physical conditions in the host galaxies of low-redshift SN~Ic-BL.
In Figure~\ref{fig:halphaveldisp}, the SN~Ic-BL hosts describe an $M$--$\sigma_{\rm gas}$ relation with an offset to high gas velocity dispersion measured from the H$\alpha$ emission lines in the SDSS fiber spectra.
The SN~Ic-BL host distribution is significantly different from the SN~Ib/Ic (\KSnonomasspegsigmahalphasnCCohfullresidualGloIbIcIcBL) and the SN~II (\KSnonomasspegsigmahalphasnCCohfullresidualGloIIIcBL) host distributions. 

The comparatively high gas velocity dispersions of SN~Ic-BL host galaxies may have one or more physical explanations. 
Rotationally supported galaxies having more compact and dense mass configurations are expected to have higher velocity dispersions.
Alternatively, recent analysis of the gas kinematics of a sample of $z \approx 0.1$--0.3 compact, highly star-forming galaxies finds complex profiles for the strong emission lines, with several narrow ($\sigma_{\rm gas} \approx 10$--120\,km\,s$^{-1}$) and broad ($\sigma_{\rm gas} \approx 100$--250\,km\,s$^{-1}$) components \citep{amorinvilchez12} that may be due to strong stellar winds, or emission from expanding
SN remnants.

In  Figures~\ref{fig:icblmosaic} and \ref{fig:othermosaic}, we show images of core-collapse SN 
host galaxies and the measured H$\alpha$ velocity dispersions juxtaposed with control
samples of SDSS star-forming galaxies having similar parameters.  These demonstrate the high velocity 
dispersions of SN~Ic-BL host galaxies.

\begin{figure*}[h!]
\centering
\subfigure{\includegraphics[angle=0,width=5in]{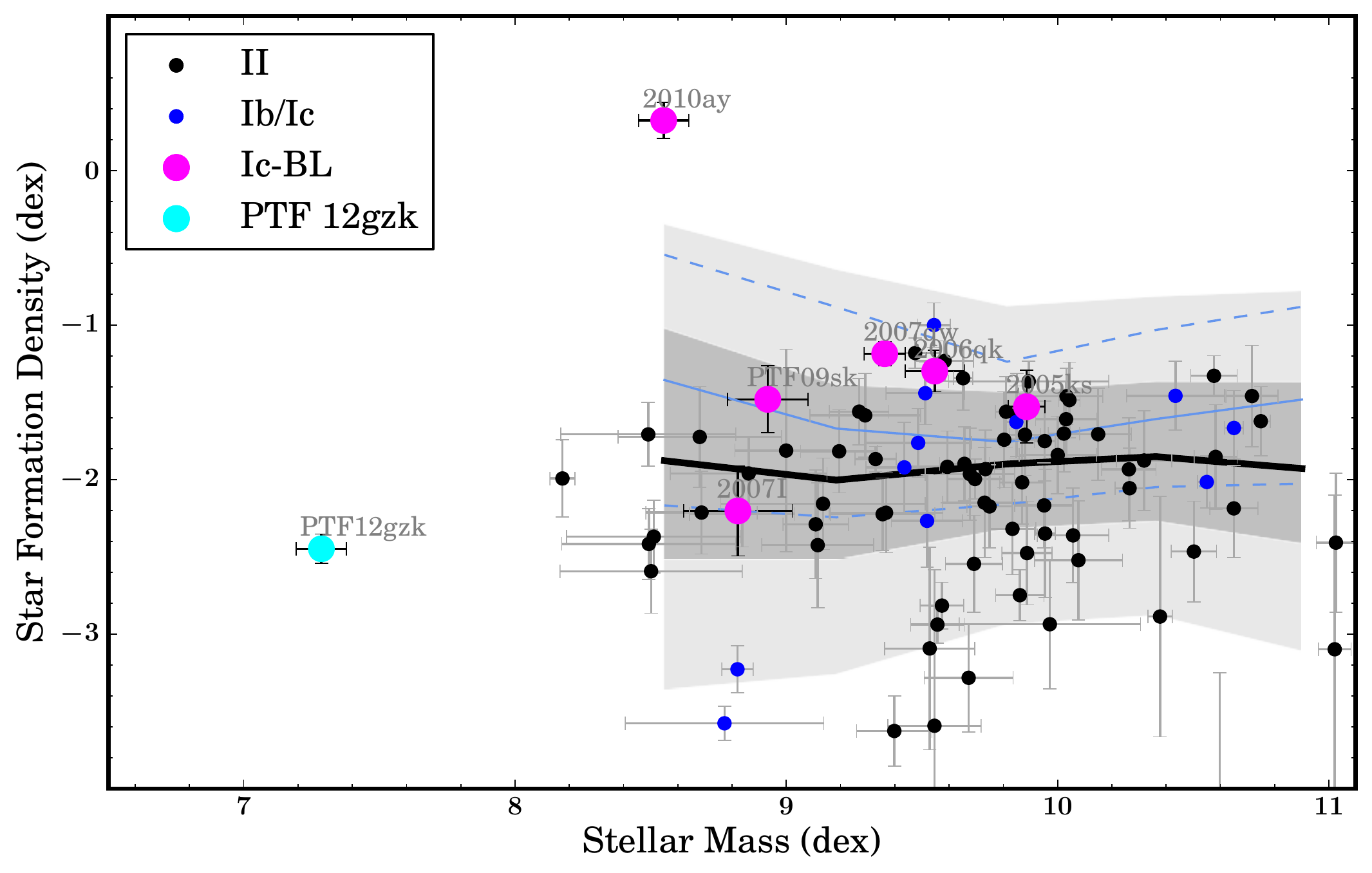}}
\subfigure{\includegraphics[angle=0,width=5in]{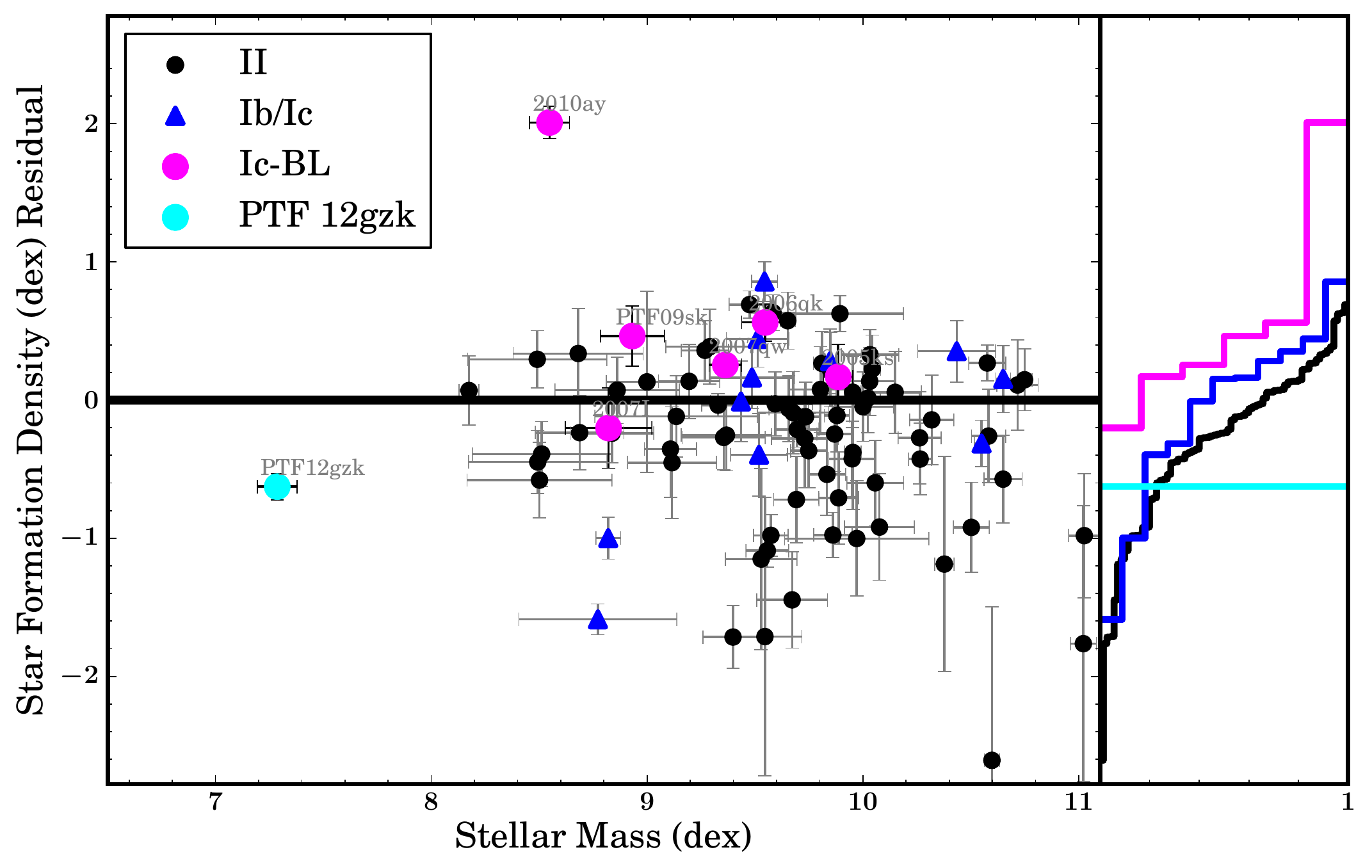}}
\caption{As in Figure~\ref{fig:massmassdensity}, but showing host galaxy hybrid spectroscopic star-formation density against stellar mass.
The star-formation density $\Sigma_{\rm SFR}$ plotted here is calculated using an SFR estimate using
the SDSS fiber spectrum and by modeling the host light outside of the 3$''$ fiber.
The SN~Ic-BL host galaxies show high $\Sigma_{\rm SFR}$ for their stellar masses, which is consistent with the pattern observed with $\Sigma_{\rm SFR}$ estimated from photometry alone.
}
\label{fig:sfrspecdensity}
\end{figure*}

\begin{figure*}[h!]
\centering
\subfigure{\includegraphics[angle=0,width=5in]{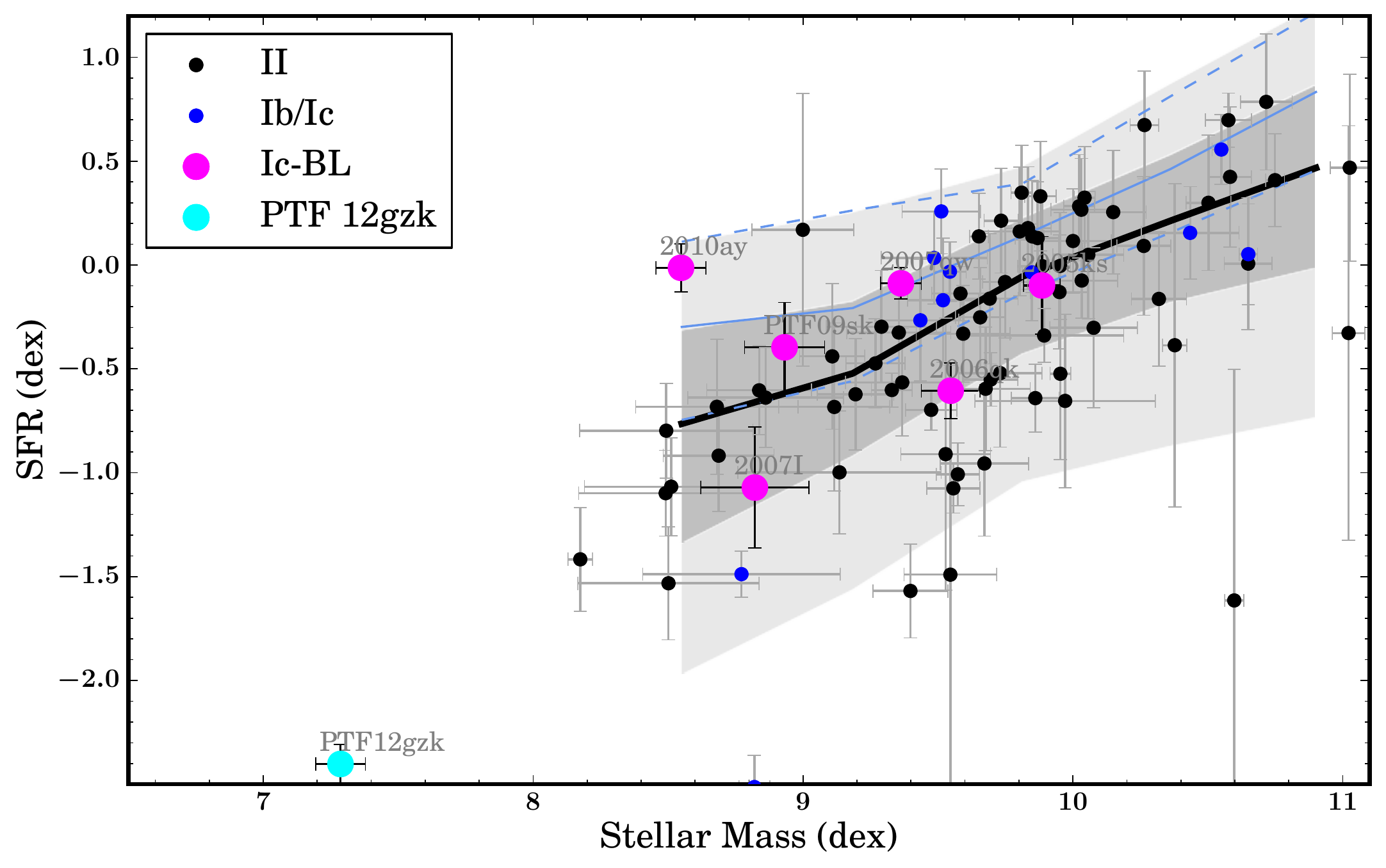}}
\subfigure{\includegraphics[angle=0,width=5in]{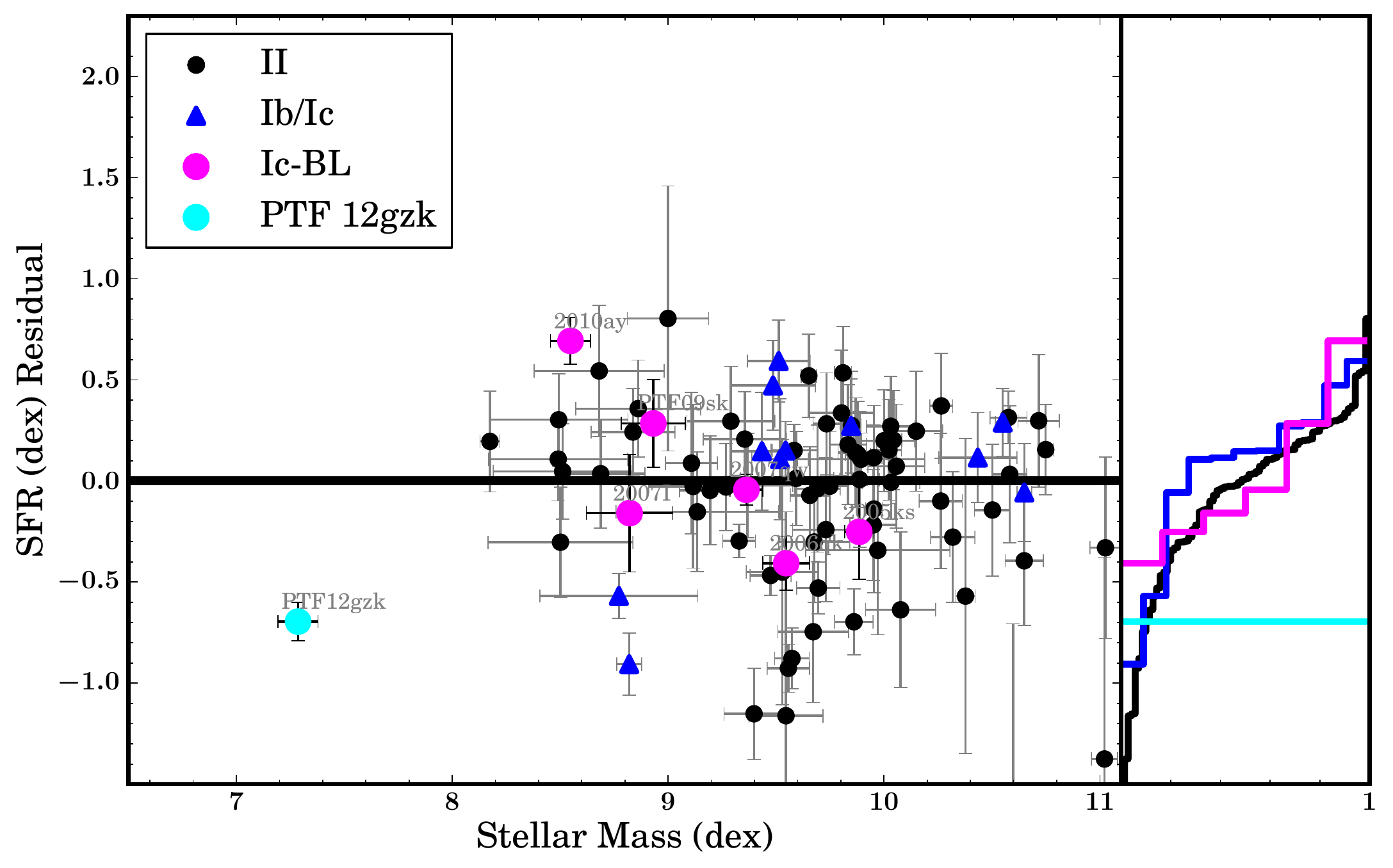}}
\caption{As in Figure~\ref{fig:massmassdensity}, but showing SFR against host stellar mass for $z < 0.2$ core-collapse SN.
The host galaxies of SN~Ic-BL exhibit an approximate agreement with the SDSS $M$--SFR relation.
}
\label{fig:msfr}
\end{figure*}

\subsection{Host-Galaxy Sizes} In Figures~\ref{fig:masshl} and~\ref{fig:sfrhl}, we plot galaxy half-light radius $r_{50}$ (in kpc) against $M$ and SFR estimated from their broadband magnitudes.
These show that $z < 0.2$ SN~Ic-BL and $z < 1.2$~LGRB host galaxies are compact for their stellar masses and SFRs, compared with the low-redshift ($z < 0.2$) SDSS star-forming galaxy population. In contrast, SN~Ib/Ic (with slower ejecta speeds) and SN~II show no preference for galaxies that are relatively compact in size. 

\subsection{Hybrid $\Sigma_{\rm SFR}$ Estimates}

In  Figure~\ref{fig:sfrspecdensity}, we plot $\Sigma_{\rm SFR}$ against $M$
 for the sample of nearby ($z < 0.2$) core-collapse SN 
galaxies with SDSS spectra. 
Here we calculate $\Sigma_{\rm SFR}$ using the SFR estimated by the MPA-JHU group 
from the fiber spectrum and by modeling the galaxy {\it ugriz} light outside of the 3$''$ fiber. 
These SFR measurements may be expected to be more accurate and precise than the SFR estimates 
that are based on galaxy broadband photometry alone, and provide evidence that the 
preference among SN~Ic-BL and LGRBs for high $\Sigma_{\rm SFR}$ seen in Figure~\ref{fig:masssfrdensity}
galaxies is robust. The SN~Ic-BL ($n=$ \KSnonomasspegsfrmpadensityrsnCCohfullnumIcBL) and SN~II ($n=$ \KSnonomasspegsfrmpadensityrsnCCohfullnumII) 
distributions show significantly different ($p=$ \KSnonomasspegsfrmpadensityrsnCCohfullresidualRelIIIcBL) residual distributions from 
their predicted $\Sigma_{\rm SFR}$ values.

\subsection{$M$--SFR Relation}
 Figure~\ref{fig:msfr} shows the relationship between $M$ and SFR for 
the core-collapse host galaxies and the SDSS star-forming population. 
This shows that host galaxies of SN~Ic-BL are not substantially more 
strongly star forming for their stellar masses than other core-collapse hosts, and suggests that relatively compact host sizes
may primarily account for their high star-formation-rate densities.

\subsection{$M$--$Z$ Relation} 
Given the existing evidence that $z \lesssim 0.3$ SN~Ic-BL \citep{kelkir12,sanderssoderberg12} and  LGRBs \citep{mod08,grahamfruchter13} prefer metal-poor environments, a reasonable question is whether the high $\Sigma_{M}$ and $\Sigma_{\rm SFR}$ host galaxies in our sample having SDSS spectra describe a metal-poor mass-metallicity ($M$--$Z$) relation. 
Kelly et al. (2014, in prep.) report no significant evidence that the SN~Ic-BL offset distribution ($n=$ \KSnonomasspegtohfoursnCCohfullnumIcBL) differs from the SN~Ib/Ic (\KSnonomasspegtohfoursnCCohfullresidualGloIbIcIcBL; $n=$ \KSnonomasspegtohfoursnCCohfullnumIbIc) or the SN~II
(\KSnonomasspegtohfoursnCCohfullresidualGloIIIcBL; $n=$ \KSnonomasspegtohfoursnCCohfullnumII) distributions.

\section{Discussion}
\label{sec:discussion}
We have found that low-redshift SN~Ic-BL and $z<1.2$ LGRB host galaxies show stellar-mass and star-formation densities high compared with those of low-redshift galaxies having similar stellar masses. 
Core-collapse SN with more slowly expanding ejecta, however, exhibit no preference for
galaxies having overdense stellar-mass distributions and star formation. 
SN~Ic-BL host galaxies exhibit high gas velocity dispersions for their stellar masses, providing evidence for
exceptional conditions in their hosts.

Given the effect of stellar metallicity to strengthen line-driven winds and remove the angular momentum of 
massive stars, studies of SN and LGRB host galaxies have often attempted to explain environmental patterns in terms of chemical abundance. 
However, across the stellar-mass range populated by SN~Ic-BL and LGRB hosts, SDSS star-forming galaxies with high stellar-mass and star-formation-rate densities are not, on average, more metal poor than 
less dense galaxies having similar stellar masses (see Figure~\ref{fig:sdsspop}). 
Therefore, a preference other than for low metallicity environments must be responsible for the 
overrepresentation of SN and LGRBs in galaxies with high stellar-mass and star-formation-rate densities.
Furthermore, the $M$--$Z$ relation of SN~Ic-BL host galaxies, which have high stellar-mass and star-formation densities, is not significantly metal poor compared to the SDSS $M$--$Z$ relation (Kelly et al. 2014, in prep.).  

In fact, a preference of high-velocity explosions for rapidly star-forming or overdense regions may help explain the observed association of LGRBs with the brightest regions of their host galaxies \citep{fru06}.
An initial suggestion was that a strong association with the brightest regions would be expected for LGRBs if they occur preferentially in low-mass, metal-poor galaxies.
If the most massive stars form in OB associations, then these will 
be more likely to correspond to the peaks of the light distribution of low-mass, metal-poor
galaxies. OB associations in more massive, metal-rich spirals may instead be
outranked in brightness by, for example, the nucleus. While such a low-metallicity preference may contribute, we suggest that an affinity to high star-formation density is important. 

A plausible explanation for the association of LGRBs with regions of higher star-formation 
density is the formation efficiency of young, bound star clusters. 
Observations of extragalactic star clusters have found evidence that bound-cluster formation efficiency increases with the star-formation density \citep{goddard10,silvavillaadamo13}.
Binary systems may be created more frequently in bound clusters, and those that form are expected to become progressively tighter through dynamical interactions with other members of the cluster \citep{heggie75,hutmcmillan92}.
Therefore, massive stars that form in dense star clusters are more likely to 
be in tight binary systems. 
Interacting massive binaries are candidate progenitor systems for
SN~Ic-BL and LGRBs, because mass transfer or common-envelope evolution leading (in some cases) to a merger
can yield a rapidly rotating star whose outer envelope is not composed of H.

If SN~Ic-BL and LGRBs explode from stars that are more massive than the progenitors of SN~Ib/Ic and SN~II, an IMF that becomes top-heavy in dense, highly star-forming regions
provides an alternative explanation for the patterns we observe.
Indirect evidence from the absorption features of  $\lesssim0.3\,{\rm M}_{\odot}$ stars suggests instead that the 
IMF may be bottom-heavy in elliptical galaxies with high stellar velocity dispersions and [Mg/Fe], which are thought to have had high star formation densities during their formation epoch \citep{vandokkumconroy10,conroyvandokkum12}.
This suggests that the IMF may be bottom heavy in dense, highly star-forming regions, while the opposite trend would be required to explain our observations. 
An improved census of massive stars across low-redshift environments may help to 
address more directly whether the patterns we have found can be explained by variation in the upper end of the IMF.

The median redshift ($z_{\rm m}=\lgrbmedian$) of the $z < 1.2$ LGRB sample is
substantially higher than those of the core-collapse SN sample ($z_{\rm m} \approx $ \ccmedian) and the 
SDSS sample of galaxy spectra ($z_{\rm m} \approx 0.08$).
The \mbox{$M$--$\Sigma_{M}$} and \mbox{$M$--$\Sigma_{\rm SFR}$} relations among star-forming galaxies may 
evolve toward lower $\Sigma_{M}$ and $\Sigma_{\rm SFR}$ densities 
from $z \approx \lgrbmedian$~to the low-redshift universe.
As we have shown, however, the low-redshift SN~Ic-BL host galaxy population 
exhibits the same preference for high stellar-mass and star-formation density galaxies as
LGRBs. 
Additionally, if stellar-mass and star-formation-rate density are proxies for the star-forming 
conditions that promote the formation of SN~Ic-BL progenitors, then we expect that these
same physical conditions exist in high-redshift LGRB host galaxies having similar positions in the
\mbox{$M$--$\Sigma_{M}$} and \mbox{$M$--$\Sigma_{\rm SFR}$} planes.

Given the evidence that the progenitors of high-velocity ejecta core-collapse explosions form more efficiently in galaxies that 
have high stellar-mass and star-formation densities, a reasonable expectation is that LGRBs will be improved tracers of star formation at earlier epochs where galaxies have higher densities of stellar mass and ongoing star formation than in the low-redshift universe \citep{trujilloforster06}.

LGRBs may be the most powerful probes of star formation at the highest redshifts.
The extreme luminosities of LGRBs allow detection at the earliest epochs of star formation ($z \gtrsim 8$),
where optical and infrared instruments are not yet able to detect sub-$L_{*}$ galaxies.
The evidence we have found that the formation of LGRB progenitors depends sensitively on 
star-forming conditions suggests that they will be critical tools for investigating early
star formation in detail. 

\section{Conclusions}
\label{sec:conclusions}
We have studied the host galaxies of a sample of 245 low-redshift
core-collapse SN, including 17 SN~Ic-BL discovered by galaxy-untargeted searches
and 15 optically luminous and dust-obscured $z<1.2$ LGRBs. 
We have used the uniform fiber spectra and photometry of the SDSS
to measure the properties of the core-collapse SN host-galaxy 
sample, as well as characterize the low-redshift star-forming galaxy population.
Published multi-band photometry as well as {\it HST}
imaging were used to study the $z < 1.2$ LGRB host galaxies.

The outflowing ejecta of SN Ic-BL, from the wide features of their spectra, and of LGRBs, through their 
association with SN Ic-BL and $\gamma$-ray emission, are inferred to expand with 
high velocities ($\sim$\,20,000--30,000\,km\,s$^{-1}$).
We have found that these core-collapse explosions in which a significant fraction of the ejecta
is moving at high velocity 
prefer galaxies with high 
stellar-mass and star-formation densities, 
when compared to galaxies having similar stellar masses.
The core-collapse SN in our sample having typical velocities of their ejecta, in contrast, are found approximately equally in galaxies with comparatively low and high stellar-mass and star-formation densities.
From the widths of H$\alpha$ emission lines, we find that the hosts of SN~Ic-BL have exceptionally high gas velocity dispersions when compared with star-forming galaxies having similar stellar masses. 

While early LGRB host environment analyses helped to establish the connection between LGRBs and massive stars (e.g., \citealt{bloom02}),
subsequent analysis has focused on the effect
of metal abundance on the creation of their progenitors (e.g., \citealt{mod08}),
or sought to explain environmental patterns in terms of a preference for metal-poor
galaxies (e.g., \citealt{fru06}; \citealt{svensson10}; \citealt{grahamfruchter13}).
Here we have shown that the preference of low-redshift SN~Ic-BL and $z < 1.2$ 
LGRBs for galaxies with high stellar-mass and star-formation-rate densities, 
compared to star-forming galaxies of similar stellar mass, 
is separate from a preference for low metal abundances.

Observations of extragalactic star formation have suggested that dense stellar clusters may form with
greater efficiency in regions of dense star formation \citep{goddard10,silvavillaadamo13}. 
A prospective explanation for the host-galaxy patterns we find is the efficient formation of tight massive binary progenitor systems in such densely star-forming environments. 
Alternatively, if SN Ic-BL and LGRB progenitors have greater stellar masses than those of core-collapse SN with typical ejecta velocities, 
a top-heavy IMF in galaxies with high stellar-mass density and star-formation density could account for these same patterns.
However, such a top-heavy IMF would need to be reconciled with evidence instead for a bottom-heavy IMF in massive galaxies that were densely star forming while assembling their mass \citep{vandokkumconroy10,conroyvandokkum12}. 
The construction of ground-based telescopes with large aperture (e.g., the Thirty Meter Telescope; Giant Magellan Telescope) and the James Webb Space Telescope will make possible observations of galaxies formed shortly after reionization, but the faintest galaxies will remain beyond their sensitivity. LGRBs, whose $\gamma$-ray emission is currently visible to $z \approx 8$ \citep{tanvir09}, may provide an effective approach to probing the detailed star-forming conditions in these early galaxies. 

\acknowledgements

We thank Sandra Savaglio who, as referee, provided insightful comments and suggestions.
We additionally acknowledge useful discussions about measurements and host galaxies with 
Jarle Brinchmann, Ori Fox, David Elbaz, John Graham, Matt Lehnert, Steven Stahler, 
Paul Crowther, and Josh Bloom. 
A.V.F.'s group at UC Berkeley has received generous financial assistance
from the Christopher R. Redlich Fund, the TABASGO Foundation, Weldon Wood, 
and NSF grant AST-1211916, as well as from NASA/{\it HST} grant AR-12850 from the
Space Telescope Science Institute, which is operated by the Association 
of Universities for Research in Astronomy (AURA), Inc.,
under NASA contract NAS5-26555. This research has made use of the GHostS database (www.grbhosts.org), which is partly funded by NASA/{\it Spitzer} RSA Agreement No. 1287913.

\clearpage
\begin{table*}[htp!]
\centering
\scriptsize
\caption{Properties of Host Galaxies from Photometry and Imaging}
\begin{tabular}{lccccccc}
\hline
Name & {\it z} & Type & Mass & SFR (phot) & $\Sigma_{M}$ & $\Sigma_{\rm SFR}$ & $r_{50}$\\
 &  &  & (log M$_{\odot}$) & (log M$_{\odot}$ yr$^{-1}$) & (log M$_{\odot}$ kpc$^{-2}$) & (log M$_{\odot}$ yr$^{-1}$ kpc$^{-2}$) & (kpc)\\
\hline
GRB010921 & 0.451 & GRB & 9.69$\pm$0.09 & 0.35$\pm$0.19 & 8.53$\pm$0.09 & -0.81$\pm$0.19 & 2.05$\pm$0.02\\
GRB011121 & 0.362 & GRB & 9.66$\pm$0.17 & 0.32$\pm$0.18 & 6.86$\pm$0.17 & -2.48$\pm$0.18 & 11.04$\pm$0.64\\
GRB021211 & 1.006 & GRB & 8.62$\pm$0.54 & -0.59$\pm$0.20 & 7.94$\pm$0.54 & -1.26$\pm$0.20 & 0.93$\pm$0.02\\
GRB030329 & 0.168 & GRB & 8.03$\pm$0.12 & -0.91$\pm$0.20 & 7.99$\pm$0.12 & -0.94$\pm$0.20 & 0.67$\pm$0.03\\
GRB050416A & 0.653 & GRB & 9.19$\pm$0.19 & 0.18$\pm$0.16 & 7.99$\pm$0.19 & -1.02$\pm$0.16 & 2.17$\pm$0.12\\
GRB051022 & 0.807 & GRB & 9.72$\pm$0.05 & 1.25$\pm$0.09 & 8.01$\pm$0.05 & -0.45$\pm$0.09 & 4.88$\pm$0.44\\
GRB060218 & 0.034 & GRB & 7.72$\pm$0.19 & -1.11$\pm$0.24 & 7.96$\pm$0.19 & -0.87$\pm$0.24 & 0.31$\pm$0.00\\
GRB090417B & 0.345 & GRB & 10.14$\pm$0.14 & -0.27$\pm$0.29 & 8.41$\pm$0.14 & -2.00$\pm$0.29 & 3.45$\pm$0.06\\
GRB970508 & 0.835 & GRB & 8.62$\pm$0.24 & -0.26$\pm$0.17 & 8.46$\pm$0.24 & -0.42$\pm$0.17 & 0.68$\pm$0.01\\
GRB970828 & 0.960 & GRB & 9.62$\pm$0.52 & 0.23$\pm$0.25 & 8.65$\pm$0.52 & -0.73$\pm$0.25 & 1.66$\pm$0.80\\
GRB980613 & 1.097 & GRB & 9.00$\pm$0.34 & 0.68$\pm$0.18 & 8.42$\pm$0.34 & 0.10$\pm$0.18 & 1.03$\pm$0.19\\
GRB980703 & 0.966 & GRB & 10.00$\pm$0.15 & 0.98$\pm$0.16 & 9.06$\pm$0.15 & 0.03$\pm$0.16 & 1.37$\pm$0.04\\
GRB990712 & 0.433 & GRB & 9.37$\pm$0.05 & -0.22$\pm$0.09 & 8.37$\pm$0.05 & -1.21$\pm$0.09 & 2.25$\pm$0.08\\
GRB991208 & 0.706 & GRB & 8.91$\pm$0.25 & -0.12$\pm$0.17 & 9.09$\pm$0.25 & 0.06$\pm$0.17 & 0.33$\pm$0.02\\
PTF 09awk & 0.062 & Ib & 9.53$\pm$0.07 & 0.16$\pm$0.12 & 8.56$\pm$0.07 & -0.81$\pm$0.12 & 1.61$\pm$0.05\\
PTF 09axc & 0.115 & II & 9.99$\pm$0.06 & -0.19$\pm$0.18 & 8.87$\pm$0.06 & -1.32$\pm$0.18 & 1.51$\pm$0.07\\
PTF 09axi & 0.064 & II & 9.33$\pm$0.09 & -0.28$\pm$0.15 & 7.42$\pm$0.09 & -2.18$\pm$0.15 & 4.10$\pm$0.15\\
PTF 09bce & 0.023 & II & 10.95$\pm$0.03 & 0.70$\pm$0.07 & 9.17$\pm$0.03 & -1.08$\pm$0.07 & 4.46$\pm$0.03\\
PTF 09bw & 0.150 & II & 9.55$\pm$0.17 & -0.26$\pm$0.36 & 7.90$\pm$0.17 & -1.90$\pm$0.36 & 2.99$\pm$0.40\\
PTF 09cjq & 0.019 & II & 10.40$\pm$0.03 & 0.30$\pm$0.07 & 8.10$\pm$0.03 & -2.00$\pm$0.07 & 7.32$\pm$0.03\\
PTF 09ct & 0.150 & II & 9.95$\pm$0.14 & 0.17$\pm$0.26 & 8.46$\pm$0.14 & -1.33$\pm$0.26 & 2.76$\pm$0.38\\
PTF 09cu & 0.057 & II & 10.47$\pm$0.04 & 0.68$\pm$0.08 & 8.19$\pm$0.04 & -1.60$\pm$0.08 & 6.63$\pm$0.06\\
PTF 09dfk & 0.016 & Ib & 8.96$\pm$0.31 & -0.82$\pm$0.34 & 8.49$\pm$0.31 & -1.29$\pm$0.34 & 0.76$\pm$0.02\\
PTF 09djl & 0.184 & II & 10.09$\pm$0.15 & -0.04$\pm$0.36 & 9.47$\pm$0.15 & -0.65$\pm$0.36 & 1.03$\pm$0.21\\
PTF 09dra & 0.077 & II & 10.57$\pm$0.07 & 0.81$\pm$0.08 & 7.80$\pm$0.07 & -1.96$\pm$0.08 & 15.03$\pm$0.22\\
PTF 09due & 0.029 & II & 10.25$\pm$0.03 & 0.70$\pm$0.07 & 7.93$\pm$0.03 & -1.62$\pm$0.07 & 7.30$\pm$0.03\\
PTF 09dzt & 0.087 & Ic & 9.98$\pm$0.06 & 0.44$\pm$0.16 & 7.18$\pm$0.06 & -2.37$\pm$0.16 & 12.35$\pm$0.46\\
PTF 09ebq & 0.024 & II & 9.75$\pm$0.04 & 0.02$\pm$0.08 & 8.51$\pm$0.04 & -1.22$\pm$0.08 & 2.04$\pm$0.02\\
PTF 09fbf & 0.021 & II & 10.05$\pm$0.03 & 0.60$\pm$0.07 & 7.97$\pm$0.03 & -1.48$\pm$0.07 & 8.30$\pm$0.04\\
PTF 09fmk & 0.063 & II & \nodata & \nodata & \nodata & \nodata & 5.24$\pm$0.14\\
PTF 09foy & 0.060 & II & 10.20$\pm$0.09 & 0.52$\pm$0.13 & 7.67$\pm$0.09 & -2.01$\pm$0.13 & 7.46$\pm$0.09\\
PTF 09g & 0.040 & II & 9.55$\pm$0.04 & 0.18$\pm$0.08 & 7.84$\pm$0.04 & -1.53$\pm$0.08 & 2.98$\pm$0.02\\
PTF 09gof & 0.103 & II & 10.06$\pm$0.06 & 0.60$\pm$0.20 & 7.56$\pm$0.06 & -1.91$\pm$0.20 & 8.63$\pm$0.24\\
PTF 09hdo & 0.047 & II & 10.75$\pm$0.03 & 0.70$\pm$0.07 & 8.61$\pm$0.03 & -1.44$\pm$0.07 & 5.63$\pm$0.03\\
PTF 09hzg & 0.028 & II & 10.55$\pm$0.03 & -0.40$\pm$0.07 & 8.38$\pm$0.03 & -2.57$\pm$0.07 & 9.06$\pm$0.05\\
PTF 09iex & 0.020 & II & 8.36$\pm$0.47 & -1.10$\pm$0.50 & 6.98$\pm$0.47 & -2.48$\pm$0.50 & 2.42$\pm$0.14\\
PTF 09ige & 0.064 & II & 9.76$\pm$0.07 & 0.32$\pm$0.20 & 7.61$\pm$0.07 & -1.83$\pm$0.20 & 5.51$\pm$0.08\\
PTF 09igz & 0.086 & II & 9.40$\pm$0.08 & -0.16$\pm$0.15 & 7.72$\pm$0.08 & -1.84$\pm$0.15 & 2.89$\pm$0.17\\
PTF 09ism & 0.029 & II & 9.56$\pm$0.19 & -0.16$\pm$0.19 & 7.83$\pm$0.19 & -1.89$\pm$0.19 & 3.34$\pm$0.07\\
PTF 09ps & 0.106 & Ic & 9.29$\pm$0.08 & -0.18$\pm$0.15 & 8.20$\pm$0.08 & -1.27$\pm$0.15 & 2.80$\pm$0.14\\
PTF 09q & 0.090 & Ic & 10.83$\pm$0.09 & 0.86$\pm$0.12 & 8.23$\pm$0.09 & -1.74$\pm$0.12 & 8.95$\pm$0.11\\
PTF 09r & 0.027 & II & 9.01$\pm$0.23 & -1.16$\pm$0.46 & 8.45$\pm$0.23 & -1.72$\pm$0.46 & 1.49$\pm$0.02\\
PTF 09sh & 0.038 & II & 9.95$\pm$0.03 & 0.40$\pm$0.07 & 7.97$\pm$0.03 & -1.58$\pm$0.07 & 4.76$\pm$0.08\\
PTF 09sk & 0.035 & Ic-BL & 8.85$\pm$0.14 & -0.38$\pm$0.23 & 7.79$\pm$0.14 & -1.45$\pm$0.23 & 2.36$\pm$0.05\\
PTF 09t & 0.039 & II & 9.60$\pm$0.03 & 0.34$\pm$0.10 & 7.69$\pm$0.03 & -1.57$\pm$0.10 & 5.14$\pm$0.04\\
PTF 09tm & 0.035 & II & 10.35$\pm$0.03 & 0.30$\pm$0.07 & 8.62$\pm$0.03 & -1.43$\pm$0.07 & 4.80$\pm$0.02\\
PTF 09uj & 0.065 & II & 9.78$\pm$0.08 & 0.20$\pm$0.19 & 7.69$\pm$0.08 & -1.89$\pm$0.19 & 5.75$\pm$0.10\\
PTF 10aavz & 0.062 & Ic-BL & 9.09$\pm$0.11 & -0.55$\pm$0.17 & 8.00$\pm$0.11 & -1.64$\pm$0.17 & 2.33$\pm$0.13\\
PTF 10bau & 0.026 & II & 10.35$\pm$0.03 & 0.40$\pm$0.07 & 8.32$\pm$0.03 & -1.63$\pm$0.07 & 4.62$\pm$0.02\\
PTF 10bhu & 0.036 & Ic & 9.44$\pm$0.14 & -0.05$\pm$0.17 & 7.78$\pm$0.14 & -1.71$\pm$0.17 & 3.63$\pm$0.04\\
PTF 10bip & 0.051 & Ic & 9.10$\pm$0.10 & -0.43$\pm$0.21 & 8.10$\pm$0.10 & -1.43$\pm$0.21 & 2.01$\pm$0.06\\
PTF 10con & 0.033 & II & \nodata & \nodata & \nodata & \nodata & 0.37$\pm$62.69\\
PTF 10cqh & 0.041 & II & 10.77$\pm$0.04 & 0.97$\pm$0.08 & 8.51$\pm$0.04 & -1.29$\pm$0.08 & 6.80$\pm$0.04\\
PTF 10cwx & 0.073 & II & 9.65$\pm$0.08 & 0.08$\pm$0.18 & 7.80$\pm$0.08 & -1.78$\pm$0.18 & 3.68$\pm$0.30\\
PTF 10cxq & 0.047 & II & 8.93$\pm$0.11 & -0.13$\pm$0.23 & 7.38$\pm$0.11 & -1.68$\pm$0.23 & 5.86$\pm$0.14\\
\hline
\end{tabular}
\end{table*}\begin{table*}[htp!]
\centering
\scriptsize\begin{tabular}{lccccccc}
\hline
Name & {\it z} & Type & Mass & SFR (phot) & $\Sigma_{M}$ & $\Sigma_{\rm SFR}$ & $r_{50}$\\
 &  &  & (log M$_{\odot}$) & (log M$_{\odot}$ yr$^{-1}$) & (log M$_{\odot}$ kpc$^{-2}$) & (log M$_{\odot}$ yr$^{-1}$ kpc$^{-2}$) & (kpc)\\
\hline
PTF 10cxx & 0.034 & II & 10.27$\pm$0.13 & 0.24$\pm$0.17 & 8.88$\pm$0.13 & -1.15$\pm$0.17 & 2.62$\pm$0.02\\
PTF 10czn & 0.045 & II & 10.45$\pm$0.04 & 0.96$\pm$0.09 & 7.74$\pm$0.04 & -1.75$\pm$0.09 & 9.71$\pm$0.05\\
PTF 10dk & 0.074 & II & 8.54$\pm$0.19 & -1.06$\pm$0.18 & 7.64$\pm$0.19 & -1.96$\pm$0.18 & 1.47$\pm$0.51\\
PTF 10hv & 0.052 & II & 11.22$\pm$0.04 & 1.60$\pm$0.07 & 9.17$\pm$0.04 & -0.46$\pm$0.07 & 6.85$\pm$0.11\\
PTF 10qts & 0.091 & Ic-BL & 9.60$\pm$0.09 & -0.91$\pm$0.30 & 8.83$\pm$0.09 & -1.68$\pm$0.30 & 1.17$\pm$0.36\\
PTF 10s & 0.051 & II & 9.62$\pm$0.09 & -0.09$\pm$0.15 & 7.98$\pm$0.09 & -1.73$\pm$0.15 & 2.97$\pm$0.06\\
PTF 11cgx & 0.033 & II & 9.92$\pm$0.13 & 0.41$\pm$0.17 & 8.19$\pm$0.13 & -1.32$\pm$0.17 & 3.48$\pm$0.02\\
PTF 11cwi & 0.056 & II & 10.56$\pm$0.04 & 0.51$\pm$0.08 & 8.54$\pm$0.04 & -1.51$\pm$0.08 & 4.63$\pm$0.08\\
PTF 11dad & 0.072 & II & 9.97$\pm$0.08 & 0.33$\pm$0.15 & 7.93$\pm$0.08 & -1.71$\pm$0.15 & 7.48$\pm$0.15\\
PTF 11dqk & 0.036 & II & 9.85$\pm$0.03 & 0.45$\pm$0.13 & 7.95$\pm$0.03 & -1.45$\pm$0.13 & 3.93$\pm$0.02\\
PTF 11dqr & 0.082 & II & 9.34$\pm$0.10 & -0.21$\pm$0.19 & 6.64$\pm$0.10 & -2.91$\pm$0.19 & 10.25$\pm$0.98\\
PTF 11dsb & 0.190 & II & 9.70$\pm$0.08 & 0.23$\pm$0.16 & 7.28$\pm$0.08 & -2.18$\pm$0.16 & 7.42$\pm$0.57\\
PTF 11dtd & 0.040 & II & 10.25$\pm$0.03 & 0.80$\pm$0.07 & 7.75$\pm$0.03 & -1.70$\pm$0.07 & 9.01$\pm$0.06\\
PTF 11ecp & 0.034 & II & 10.00$\pm$0.03 & 0.60$\pm$0.07 & 8.05$\pm$0.03 & -1.35$\pm$0.07 & 6.36$\pm$0.04\\
PTF 11ekj & 0.043 & II & 9.92$\pm$0.12 & -0.35$\pm$0.15 & 10.10$\pm$0.12 & -0.18$\pm$0.15 & 0.43$\pm$0.08\\
PTF 11emc & 0.082 & II & 9.22$\pm$0.09 & -0.38$\pm$0.18 & 7.68$\pm$0.09 & -1.92$\pm$0.18 & 3.97$\pm$0.23\\
PTF 11epi & 0.032 & II & 11.00$\pm$0.03 & 0.10$\pm$0.07 & 8.48$\pm$0.03 & -2.42$\pm$0.07 & 11.21$\pm$0.06\\
PTF 11ftr & 0.018 & II & 7.80$\pm$0.32 & -0.84$\pm$0.36 & 7.31$\pm$0.32 & -1.33$\pm$0.36 & 0.84$\pm$0.01\\
PTF 11fuu & 0.097 & IIn & 10.34$\pm$0.05 & 0.47$\pm$0.09 & 8.56$\pm$0.05 & -1.30$\pm$0.09 & 4.02$\pm$0.08\\
PTF 11fuv & 0.030 & II & 11.15$\pm$0.03 & 1.20$\pm$0.07 & 8.76$\pm$0.03 & -1.19$\pm$0.07 & 7.24$\pm$0.04\\
PTF 11gdz & 0.013 & II & 10.39$\pm$0.04 & 0.38$\pm$0.08 & 9.40$\pm$0.04 & -0.61$\pm$0.08 & 1.29$\pm$0.01\\
PTF 11hyg & 0.030 & Ic & 10.84$\pm$0.04 & 1.02$\pm$0.07 & 8.32$\pm$0.04 & -1.50$\pm$0.07 & 7.81$\pm$0.03\\
PTF 11iqb & 0.013 & IIn & 11.59$\pm$0.32 & 1.58$\pm$0.35 & 9.34$\pm$0.32 & -0.68$\pm$0.35 & 5.93$\pm$0.03\\
PTF 11ixk & 0.021 & Ic & 10.26$\pm$0.04 & 0.60$\pm$0.07 & 8.19$\pm$0.04 & -1.46$\pm$0.07 & 5.48$\pm$0.02\\
PTF 11izq & 0.062 & Ib & 9.23$\pm$0.09 & -0.33$\pm$0.17 & 7.24$\pm$0.09 & -2.31$\pm$0.17 & 4.66$\pm$0.18\\
PTF 11jgp & 0.070 & II & 9.91$\pm$0.04 & 0.54$\pm$0.10 & 7.95$\pm$0.04 & -1.42$\pm$0.10 & 4.78$\pm$0.07\\
PTF 11kjk & 0.067 & II & 9.12$\pm$0.10 & -0.17$\pm$0.20 & 7.44$\pm$0.10 & -1.85$\pm$0.20 & 5.67$\pm$0.25\\
PTF 11klg & 0.027 & Ic & 10.62$\pm$0.19 & 0.44$\pm$0.22 & 8.56$\pm$0.19 & -1.62$\pm$0.22 & 5.15$\pm$0.04\\
PTF 11kqn & 0.066 & II & 11.26$\pm$0.05 & 1.09$\pm$0.07 & 8.89$\pm$0.05 & -1.29$\pm$0.07 & 8.06$\pm$0.08\\
PTF 11ktr & 0.106 & II & 10.45$\pm$0.04 & 0.80$\pm$0.07 & 8.03$\pm$0.04 & -1.63$\pm$0.07 & 8.32$\pm$0.18\\
PTF 11mhh & 0.039 & II & 9.96$\pm$0.15 & 0.22$\pm$0.15 & 7.95$\pm$0.15 & -1.79$\pm$0.15 & 4.45$\pm$0.07\\
PTF 11mmk & 0.049 & II & 9.55$\pm$0.11 & -0.16$\pm$0.14 & 8.10$\pm$0.11 & -1.60$\pm$0.14 & 3.75$\pm$0.07\\
PTF 11mpv & 0.043 & II & 9.26$\pm$0.09 & -0.14$\pm$0.12 & 8.18$\pm$0.09 & -1.22$\pm$0.12 & 2.58$\pm$0.04\\
PTF 11pdj & 0.024 & II & 10.85$\pm$0.03 & -0.20$\pm$0.07 & 8.96$\pm$0.03 & -2.09$\pm$0.07 & 9.74$\pm$0.07\\
PTF 11qcc & 0.043 & II & 9.73$\pm$0.12 & -0.11$\pm$0.15 & 8.19$\pm$0.12 & -1.65$\pm$0.15 & 6.11$\pm$0.12\\
PTF 11qcm & 0.050 & II & 10.60$\pm$0.19 & 0.61$\pm$0.19 & 8.43$\pm$0.19 & -1.57$\pm$0.19 & 7.71$\pm$0.05\\
PTF 11qgw & 0.027 & II & 8.98$\pm$0.19 & -0.27$\pm$0.29 & 7.16$\pm$0.19 & -2.09$\pm$0.29 & 6.88$\pm$0.10\\
PTF 11qju & 0.028 & II & 9.12$\pm$0.18 & -0.41$\pm$0.23 & 10.24$\pm$0.18 & 0.71$\pm$0.23 & 0.15$\pm$0.34\\
PTF 11qux & 0.041 & II & \nodata & \nodata & \nodata & \nodata & 0.66$\pm$0.01\\
PTF 12boj & 0.037 & II & 10.58$\pm$0.16 & 0.57$\pm$0.17 & 8.53$\pm$0.16 & -1.48$\pm$0.17 & 4.47$\pm$0.02\\
PTF 12bpy & 0.060 & II & 9.36$\pm$0.14 & -0.15$\pm$0.22 & 8.70$\pm$0.14 & -0.80$\pm$0.22 & 0.96$\pm$0.04\\
PTF 12bwq & 0.040 & Ib & 9.53$\pm$0.22 & -0.11$\pm$0.27 & 7.07$\pm$0.22 & -2.57$\pm$0.27 & 9.62$\pm$0.21\\
PTF 12cdc & 0.070 & II & 10.78$\pm$0.11 & 0.77$\pm$0.14 & 7.77$\pm$0.11 & -2.25$\pm$0.14 & 15.31$\pm$0.16\\
PTF 12cde & 0.013 & Ib/c & 8.36$\pm$0.32 & -1.17$\pm$0.38 & 7.03$\pm$0.32 & -2.50$\pm$0.38 & 3.61$\pm$0.12\\
PTF 12cgb & 0.026 & II & 9.27$\pm$0.04 & -0.14$\pm$0.12 & 7.99$\pm$0.04 & -1.42$\pm$0.12 & 2.61$\pm$0.01\\
PTF 12dke & 0.067 & II & 9.91$\pm$0.08 & 0.46$\pm$0.15 & 7.40$\pm$0.08 & -2.05$\pm$0.15 & 11.27$\pm$0.20\\
PTF 12eje & 0.078 & II & 9.92$\pm$0.07 & 0.49$\pm$0.18 & 7.82$\pm$0.07 & -1.61$\pm$0.18 & 5.60$\pm$0.07\\
PTF 12fes & 0.036 & Ib & 10.55$\pm$0.04 & 0.99$\pm$0.07 & 7.98$\pm$0.04 & -1.58$\pm$0.07 & 8.51$\pm$0.07\\
PTF 12gcx & 0.045 & II & 9.74$\pm$0.04 & 0.48$\pm$0.08 & 7.42$\pm$0.04 & -1.84$\pm$0.08 & 7.16$\pm$0.09\\
PTF 12gvr & 0.056 & Ib/c & 10.60$\pm$0.03 & 1.08$\pm$0.08 & 8.07$\pm$0.03 & -1.46$\pm$0.08 & 9.24$\pm$0.04\\
PTF 12gzk & 0.014 & 12gzk & 7.47$\pm$0.11 & -1.46$\pm$0.38 & 7.43$\pm$0.11 & -1.50$\pm$0.38 & 0.54$\pm$0.05\\
PTF 12jje & 0.042 & II & 10.01$\pm$0.12 & 0.27$\pm$0.16 & 7.89$\pm$0.12 & -1.84$\pm$0.16 & 5.37$\pm$0.08\\
PTF 12ne & 0.033 & II & 9.91$\pm$0.13 & -0.06$\pm$0.17 & 8.62$\pm$0.13 & -1.35$\pm$0.17 & 1.82$\pm$0.02\\
PTF 13bvn & 0.004 & Ib & 11.10$\pm$0.03 & 1.00$\pm$0.07 & 9.66$\pm$0.03 & -0.44$\pm$0.07 & 2.68$\pm$0.02\\
PTF 13c & 0.011 & II & 9.04$\pm$0.04 & -0.70$\pm$0.07 & 8.00$\pm$0.04 & -1.74$\pm$0.07 & 1.56$\pm$0.01\\
\hline
\end{tabular}
\end{table*}\begin{table*}[htp!]
\centering
\scriptsize\begin{tabular}{lccccccc}
\hline
Name & {\it z} & Type & Mass & SFR (phot) & $\Sigma_{M}$ & $\Sigma_{\rm SFR}$ & $r_{50}$\\
 &  &  & (log M$_{\odot}$) & (log M$_{\odot}$ yr$^{-1}$) & (log M$_{\odot}$ kpc$^{-2}$) & (log M$_{\odot}$ yr$^{-1}$ kpc$^{-2}$) & (kpc)\\
\hline
PTF 13cab & 0.030 & Ib & 9.32$\pm$0.30 & -0.28$\pm$0.37 & 7.72$\pm$0.30 & -1.88$\pm$0.37 & 4.39$\pm$0.06\\
PTF 13cac & 0.030 & II & 8.91$\pm$0.31 & -0.87$\pm$0.37 & 7.38$\pm$0.31 & -2.40$\pm$0.37 & 3.97$\pm$0.13\\
PTF 13cbf & 0.039 & Ic & 9.40$\pm$0.03 & -0.00$\pm$0.07 & 7.72$\pm$0.03 & -1.68$\pm$0.07 & 5.78$\pm$0.06\\
PTF 13d & 0.024 & II & 10.25$\pm$0.04 & 0.60$\pm$0.07 & 8.07$\pm$0.04 & -1.58$\pm$0.07 & 6.27$\pm$0.02\\
SN 1999ap & 0.040 & II        & 9.35$\pm$0.12 & -0.04$\pm$0.18 & 7.33$\pm$0.12 & -2.06$\pm$0.18 & 5.25$\pm$0.10\\
SN 1999as & 0.127 & Ic pec    & \nodata & \nodata & \nodata & \nodata & 9.02$\pm$1.05\\
SN 1999bc & 0.021 & Ic        & 10.46$\pm$0.04 & 0.62$\pm$0.08 & 8.18$\pm$0.04 & -1.66$\pm$0.08 & 6.74$\pm$0.03\\
SN 1999bd & 0.151 & II        & 9.66$\pm$0.09 & 0.07$\pm$0.16 & 8.58$\pm$0.09 & -1.01$\pm$0.16 & 2.86$\pm$0.28\\
SN 2001bk & 0.043 & II        & 8.82$\pm$0.13 & -0.79$\pm$0.19 & 7.58$\pm$0.13 & -2.02$\pm$0.19 & 3.13$\pm$0.15\\
SN 2001fb & 0.032 & II        & 9.95$\pm$0.03 & 0.60$\pm$0.07 & 8.53$\pm$0.03 & -0.82$\pm$0.07 & 3.46$\pm$0.02\\
SN 2001ij & 0.038 & II P      & 10.20$\pm$0.03 & 0.40$\pm$0.07 & 8.22$\pm$0.03 & -1.59$\pm$0.07 & 5.55$\pm$0.04\\
SN 2002dg & 0.047 & Ib        & 9.10$\pm$0.11 & 0.06$\pm$0.16 & 7.10$\pm$0.11 & -1.94$\pm$0.16 & 4.31$\pm$0.07\\
SN 2002ew & 0.030 & II        & 9.36$\pm$0.04 & -0.08$\pm$0.11 & 7.95$\pm$0.04 & -1.49$\pm$0.11 & 2.76$\pm$0.02\\
SN 2002fa & 0.060 & II        & 10.12$\pm$0.09 & 0.59$\pm$0.20 & 7.43$\pm$0.09 & -2.10$\pm$0.20 & 9.16$\pm$0.11\\
SN 2002fu & 0.091 & II        & 9.86$\pm$0.07 & 0.25$\pm$0.14 & 7.90$\pm$0.07 & -1.72$\pm$0.14 & 9.06$\pm$0.21\\
SN 2002hj & 0.024 & II        & 9.60$\pm$0.20 & -0.02$\pm$0.26 & 7.75$\pm$0.20 & -1.87$\pm$0.26 & 4.04$\pm$0.05\\
SN 2002ik & 0.032 & II P      & 10.55$\pm$0.03 & 0.70$\pm$0.07 & 8.52$\pm$0.03 & -1.33$\pm$0.07 & 5.92$\pm$0.03\\
SN 2002in & 0.076 & II        & 9.11$\pm$0.09 & -0.48$\pm$0.19 & 7.64$\pm$0.09 & -1.96$\pm$0.19 & 4.15$\pm$0.33\\
SN 2002ip & 0.079 & II        & 8.90$\pm$0.10 & -0.61$\pm$0.19 & 6.76$\pm$0.10 & -2.75$\pm$0.19 & 5.55$\pm$0.80\\
SN 2002iq & 0.056 & II        & 9.38$\pm$0.06 & 0.13$\pm$0.14 & 7.81$\pm$0.06 & -1.44$\pm$0.14 & 3.98$\pm$0.18\\
SN 2002jl & 0.064 & II        & 8.01$\pm$0.23 & -1.23$\pm$0.22 & 7.30$\pm$0.23 & -1.93$\pm$0.22 & 1.37$\pm$0.33\\
SN 2003cv & 0.028 & II pec    & 8.31$\pm$0.20 & -0.96$\pm$0.27 & 7.40$\pm$0.20 & -1.86$\pm$0.27 & 1.90$\pm$0.05\\
SN 2003dq & 0.046 & II        & 9.01$\pm$0.12 & -0.43$\pm$0.16 & 7.59$\pm$0.12 & -1.85$\pm$0.16 & 2.95$\pm$0.12\\
SN 2003kj & 0.100 & II        & 8.77$\pm$0.18 & -0.79$\pm$0.18 & 7.14$\pm$0.18 & -2.43$\pm$0.18 & 3.04$\pm$0.78\\
SN 2004cm & 0.004 & II P      & 9.65$\pm$0.03 & 0.00$\pm$0.07 & 8.51$\pm$0.03 & -1.14$\pm$0.07 & 1.75$\pm$0.01\\
SN 2004gy & 0.027 & II        & 8.26$\pm$0.20 & -1.02$\pm$0.26 & 7.44$\pm$0.20 & -1.83$\pm$0.26 & 1.63$\pm$0.05\\
SN 2004ht & 0.067 & II        & 10.30$\pm$0.07 & 0.60$\pm$0.09 & 7.97$\pm$0.07 & -1.72$\pm$0.09 & 7.58$\pm$0.09\\
SN 2004hv & 0.061 & II        & 8.82$\pm$0.13 & -0.70$\pm$0.19 & 6.90$\pm$0.13 & -2.62$\pm$0.19 & 5.27$\pm$0.98\\
SN 2004hx & 0.014 & II        & 8.69$\pm$0.31 & -1.11$\pm$0.35 & 7.72$\pm$0.31 & -2.08$\pm$0.35 & 1.63$\pm$0.02\\
SN 2004ic & 0.093 & II        & 10.92$\pm$0.06 & 1.03$\pm$0.08 & 7.80$\pm$0.06 & -2.09$\pm$0.08 & 15.04$\pm$0.30\\
SN 2005bn & 0.028 & II        & 9.14$\pm$0.20 & -0.44$\pm$0.28 & 8.30$\pm$0.20 & -1.28$\pm$0.28 & 1.38$\pm$0.01\\
SN 2005fq & 0.140 & II        & 8.83$\pm$0.14 & -0.54$\pm$0.18 & 7.46$\pm$0.14 & -1.91$\pm$0.18 & 3.35$\pm$1.08\\
SN 2005gi & 0.050 & II        & 9.12$\pm$0.14 & -0.39$\pm$0.16 & 7.10$\pm$0.14 & -2.41$\pm$0.16 & 4.74$\pm$0.26\\
SN 2005hl & 0.020 & Ib        & 10.40$\pm$0.04 & 0.40$\pm$0.07 & 8.91$\pm$0.04 & -1.09$\pm$0.07 & 2.94$\pm$0.01\\
SN 2005hm & 0.030 & Ib        & 8.61$\pm$0.30 & -1.53$\pm$0.34 & 8.28$\pm$0.30 & -1.87$\pm$0.34 & 0.94$\pm$0.19\\
SN 2005kb & 0.015 & II        & 9.35$\pm$0.31 & -0.43$\pm$0.39 & 8.19$\pm$0.31 & -1.59$\pm$0.39 & 4.32$\pm$0.05\\
SN 2005kr & 0.130 & Ic-BL & 8.63$\pm$0.16 & -0.62$\pm$0.20 & 8.36$\pm$0.16 & -0.88$\pm$0.20 & 0.57$\pm$0.30\\
SN 2005ks & 0.100 & Ic-BL & 9.88$\pm$0.08 & 0.27$\pm$0.14 & 8.44$\pm$0.08 & -1.17$\pm$0.14 & 3.70$\pm$0.08\\
SN 2005lb & 0.030 & II        & 8.75$\pm$0.30 & -0.94$\pm$0.32 & 7.47$\pm$0.30 & -2.22$\pm$0.32 & 3.95$\pm$0.37\\
SN 2005lc & 0.010 & II        & 8.21$\pm$0.33 & -1.34$\pm$0.39 & 7.41$\pm$0.33 & -2.13$\pm$0.39 & 1.32$\pm$0.06\\
SN 2005lm & 0.080 & II        & 9.23$\pm$0.08 & -0.24$\pm$0.16 & 8.01$\pm$0.08 & -1.46$\pm$0.16 & 1.83$\pm$0.07\\
SN 2005mk & 0.150 & II        & 9.44$\pm$0.09 & -0.16$\pm$0.17 & 7.50$\pm$0.09 & -2.09$\pm$0.17 & 4.13$\pm$0.64\\
SN 2005mn & 0.050 & Ib        & 9.61$\pm$0.16 & 0.17$\pm$0.22 & 7.46$\pm$0.16 & -1.97$\pm$0.22 & 11.24$\pm$0.26\\
SN 2006L & 0.039 & IIn       & 8.28$\pm$0.16 & -1.19$\pm$0.19 & 7.63$\pm$0.16 & -1.84$\pm$0.19 & 1.18$\pm$0.08\\
SN 2006M & 0.015 & IIn       & 9.05$\pm$0.03 & -0.42$\pm$0.08 & 7.73$\pm$0.03 & -1.74$\pm$0.08 & 1.93$\pm$0.01\\
SN 2006ad & 0.030 & II        & 8.97$\pm$0.05 & -1.27$\pm$0.10 & 11.66$\pm$0.05 & 1.42$\pm$0.10 & 0.02$\pm$0.02\\
SN 2006ag & 0.035 & IIn       & 8.05$\pm$0.16 & -0.63$\pm$0.18 & 6.91$\pm$0.16 & -1.77$\pm$0.18 & 1.74$\pm$0.08\\
SN 2006aj & 0.033 & Ic-BL & 8.08$\pm$0.17 & -1.31$\pm$0.25 & 8.01$\pm$0.17 & -1.38$\pm$0.25 & 0.46$\pm$0.10\\
SN 2006bj & 0.038 & II        & 9.38$\pm$0.15 & -0.08$\pm$0.19 & 7.32$\pm$0.15 & -2.15$\pm$0.19 & 4.80$\pm$0.07\\
SN 2006cu & 0.029 & IIn       & 10.15$\pm$0.03 & 0.50$\pm$0.07 & 7.79$\pm$0.03 & -1.86$\pm$0.07 & 7.15$\pm$0.02\\
SN 2006cv & 0.100 & IIn       & 10.33$\pm$0.05 & -0.60$\pm$0.07 & 8.61$\pm$0.05 & -2.32$\pm$0.07 & 3.18$\pm$0.22\\
SN 2006cw & 0.065 & II        & 10.02$\pm$0.06 & 0.50$\pm$0.09 & 7.82$\pm$0.06 & -1.70$\pm$0.09 & 6.72$\pm$0.11\\
SN 2006cy & 0.036 & IIn       & 10.21$\pm$0.04 & 0.69$\pm$0.08 & 7.88$\pm$0.04 & -1.63$\pm$0.08 & 6.07$\pm$0.03\\
SN 2006db & 0.023 & IIn       & 8.69$\pm$0.19 & -0.59$\pm$0.24 & 7.40$\pm$0.19 & -1.88$\pm$0.24 & 2.39$\pm$0.04\\
SN 2006fg & 0.030 & II        & 7.99$\pm$0.30 & -1.53$\pm$0.32 & 8.17$\pm$0.30 & -1.35$\pm$0.32 & 0.34$\pm$0.03\\
\hline
\end{tabular}
\end{table*}\begin{table*}[htp!]
\centering
\scriptsize\begin{tabular}{lccccccc}
\hline
Name & {\it z} & Type & Mass & SFR (phot) & $\Sigma_{M}$ & $\Sigma_{\rm SFR}$ & $r_{50}$\\
 &  &  & (log M$_{\odot}$) & (log M$_{\odot}$ yr$^{-1}$) & (log M$_{\odot}$ kpc$^{-2}$) & (log M$_{\odot}$ yr$^{-1}$ kpc$^{-2}$) & (kpc)\\
\hline
SN 2006fo & 0.021 & Ib & 10.25$\pm$0.03 & 0.50$\pm$0.07 & 8.53$\pm$0.03 & -1.22$\pm$0.07 & 3.18$\pm$0.01\\
SN 2006fq & 0.070 & II P      & 9.96$\pm$0.04 & 0.50$\pm$0.07 & 8.13$\pm$0.04 & -1.34$\pm$0.07 & 4.61$\pm$0.04\\
SN 2006gd & 0.150 & II P      & 10.97$\pm$0.07 & 0.86$\pm$0.16 & 8.23$\pm$0.07 & -1.89$\pm$0.16 & 15.45$\pm$0.42\\
SN 2006gy & 0.019 & IIn       & 11.17$\pm$0.32 & 0.54$\pm$0.35 & 9.36$\pm$0.32 & -1.27$\pm$0.35 & 4.74$\pm$0.03\\
SN 2006ho & 0.110 & II        & 11.25$\pm$0.03 & 0.80$\pm$0.07 & 9.01$\pm$0.03 & -1.44$\pm$0.07 & 6.09$\pm$0.09\\
SN 2006ic & 0.040 & II        & 10.45$\pm$0.04 & 0.70$\pm$0.07 & 8.17$\pm$0.04 & -1.58$\pm$0.07 & 5.95$\pm$0.04\\
SN 2006ih & 0.130 & II        & 8.95$\pm$0.18 & -0.30$\pm$0.20 & 9.55$\pm$0.18 & 0.30$\pm$0.20 & 0.24$\pm$0.09\\
SN 2006ii & 0.030 & II        & 9.44$\pm$0.30 & -0.32$\pm$0.31 & 7.51$\pm$0.30 & -2.25$\pm$0.31 & 4.11$\pm$0.07\\
SN 2006ij & 0.040 & II        & 9.81$\pm$0.11 & 0.28$\pm$0.16 & 7.74$\pm$0.11 & -1.79$\pm$0.16 & 6.70$\pm$0.06\\
SN 2006ip & 0.030 & Ic & 9.64$\pm$0.19 & 0.15$\pm$0.19 & 7.87$\pm$0.19 & -1.61$\pm$0.19 & 3.61$\pm$0.03\\
SN 2006ir & 0.020 & Ic & 8.77$\pm$0.32 & -0.85$\pm$0.35 & 7.11$\pm$0.32 & -2.51$\pm$0.35 & 3.49$\pm$0.05\\
SN 2006iw & 0.030 & II        & 9.67$\pm$0.29 & -0.02$\pm$0.31 & 8.06$\pm$0.29 & -1.63$\pm$0.31 & 4.19$\pm$0.09\\
SN 2006ix & 0.080 & II        & 9.65$\pm$0.11 & -0.02$\pm$0.19 & 7.79$\pm$0.11 & -1.88$\pm$0.19 & 7.91$\pm$0.23\\
SN 2006kh & 0.060 & II        & 9.46$\pm$0.08 & -0.23$\pm$0.13 & 8.97$\pm$0.08 & -0.72$\pm$0.13 & 0.81$\pm$0.02\\
SN 2006kn & 0.120 & II        & 9.51$\pm$0.07 & 0.10$\pm$0.19 & 7.30$\pm$0.07 & -2.12$\pm$0.19 & 10.94$\pm$0.58\\
SN 2006lc & 0.016 & Ib & 10.65$\pm$0.03 & 0.40$\pm$0.07 & 8.59$\pm$0.03 & -1.66$\pm$0.07 & 5.27$\pm$0.02\\
SN 2006lh & 0.032 & II        & 7.95$\pm$0.18 & -1.41$\pm$0.22 & 7.71$\pm$0.18 & -1.65$\pm$0.22 & 0.62$\pm$0.06\\
SN 2006ls & 0.140 & I pec     & 9.88$\pm$0.08 & 0.25$\pm$0.16 & 7.33$\pm$0.08 & -2.30$\pm$0.16 & 11.10$\pm$0.56\\
SN 2006lt & 0.015 & Ib & 8.82$\pm$0.34 & -1.01$\pm$0.35 & 9.60$\pm$0.34 & -0.23$\pm$0.35 & 0.18$\pm$0.08\\
SN 2006nq & 0.025 & II        & 9.38$\pm$0.19 & -0.11$\pm$0.19 & 7.40$\pm$0.19 & -2.09$\pm$0.19 & 4.86$\pm$0.05\\
SN 2006ns & 0.120 & II        & 9.81$\pm$0.04 & 0.54$\pm$0.15 & 7.77$\pm$0.04 & -1.50$\pm$0.15 & 6.99$\pm$0.14\\
SN 2006nx & 0.050 & Ic-BL & 8.57$\pm$0.17 & -0.98$\pm$0.25 & 7.40$\pm$0.17 & -2.16$\pm$0.25 & 1.79$\pm$0.20\\
SN 2006ny & 0.080 & II P      & 10.22$\pm$0.05 & 0.18$\pm$0.09 & 8.41$\pm$0.05 & -1.63$\pm$0.09 & 7.10$\pm$0.15\\
SN 2006qk & 0.060 & Ic-BL & 9.61$\pm$0.13 & -0.31$\pm$0.17 & 8.89$\pm$0.13 & -1.03$\pm$0.17 & 1.71$\pm$0.05\\
SN 2006rc & 0.080 & IIn       & 9.88$\pm$0.11 & 0.25$\pm$0.17 & 7.70$\pm$0.11 & -1.93$\pm$0.17 & 6.60$\pm$0.15\\
SN 2006rq & 0.070 & II        & 10.86$\pm$0.04 & 0.02$\pm$0.07 & 8.67$\pm$0.04 & -2.17$\pm$0.07 & 8.67$\pm$0.08\\
SN 2006ru & 0.020 & II        & 11.15$\pm$0.03 & 0.10$\pm$0.07 & 10.25$\pm$0.03 & -0.80$\pm$0.07 & 1.63$\pm$0.00\\
SN 2006ry & 0.060 & II        & 10.95$\pm$0.03 & \nodata & 9.51$\pm$0.03 & \nodata & 4.05$\pm$0.07\\
SN 2006tf & 0.074 & IIn       & 8.16$\pm$0.17 & -0.97$\pm$0.21 & 8.84$\pm$0.17 & -0.29$\pm$0.21 & 0.61$\pm$1.85\\
SN 2006th & 0.140 & II        & \nodata & \nodata & \nodata & \nodata & 4.24$\pm$0.97\\
SN 2007I & 0.022 & Ic-BL & 8.92$\pm$0.21 & -0.75$\pm$0.26 & 7.78$\pm$0.21 & -1.88$\pm$0.26 & 1.92$\pm$0.05\\
SN 2007bg & 0.034 & Ic-BL & 7.99$\pm$0.23 & -2.03$\pm$0.33 & 8.21$\pm$0.23 & -1.81$\pm$0.33 & 0.39$\pm$0.16\\
SN 2007bo & 0.040 & II        & 9.13$\pm$0.21 & -0.33$\pm$0.28 & 7.35$\pm$0.21 & -2.10$\pm$0.28 & 3.61$\pm$0.08\\
SN 2007bp & 0.030 & II        & 10.80$\pm$0.14 & 1.00$\pm$0.22 & 7.99$\pm$0.14 & -1.81$\pm$0.22 & 10.71$\pm$0.06\\
SN 2007bt & 0.040 & IIn       & 9.65$\pm$0.22 & -0.03$\pm$0.29 & 7.62$\pm$0.22 & -2.07$\pm$0.29 & 4.85$\pm$0.06\\
SN 2007bu & 0.030 & II        & 8.46$\pm$0.30 & -1.26$\pm$0.34 & 7.67$\pm$0.30 & -2.05$\pm$0.34 & 1.54$\pm$0.10\\
SN 2007bv & 0.050 & II        & 11.34$\pm$0.09 & 1.10$\pm$0.15 & 8.65$\pm$0.09 & -1.59$\pm$0.15 & 11.25$\pm$0.09\\
SN 2007bw & 0.140 & IIn       & 9.66$\pm$0.08 & 0.40$\pm$0.15 & 7.45$\pm$0.08 & -1.81$\pm$0.15 & 5.43$\pm$0.28\\
SN 2007bx & 0.020 & II        & 8.16$\pm$0.48 & -1.24$\pm$0.51 & 7.19$\pm$0.48 & -2.21$\pm$0.51 & 2.16$\pm$0.14\\
SN 2007by & 0.040 & II        & 10.23$\pm$0.13 & 0.42$\pm$0.19 & 8.00$\pm$0.13 & -1.81$\pm$0.19 & 6.11$\pm$0.04\\
SN 2007ce & 0.046 & Ic-BL & 8.08$\pm$0.13 & -0.60$\pm$0.16 & 8.03$\pm$0.13 & -0.66$\pm$0.16 & 0.45$\pm$0.19\\
SN 2007dp & 0.030 & II        & 9.14$\pm$0.28 & -0.52$\pm$0.33 & 8.03$\pm$0.28 & -1.63$\pm$0.33 & 1.77$\pm$0.02\\
SN 2007dq & 0.050 & II        & 8.71$\pm$0.20 & -1.01$\pm$0.28 & 7.14$\pm$0.20 & -2.58$\pm$0.28 & 3.42$\pm$0.39\\
SN 2007dw & 0.050 & II        & 10.07$\pm$0.06 & 0.59$\pm$0.09 & 7.80$\pm$0.06 & -1.69$\pm$0.09 & 6.10$\pm$0.04\\
SN 2007dy & 0.040 & Ib        & 9.21$\pm$0.22 & -0.32$\pm$0.27 & 7.29$\pm$0.22 & -2.24$\pm$0.27 & 3.71$\pm$0.12\\
SN 2007eb & 0.040 & Ic-BL & 8.50$\pm$0.22 & -0.78$\pm$0.30 & 7.22$\pm$0.22 & -2.06$\pm$0.30 & 3.31$\pm$0.16\\
SN 2007ed & 0.070 & II        & 10.69$\pm$0.11 & 0.54$\pm$0.15 & 8.71$\pm$0.11 & -1.44$\pm$0.15 & 4.87$\pm$0.08\\
SN 2007eh & 0.010 & II        & 8.04$\pm$0.32 & -1.32$\pm$0.39 & 9.05$\pm$0.32 & -0.32$\pm$0.39 & 0.15$\pm$0.06\\
SN 2007el & 0.030 & II        & 9.53$\pm$0.30 & -0.13$\pm$0.34 & 7.21$\pm$0.30 & -2.45$\pm$0.34 & 6.23$\pm$0.17\\
SN 2007em & 0.030 & II        & 7.63$\pm$0.31 & -1.54$\pm$0.36 & 7.00$\pm$0.31 & -2.17$\pm$0.36 & 1.21$\pm$0.25\\
SN 2007eq & 0.030 & Ib/c      & 8.49$\pm$0.30 & -1.14$\pm$0.32 & 7.47$\pm$0.30 & -2.17$\pm$0.32 & 2.80$\pm$0.16\\
SN 2007er & 0.070 & II        & 9.62$\pm$0.12 & 0.08$\pm$0.16 & 7.32$\pm$0.12 & -2.22$\pm$0.16 & 6.69$\pm$0.19\\
SN 2007es & 0.030 & II        & 10.94$\pm$0.04 & 0.99$\pm$0.08 & 8.56$\pm$0.04 & -1.39$\pm$0.08 & 7.46$\pm$0.08\\
SN 2007et & 0.040 & II        & 10.00$\pm$0.04 & 0.41$\pm$0.08 & 7.71$\pm$0.04 & -1.89$\pm$0.08 & 6.38$\pm$0.03\\
SN 2007eu & 0.040 & II        & 9.33$\pm$0.05 & -0.03$\pm$0.13 & 7.40$\pm$0.05 & -1.96$\pm$0.13 & 3.92$\pm$0.06\\
\hline
\end{tabular}
\end{table*}\begin{table*}[htp!]
\centering
\scriptsize\begin{tabular}{lccccccc}
\hline
Name & {\it z} & Type & Mass & SFR (phot) & $\Sigma_{M}$ & $\Sigma_{\rm SFR}$ & $r_{50}$\\
 &  &  & (log M$_{\odot}$) & (log M$_{\odot}$ yr$^{-1}$) & (log M$_{\odot}$ kpc$^{-2}$) & (log M$_{\odot}$ yr$^{-1}$ kpc$^{-2}$) & (kpc)\\
\hline
SN 2007ew & 0.030 & II        & 9.38$\pm$0.30 & -0.40$\pm$0.36 & 7.54$\pm$0.30 & -2.24$\pm$0.36 & 4.78$\pm$0.14\\
SN 2007fa & 0.060 & II        & 11.07$\pm$0.09 & 1.05$\pm$0.10 & 8.54$\pm$0.09 & -1.48$\pm$0.10 & 8.70$\pm$0.10\\
SN 2007fe & 0.030 & II        & 9.39$\pm$0.21 & -0.38$\pm$0.23 & 7.89$\pm$0.21 & -1.88$\pm$0.23 & 2.63$\pm$0.02\\
SN 2007ff & 0.050 & Ic & 10.45$\pm$0.13 & 0.64$\pm$0.18 & 8.09$\pm$0.13 & -1.72$\pm$0.18 & 6.45$\pm$0.06\\
SN 2007fg & 0.030 & II        & 8.87$\pm$0.20 & -0.35$\pm$0.29 & 7.12$\pm$0.20 & -2.10$\pm$0.29 & 3.81$\pm$0.08\\
SN 2007fk & 0.040 & IIn       & 8.61$\pm$0.22 & -1.05$\pm$0.41 & 7.04$\pm$0.22 & -2.62$\pm$0.41 & 3.64$\pm$0.20\\
SN 2007fw & 0.050 & IIn       & 9.05$\pm$0.18 & -0.52$\pm$0.22 & 7.74$\pm$0.18 & -1.83$\pm$0.22 & 3.91$\pm$0.15\\
SN 2007fy & 0.050 & II        & 9.96$\pm$0.04 & 0.61$\pm$0.07 & 8.00$\pm$0.04 & -1.35$\pm$0.07 & 5.62$\pm$0.04\\
SN 2007fz & 0.014 & II        & 8.44$\pm$0.31 & -1.22$\pm$0.32 & 7.85$\pm$0.31 & -1.81$\pm$0.32 & 1.54$\pm$0.01\\
SN 2007gh & 0.020 & II        & 9.45$\pm$0.30 & -0.30$\pm$0.33 & 8.70$\pm$0.30 & -1.05$\pm$0.33 & 1.10$\pm$0.01\\
SN 2007gl & 0.030 & Ic & 9.90$\pm$0.17 & 0.20$\pm$0.17 & 7.39$\pm$0.17 & -2.31$\pm$0.17 & 7.37$\pm$0.07\\
SN 2007gm & 0.030 & II        & 9.13$\pm$0.31 & -0.51$\pm$0.43 & 8.05$\pm$0.31 & -1.59$\pm$0.43 & 1.48$\pm$0.03\\
SN 2007gs & 0.040 & II        & 9.57$\pm$0.13 & -0.01$\pm$0.12 & 7.40$\pm$0.13 & -2.18$\pm$0.12 & 6.47$\pm$0.12\\
SN 2007gy & 0.040 & IIn       & 9.34$\pm$0.22 & -0.22$\pm$0.23 & 7.26$\pm$0.22 & -2.29$\pm$0.23 & 4.40$\pm$0.18\\
SN 2007gz & 0.050 & II        & 9.80$\pm$0.18 & 0.01$\pm$0.19 & 9.01$\pm$0.18 & -0.78$\pm$0.19 & 1.01$\pm$0.01\\
SN 2007hb & 0.022 & Ic & 10.55$\pm$0.03 & 0.80$\pm$0.07 & 8.51$\pm$0.03 & -1.24$\pm$0.07 & 4.87$\pm$0.01\\
SN 2007hi & 0.070 & II        & 8.11$\pm$0.35 & -1.49$\pm$0.26 & 10.20$\pm$0.35 & 0.59$\pm$0.26 & 0.06$\pm$0.12\\
SN 2007hn & 0.030 & Ic & 9.86$\pm$0.30 & -0.03$\pm$0.30 & 8.29$\pm$0.30 & -1.60$\pm$0.30 & 3.30$\pm$0.04\\
SN 2007hs & 0.070 & II        & 8.68$\pm$0.20 & -1.11$\pm$0.28 & 7.65$\pm$0.20 & -2.15$\pm$0.28 & 2.50$\pm$0.75\\
SN 2007hw & 0.080 & II        & 10.85$\pm$0.03 & 1.00$\pm$0.07 & 8.55$\pm$0.03 & -1.30$\pm$0.07 & 7.15$\pm$0.06\\
SN 2007ib & 0.030 & II        & 9.52$\pm$0.18 & -0.02$\pm$0.30 & 7.69$\pm$0.18 & -1.86$\pm$0.30 & 4.31$\pm$0.03\\
SN 2007iu & 0.090 & II        & 9.31$\pm$0.13 & -0.35$\pm$0.15 & 6.59$\pm$0.13 & -3.07$\pm$0.15 & 9.77$\pm$0.24\\
SN 2007ja & 0.090 & II P      & 10.65$\pm$0.03 & 1.00$\pm$0.07 & 8.22$\pm$0.03 & -1.43$\pm$0.07 & 7.78$\pm$0.08\\
SN 2007jf & 0.070 & II P      & 9.65$\pm$0.13 & -0.03$\pm$0.21 & 7.78$\pm$0.13 & -1.90$\pm$0.21 & 5.06$\pm$0.19\\
SN 2007jm & 0.090 & II n      & 9.77$\pm$0.10 & 0.10$\pm$0.16 & 8.34$\pm$0.10 & -1.34$\pm$0.16 & 4.36$\pm$0.19\\
SN 2007jn & 0.060 & II        & 8.99$\pm$0.14 & -0.38$\pm$0.21 & 7.54$\pm$0.14 & -1.84$\pm$0.21 & 4.79$\pm$0.24\\
SN 2007kw & 0.070 & II        & 10.82$\pm$0.06 & 0.72$\pm$0.08 & 8.67$\pm$0.06 & -1.44$\pm$0.08 & 6.24$\pm$0.07\\
SN 2007ky & 0.070 & II        & 10.85$\pm$0.03 & 0.90$\pm$0.07 & 8.02$\pm$0.03 & -1.93$\pm$0.07 & 14.62$\pm$0.12\\
SN 2007kz & 0.130 & II        & 11.20$\pm$0.05 & 1.36$\pm$0.09 & 8.14$\pm$0.05 & -1.70$\pm$0.09 & 16.25$\pm$0.21\\
SN 2007lb & 0.060 & II        & \nodata & \nodata & \nodata & \nodata & 1.90$\pm$0.67\\
SN 2007ld & 0.030 & II        & 7.96$\pm$0.30 & -1.35$\pm$0.34 & 6.54$\pm$0.30 & -2.77$\pm$0.34 & 2.72$\pm$0.43\\
SN 2007lj & 0.040 & II        & 8.06$\pm$0.24 & -1.27$\pm$0.28 & 8.43$\pm$0.24 & -0.90$\pm$0.28 & 0.51$\pm$0.90\\
SN 2007lz & 0.090 & II        & 9.30$\pm$0.12 & -0.39$\pm$0.20 & 7.18$\pm$0.12 & -2.51$\pm$0.20 & 5.65$\pm$0.49\\
SN 2007md & 0.050 & II        & 10.93$\pm$0.10 & 0.97$\pm$0.14 & 8.57$\pm$0.10 & -1.40$\pm$0.14 & 9.00$\pm$0.04\\
SN 2007ms & 0.040 & II pec    & 8.50$\pm$0.22 & -0.57$\pm$0.27 & 7.70$\pm$0.22 & -1.37$\pm$0.27 & 1.70$\pm$0.09\\
SN 2007nm & 0.046 & Ic        & 8.47$\pm$0.22 & -1.72$\pm$0.35 & 8.87$\pm$0.22 & -1.32$\pm$0.35 & 0.36$\pm$0.22\\
SN 2007nr & 0.140 & II P      & 9.46$\pm$0.07 & 0.00$\pm$0.18 & 7.67$\pm$0.07 & -1.79$\pm$0.18 & 3.54$\pm$0.26\\
SN 2007nw & 0.060 & II P      & 10.26$\pm$0.05 & 0.42$\pm$0.08 & 8.19$\pm$0.05 & -1.65$\pm$0.08 & 5.68$\pm$0.09\\
SN 2007ny & 0.140 & II P      & 9.11$\pm$0.18 & -0.66$\pm$0.24 & 7.52$\pm$0.18 & -2.25$\pm$0.24 & 6.39$\pm$3.82\\
SN 2007qb & 0.080 & II        & 10.27$\pm$0.07 & 0.73$\pm$0.15 & 7.53$\pm$0.07 & -2.01$\pm$0.15 & 10.93$\pm$0.18\\
SN 2007qv & 0.100 & II        & 10.51$\pm$0.10 & 0.57$\pm$0.12 & 8.32$\pm$0.10 & -1.63$\pm$0.12 & 6.88$\pm$0.17\\
SN 2007qw & 0.150 & Ic-BL & 9.40$\pm$0.05 & 0.13$\pm$0.13 & 8.31$\pm$0.05 & -0.96$\pm$0.13 & 1.84$\pm$0.22\\
SN 2007qx & 0.060 & Ib        & 10.03$\pm$0.11 & 0.34$\pm$0.15 & 7.90$\pm$0.11 & -1.80$\pm$0.15 & 6.87$\pm$0.08\\
SN 2007sd & 0.090 & II P      & 8.95$\pm$0.11 & -0.48$\pm$0.19 & 7.38$\pm$0.11 & -2.06$\pm$0.19 & 4.73$\pm$0.61\\
SN 2007sj & 0.040 & Ib/c      & 10.45$\pm$0.03 & 0.80$\pm$0.07 & 8.01$\pm$0.03 & -1.64$\pm$0.07 & 7.09$\pm$0.03\\
SN 2007sx & 0.120 & II        & 10.96$\pm$0.04 & 1.05$\pm$0.10 & 7.84$\pm$0.04 & -2.07$\pm$0.10 & 16.58$\pm$0.29\\
SN 2007sz & 0.020 & II        & 8.56$\pm$0.31 & -0.98$\pm$0.31 & 7.21$\pm$0.31 & -2.34$\pm$0.31 & 2.20$\pm$0.04\\
SN 2007tn & 0.050 & II        & 10.27$\pm$0.16 & 0.36$\pm$0.16 & 8.18$\pm$0.16 & -1.73$\pm$0.16 & 4.56$\pm$0.06\\
SN 2008bj & 0.019 & II        & 8.54$\pm$0.28 & -0.95$\pm$0.28 & 6.92$\pm$0.28 & -2.57$\pm$0.28 & 3.24$\pm$0.04\\
SN 2008fm & 0.039 & IIn       & 11.29$\pm$0.13 & 1.28$\pm$0.14 & 8.30$\pm$0.13 & -1.71$\pm$0.14 & 17.13$\pm$0.10\\
SN 2008fn & 0.030 & Ib/c      & 9.86$\pm$0.20 & -0.56$\pm$0.30 & 8.33$\pm$0.20 & -2.09$\pm$0.30 & 2.63$\pm$0.03\\
SN 2008fo & 0.030 & Ic        & 9.50$\pm$0.04 & -0.07$\pm$0.09 & 7.70$\pm$0.04 & -1.87$\pm$0.09 & 4.20$\pm$0.02\\
SN 2008fs & 0.039 & Ib/c      & 10.19$\pm$0.05 & 0.25$\pm$0.11 & 8.29$\pm$0.05 & -1.65$\pm$0.11 & 4.15$\pm$0.04\\
SN 2008fz & 0.133 & IIn       & 9.93$\pm$0.05 & 1.25$\pm$0.10 & 13.12$\pm$0.05 & 4.44$\pm$0.10 & 0.01$\pm$0.08\\
SN 2008gd & 0.059 & II        & 9.89$\pm$0.09 & 0.47$\pm$0.14 & 7.42$\pm$0.09 & -2.00$\pm$0.14 & 7.67$\pm$0.10\\
\hline
\end{tabular}
\end{table*}\begin{table*}[htp!]
\centering
\scriptsize\begin{tabular}{lccccccc}
\hline
Name & {\it z} & Type & Mass & SFR (phot) & $\Sigma_{M}$ & $\Sigma_{\rm SFR}$ & $r_{50}$\\
 &  &  & (log M$_{\odot}$) & (log M$_{\odot}$ yr$^{-1}$) & (log M$_{\odot}$ kpc$^{-2}$) & (log M$_{\odot}$ yr$^{-1}$ kpc$^{-2}$) & (kpc)\\
\hline
SN 2008iu & 0.130 & Ic-BL & 8.22$\pm$0.28 & -0.72$\pm$0.23 & 8.26$\pm$0.28 & -0.68$\pm$0.23 & 0.63$\pm$1.21\\
SN 2008iy & 0.041 & IIn       & 10.20$\pm$0.03 & -0.10$\pm$0.07 & 18.89$\pm$0.03 & 8.59$\pm$0.07 & 0.00$\pm$0.00\\
SN 2008ja & 0.069 & IIn       & 8.38$\pm$0.20 & -1.21$\pm$0.21 & 7.85$\pm$0.20 & -1.74$\pm$0.21 & 1.35$\pm$0.37\\
SN 2009W & 0.017 & II P      & 8.92$\pm$0.32 & -0.69$\pm$0.34 & 7.23$\pm$0.32 & -2.39$\pm$0.34 & 3.06$\pm$0.12\\
SN 2009bh & 0.090 & Ic        & 10.83$\pm$0.10 & 0.86$\pm$0.12 & 8.23$\pm$0.10 & -1.74$\pm$0.12 & 8.95$\pm$0.11\\
SN 2009bj & 0.027 & II        & 9.01$\pm$0.23 & -1.16$\pm$0.46 & 8.45$\pm$0.23 & -1.72$\pm$0.46 & 1.49$\pm$0.02\\
SN 2009bk & 0.039 & II        & 9.60$\pm$0.03 & 0.33$\pm$0.09 & 7.69$\pm$0.03 & -1.58$\pm$0.09 & 5.14$\pm$0.04\\
SN 2009bl & 0.040 & II        & 9.55$\pm$0.04 & 0.18$\pm$0.08 & 7.84$\pm$0.04 & -1.53$\pm$0.08 & 2.98$\pm$0.02\\
SN 2009ct & 0.060 & II        & 10.47$\pm$0.06 & 0.69$\pm$0.09 & 8.15$\pm$0.06 & -1.64$\pm$0.09 & 6.98$\pm$0.06\\
SN 2009dh & 0.060 & II P      & 7.99$\pm$0.19 & -1.54$\pm$0.25 & 7.80$\pm$0.19 & -1.73$\pm$0.25 & 0.61$\pm$0.33\\
SN 2009di & 0.130 & Ic        & 8.57$\pm$0.24 & -0.47$\pm$0.22 & 6.98$\pm$0.24 & -2.07$\pm$0.22 & 3.54$\pm$1.26\\
SN 2009dw & 0.042 & II P      & 7.97$\pm$0.27 & -1.45$\pm$0.23 & 6.35$\pm$0.27 & -3.07$\pm$0.23 & 3.51$\pm$1.06\\
SN 2009fe & 0.047 & II        & 10.80$\pm$0.03 & -0.09$\pm$0.08 & 9.59$\pm$0.03 & -1.30$\pm$0.08 & 1.82$\pm$0.02\\
SN 2009jd & 0.025 & II        & 9.40$\pm$0.19 & -0.08$\pm$0.20 & 7.16$\pm$0.19 & -2.31$\pm$0.20 & 5.67$\pm$0.07\\
SN 2009kf & 0.182 & II P      & 9.66$\pm$0.09 & 0.14$\pm$0.19 & 7.98$\pm$0.09 & -1.54$\pm$0.19 & 4.10$\pm$0.55\\
SN 2009lx & 0.027 & II P      & 10.45$\pm$0.03 & 0.51$\pm$0.07 & 8.37$\pm$0.03 & -1.57$\pm$0.07 & 4.79$\pm$0.03\\
SN 2009nn & 0.046 & IIn       & 9.76$\pm$0.05 & 0.21$\pm$0.08 & 7.79$\pm$0.05 & -1.76$\pm$0.08 & 5.20$\pm$0.06\\
SN 2009nu & 0.040 & II        & 9.24$\pm$0.22 & -0.56$\pm$0.29 & 7.02$\pm$0.22 & -2.77$\pm$0.29 & 5.84$\pm$0.28\\
SN 2010K & 0.020 & II        & 8.36$\pm$0.47 & -1.10$\pm$0.50 & 6.98$\pm$0.47 & -2.48$\pm$0.50 & 2.42$\pm$0.14\\
SN 2010Q & 0.055 & Ic        & 7.46$\pm$0.20 & -1.18$\pm$0.22 & 6.80$\pm$0.20 & -1.83$\pm$0.22 & 1.03$\pm$0.40\\
SN 2010ah & 0.050 & Ic-BL & 8.82$\pm$0.13 & -0.85$\pm$0.20 & 7.67$\pm$0.13 & -2.00$\pm$0.20 & 4.43$\pm$0.27\\
SN 2010ay & 0.067 & Ic-BL & 8.58$\pm$0.09 & 0.03$\pm$0.11 & 8.92$\pm$0.09 & 0.37$\pm$0.11 & 0.34$\pm$0.02\\
SN 2010gq & 0.018 & II        & 10.50$\pm$0.03 & 0.50$\pm$0.07 & 8.63$\pm$0.03 & -1.38$\pm$0.07 & 4.64$\pm$0.02\\
SN 2010jc & 0.024 & II P      & 10.75$\pm$0.03 & 0.70$\pm$0.07 & 7.92$\pm$0.03 & -2.13$\pm$0.07 & 11.07$\pm$0.08\\
SN 2010jy & 0.042 & IIn       & 9.11$\pm$0.12 & -0.50$\pm$0.18 & 7.58$\pm$0.12 & -2.04$\pm$0.18 & 2.73$\pm$0.11\\
SN 2010mb & 0.133 & Ic        & 9.63$\pm$0.07 & 0.14$\pm$0.16 & 7.56$\pm$0.07 & -1.93$\pm$0.16 & 8.76$\pm$0.32\\
SN 2011ak & 0.027 & II P      & 10.45$\pm$0.03 & 0.50$\pm$0.07 & 7.95$\pm$0.03 & -2.00$\pm$0.07 & 7.73$\pm$0.05\\
SN 2011an & 0.016 & IIn       & 9.61$\pm$0.04 & -0.02$\pm$0.07 & 7.23$\pm$0.04 & -2.39$\pm$0.07 & 7.18$\pm$0.04\\
SN 2011aw & 0.055 & Ib/c      & \nodata & \nodata & \nodata & \nodata & 1.67$\pm$1.37\\
SN 2011bm & 0.022 & Ic        & 9.45$\pm$0.03 & 0.10$\pm$0.07 & 7.88$\pm$0.03 & -1.47$\pm$0.07 & 2.67$\pm$0.01\\
SN 2011bn & 0.031 & II        & 11.20$\pm$0.03 & 1.20$\pm$0.07 & 8.80$\pm$0.03 & -1.20$\pm$0.07 & 8.83$\pm$0.06\\
SN 2011bs & 0.036 & II        & 7.44$\pm$0.28 & -2.12$\pm$0.26 & 8.43$\pm$0.28 & -1.12$\pm$0.26 & 0.56$\pm$0.18\\
SN 2011cl & 0.025 & II P      & \nodata & \nodata & \nodata & \nodata & 8.76$\pm$0.07\\
SN 2011cq & 0.017 & II pec    & 9.69$\pm$0.31 & 0.13$\pm$0.34 & 7.41$\pm$0.31 & -2.15$\pm$0.34 & 5.51$\pm$0.03\\
SN 2011cw & 0.040 & IIn       & \nodata & \nodata & \nodata & \nodata & 0.59$\pm$0.28\\
SN 2011cz & 0.060 & II P      & 8.24$\pm$0.22 & -1.24$\pm$0.25 & 7.32$\pm$0.22 & -2.16$\pm$0.25 & 3.03$\pm$2.16\\
SN 2011db & 0.025 & II        & 9.68$\pm$0.19 & 0.14$\pm$0.27 & 7.50$\pm$0.19 & -2.04$\pm$0.27 & 5.19$\pm$0.03\\
SN 2011en & 0.020 & II P      & 8.70$\pm$0.46 & -0.48$\pm$0.42 & 7.29$\pm$0.46 & -1.89$\pm$0.42 & 2.46$\pm$0.05\\
SN 2011eo & 0.030 & II P      & 8.02$\pm$0.30 & -1.57$\pm$0.33 & 7.80$\pm$0.30 & -1.78$\pm$0.33 & 0.62$\pm$0.07\\
SN 2011eu & 0.110 & IIn       & 9.54$\pm$0.15 & -0.38$\pm$0.34 & 9.76$\pm$0.15 & -0.17$\pm$0.34 & 1.39$\pm$0.98\\
SN 2011ev & 0.030 & II P      & 8.99$\pm$0.27 & -0.23$\pm$0.33 & 7.08$\pm$0.27 & -2.14$\pm$0.33 & 7.65$\pm$0.11\\
SN 2011ew & 0.070 & II P      & 10.19$\pm$0.04 & -0.80$\pm$0.07 & 9.92$\pm$0.04 & -1.07$\pm$0.07 & 0.98$\pm$0.78\\
SN 2011fa & 0.060 & II P      & 7.85$\pm$0.23 & -1.61$\pm$0.25 & 7.69$\pm$0.23 & -1.76$\pm$0.25 & 0.57$\pm$0.25\\
SN 2011fz & 0.016 & Ib/c      & 10.55$\pm$0.03 & 0.60$\pm$0.07 & 8.00$\pm$0.03 & -1.95$\pm$0.07 & 7.92$\pm$0.06\\
SN 2011hn & 0.014 & II P      & 10.02$\pm$0.31 & 0.15$\pm$0.33 & 8.20$\pm$0.31 & -1.67$\pm$0.33 & 6.02$\pm$0.02\\
SN 2011iw & 0.023 & IIn       & 7.92$\pm$0.22 & -1.64$\pm$0.23 & 7.97$\pm$0.22 & -1.59$\pm$0.23 & 0.48$\pm$0.05\\
SN 2011jb & 0.084 & IIn       & 9.20$\pm$0.09 & -0.46$\pm$0.18 & 8.48$\pm$0.09 & -1.18$\pm$0.18 & 1.24$\pm$0.11\\
SN 2011jj & 0.045 & II P      & 10.99$\pm$0.04 & 1.10$\pm$0.07 & 8.05$\pm$0.04 & -1.84$\pm$0.07 & 13.98$\pm$0.09\\
SN 2011ke & 0.143 & Ic        & 8.91$\pm$0.15 & -0.42$\pm$0.19 & 6.86$\pm$0.15 & -2.47$\pm$0.19 & 6.77$\pm$1.61\\
SN 2012D & 0.026 & II P      & 9.60$\pm$0.04 & 0.01$\pm$0.07 & 8.12$\pm$0.04 & -1.48$\pm$0.07 & 2.57$\pm$0.01\\
SN 2012F & 0.030 & Ib        & 7.86$\pm$0.30 & -1.63$\pm$0.36 & 7.74$\pm$0.30 & -1.75$\pm$0.36 & 0.64$\pm$0.06\\
SN 2012W & 0.018 & II        & 10.05$\pm$0.03 & 0.30$\pm$0.07 & 7.70$\pm$0.03 & -2.05$\pm$0.07 & 7.34$\pm$0.03\\
SN 2012al & 0.040 & IIn       & 9.77$\pm$0.17 & 0.01$\pm$0.26 & 7.62$\pm$0.17 & -2.13$\pm$0.26 & 6.66$\pm$0.11\\
SN 2012bg & 0.033 & II P      & 8.29$\pm$0.16 & -1.30$\pm$0.20 & 7.95$\pm$0.16 & -1.64$\pm$0.20 & 0.89$\pm$0.06\\
SN 2012br & 0.019 & II P      & 7.40$\pm$0.34 & -1.96$\pm$0.37 & 7.13$\pm$0.34 & -2.23$\pm$0.37 & 0.72$\pm$0.05\\
\hline
\end{tabular}
\end{table*}\begin{table*}[htp!]
\centering
\scriptsize\begin{tabular}{lccccccc}
\hline
Name & {\it z} & Type & Mass & SFR (phot) & $\Sigma_{M}$ & $\Sigma_{\rm SFR}$ & $r_{50}$\\
 &  &  & (log M$_{\odot}$) & (log M$_{\odot}$ yr$^{-1}$) & (log M$_{\odot}$ kpc$^{-2}$) & (log M$_{\odot}$ yr$^{-1}$ kpc$^{-2}$) & (kpc)\\
\hline
SN 2012ch & 0.009 & II P      & 8.30$\pm$0.04 & -1.20$\pm$0.07 & 7.72$\pm$0.04 & -1.78$\pm$0.07 & 1.52$\pm$0.02\\
SN 2012cr & 0.010 & II        & 9.95$\pm$0.03 & 0.00$\pm$0.07 & 8.45$\pm$0.03 & -1.50$\pm$0.07 & 3.82$\pm$0.01\\
SN 2012cz & 0.036 & IIn       & 10.35$\pm$0.03 & 0.30$\pm$0.07 & 8.66$\pm$0.03 & -1.39$\pm$0.07 & 4.85$\pm$0.02\\
SN 2012dp & 0.036 & Ib        & 10.59$\pm$0.04 & 0.82$\pm$0.12 & 8.00$\pm$0.04 & -1.76$\pm$0.12 & 8.58$\pm$0.07\\
SN 2012ed & 0.015 & II        & 7.32$\pm$0.17 & -2.01$\pm$0.42 & 7.32$\pm$0.17 & -2.01$\pm$0.42 & 0.80$\pm$0.12\\
SN 2012ex & 0.023 & Ib        & 10.45$\pm$0.03 & 0.70$\pm$0.07 & 8.48$\pm$0.03 & -1.27$\pm$0.07 & 4.07$\pm$0.01\\
SN 2012fc & 0.023 & II P      & 9.80$\pm$0.03 & 0.29$\pm$0.07 & 8.19$\pm$0.03 & -1.32$\pm$0.07 & 3.82$\pm$0.02\\
SN 2012hw & 0.038 & II P      & 9.90$\pm$0.11 & 0.19$\pm$0.15 & 7.49$\pm$0.11 & -2.22$\pm$0.15 & 6.59$\pm$0.05\\
SN 2012il & 0.175 & Ic        & 9.03$\pm$0.16 & -0.59$\pm$0.20 & 8.18$\pm$0.16 & -1.44$\pm$0.20 & 1.93$\pm$0.63\\
SN 2013an & 0.014 & II        & \nodata & \nodata & \nodata & \nodata & 0.32$\pm$0.25\\
SN 2013aw & 0.027 & II        & 9.45$\pm$0.19 & -0.30$\pm$0.24 & 7.80$\pm$0.19 & -1.95$\pm$0.24 & 6.63$\pm$0.06\\
SN 2013bn & 0.054 & Ic        & \nodata & \nodata & \nodata & \nodata & 0.78$\pm$0.13\\
SN 2013br & 0.074 & II        & 9.15$\pm$0.10 & -0.47$\pm$0.19 & 8.14$\pm$0.10 & -1.48$\pm$0.19 & 1.73$\pm$0.22\\
SN 2013bw & 0.038 & II P      & 10.60$\pm$0.03 & 1.10$\pm$0.07 & 8.15$\pm$0.03 & -1.35$\pm$0.07 & 7.61$\pm$0.03\\
SN 2013dn & 0.056 & IIn       & 10.70$\pm$0.03 & 1.20$\pm$0.07 & 8.59$\pm$0.03 & -0.91$\pm$0.07 & 5.50$\pm$0.06\\
\hline
\end{tabular}
\tablecomments{Column ``$z$'' shows the spectroscopic redshift of the SN or LGRB, while ``Type" lists the spectroscopic classification of the SN. The ``Mass" and ``SFR (phot)'' columns are, respectively, the galaxy stellar mass and SFR we estimate by fitting broadband photometry with PEGASE2 \citep{fi99} stellar population synthesis models.  Column $\Sigma_{\rm SFR}$ is the projected SFR surface density \mbox{$\Sigma_{\rm SFR} = {\rm log}_{10}({\rm SFR} \mathbin{/} 2 \mathbin{/} \pi A B)$} where SFR is the value estimated from photometry in the adjacent column, while $A$ and $B$ are the semimajor and semiminor axes (in kpc) of the 
isophotal ellipse that contains half of the galaxy $r$-band flux. The column ``$r_{50}$" shows the the weighted average of the half-light radii of the de Vaucouleurs and an exponential components fit to the galaxy light distribution by the SDSS {\tt photo} pipeline.}
\end{table*}
\clearpage
\begin{table*}[htp!]
\centering
\scriptsize
\caption{Properties of Host Galaxies with SDSS Spectra}\begin{tabular}{lccccccc}
\hline
Name & {\it z} & Type & Mass & $\sigma_{\rm vel}$ & SFR (spec) & $\Sigma_{\rm SFR}$ & Fraction deV.\\
 &  &  & (log M$_{\odot}$) & (km s$^{-1}$) & (log M$_{\odot}$ yr$^{-1}$) & (log M$_{\odot}$ yr$^{-1}$ kpc$^{-2}$) & \\
\hline
PTF 09awk & 0.062 & Ib & 9.54$\pm$0.06 & 63.7$\pm$0.7 & -0.03$\pm$0.14 & -1.00$\pm$0.14 & 1.0\\
PTF 09dra & 0.077 & II & 10.50$\pm$0.08 & 77.1$\pm$2.6 & 0.30$\pm$0.33 & -2.46$\pm$0.33 & 0.5\\
PTF 09ige & 0.064 & II & 9.73$\pm$0.06 & 44.2$\pm$1.2 & 0.21$\pm$0.25 & -1.93$\pm$0.25 & 0.2\\
PTF 09ism & 0.029 & II & 9.12$\pm$0.21 & 33.2$\pm$3.1 & -0.68$\pm$0.40 & -2.42$\pm$0.40 & 0.2\\
PTF 09sk & 0.036 & Ic-BL & 8.93$\pm$0.15 & 51.4$\pm$0.7 & -0.40$\pm$0.22 & -1.48$\pm$0.22 & 0.5\\
PTF 09uj & 0.065 & II & 9.75$\pm$0.08 & 52.7$\pm$3.4 & -0.08$\pm$0.27 & -2.17$\pm$0.27 & 0.0\\
PTF 10bau & 0.026 & II & 10.75$\pm$0.03 & 80.8$\pm$1.1 & 0.41$\pm$0.22 & -1.62$\pm$0.22 & 0.2\\
PTF 10bhu & 0.036 & Ic & 9.43$\pm$0.14 & 51.0$\pm$2.1 & -0.27$\pm$0.29 & -1.92$\pm$0.29 & 0.0\\
PTF 10con & 0.033 & II & 9.68$\pm$0.16 & 68.6$\pm$2.3 & -0.60$\pm$0.30 & -1.96$\pm$0.30 & 0.1\\
PTF 10cxx & 0.034 & II & 10.03$\pm$0.13 & 68.2$\pm$1.0 & -0.07$\pm$0.18 & -1.46$\pm$0.18 & 0.2\\
PTF 10s & 0.051 & II & 9.66$\pm$0.09 & 38.7$\pm$1.3 & -0.25$\pm$0.24 & -1.90$\pm$0.24 & 0.0\\
PTF 11cgx & 0.034 & II & 9.95$\pm$0.04 & 57.6$\pm$1.6 & -0.00$\pm$0.26 & -1.75$\pm$0.26 & 0.0\\
PTF 11cwi & 0.056 & II & 10.58$\pm$0.09 & 110.9$\pm$1.2 & 0.70$\pm$0.13 & -1.33$\pm$0.13 & 1.0\\
PTF 11dqk & 0.036 & II & 9.81$\pm$0.04 & 42.5$\pm$0.8 & 0.35$\pm$0.23 & -1.56$\pm$0.23 & 0.1\\
PTF 11dtd & 0.040 & II & 10.38$\pm$0.04 & 69.0$\pm$3.6 & -0.39$\pm$0.78 & -2.89$\pm$0.78 & 0.0\\
PTF 11ecp & 0.034 & II & 10.15$\pm$0.12 & 53.9$\pm$1.3 & 0.25$\pm$0.30 & -1.71$\pm$0.30 & 0.0\\
PTF 11gdz & 0.013 & II & 9.89$\pm$0.29 & 68.0$\pm$0.5 & -0.34$\pm$0.13 & -1.36$\pm$0.13 & 0.1\\
PTF 11jgp & 0.072 & II & 10.02$\pm$0.06 & 55.0$\pm$1.3 & 0.28$\pm$0.25 & -1.70$\pm$0.25 & 0.0\\
PTF 11mpv & 0.043 & II & 9.27$\pm$0.11 & 41.1$\pm$1.4 & -0.47$\pm$0.22 & -1.56$\pm$0.22 & 0.4\\
PTF 11qcm & 0.051 & II & 10.65$\pm$0.09 & 83.7$\pm$2.2 & 0.01$\pm$0.32 & -2.19$\pm$0.32 & 0.0\\
PTF 11qju & 0.028 & II & 9.35$\pm$0.19 & 37.6$\pm$1.3 & -0.32$\pm$0.23 & -2.22$\pm$0.23 & 0.0\\
PTF 11qux & 0.041 & II & 9.58$\pm$0.11 & 48.0$\pm$0.7 & -0.14$\pm$0.13 & -1.23$\pm$0.13 & 0.0\\
PTF 12cgb & 0.026 & II & 9.29$\pm$0.20 & 49.3$\pm$0.8 & -0.30$\pm$0.27 & -1.58$\pm$0.27 & 0.1\\
PTF 12dke & 0.067 & II & 9.83$\pm$0.09 & 57.7$\pm$3.0 & 0.18$\pm$0.30 & -2.32$\pm$0.30 & 0.1\\
PTF 12eje & 0.078 & II & 9.86$\pm$0.09 & 49.8$\pm$1.5 & -0.64$\pm$0.16 & -2.75$\pm$0.16 & 0.0\\
PTF 12gcx & 0.045 & II & 9.67$\pm$0.16 & 39.4$\pm$1.2 & -0.96$\pm$0.35 & -3.28$\pm$0.35 & 0.1\\
PTF 12gzk & 0.014 & 12gzk & 7.29$\pm$0.09 & 59.0$\pm$0.7 & -2.40$\pm$0.09 & -2.45$\pm$0.09 & 1.0\\
PTF 13c & 0.011 & II & 8.68$\pm$0.30 & 49.4$\pm$1.4 & -0.68$\pm$0.33 & -1.72$\pm$0.33 & 0.0\\
PTF 13cbf & 0.040 & Ic & 9.51$\pm$0.14 & 47.0$\pm$0.4 & 0.26$\pm$0.20 & -1.44$\pm$0.20 & 0.2\\
SN 2004hy & 0.058 & II        & 9.69$\pm$0.11 & 46.4$\pm$4.4 & -0.16$\pm$0.31 & -2.54$\pm$0.31 & 0.2\\
SN 2005hl & 0.023 & Ib        & 10.43$\pm$0.18 & 63.8$\pm$1.1 & 0.16$\pm$0.22 & -1.46$\pm$0.22 & 0.0\\
SN 2005kb & 0.015 & II        & 9.14$\pm$0.38 & 28.6$\pm$1.4 & -1.00$\pm$0.30 & -2.16$\pm$0.30 & 0.0\\
SN 2005ks & 0.099 & Ic-BL & 9.89$\pm$0.07 & 70.4$\pm$1.4 & -0.10$\pm$0.24 & -1.53$\pm$0.24 & 0.0\\
SN 2005lc & 0.014 & II        & 8.50$\pm$0.34 & 34.8$\pm$3.1 & -1.53$\pm$0.27 & -2.59$\pm$0.27 & 0.1\\
SN 2005lm & 0.085 & II        & 9.33$\pm$0.08 & 43.7$\pm$1.1 & -0.60$\pm$0.08 & -1.87$\pm$0.08 & 0.3\\
SN 2005mn & 0.047 & Ib        & 9.52$\pm$0.13 & 44.2$\pm$2.6 & -0.17$\pm$0.30 & -2.27$\pm$0.30 & 0.2\\
SN 2006M & 0.015 & IIn       & 8.86$\pm$0.29 & 43.9$\pm$2.1 & -0.64$\pm$0.24 & -1.96$\pm$0.24 & 0.1\\
SN 2006bj & 0.038 & II        & 9.40$\pm$0.14 & 43.9$\pm$2.9 & -1.57$\pm$0.23 & -3.63$\pm$0.23 & 0.2\\
SN 2006cw & 0.061 & II        & 9.87$\pm$0.08 & 49.7$\pm$1.6 & 0.13$\pm$0.27 & -2.02$\pm$0.27 & 0.0\\
SN 2006db & 0.023 & IIn       & 8.69$\pm$0.21 & 46.4$\pm$3.0 & -0.92$\pm$0.27 & -2.21$\pm$0.27 & 0.2\\
SN 2006fo & 0.021 & Ib & 10.65$\pm$0.03 & 61.3$\pm$0.8 & 0.05$\pm$0.24 & -1.67$\pm$0.24 & 0.1\\
SN 2006fq & 0.068 & II P      & 10.04$\pm$0.04 & 46.1$\pm$0.4 & 0.32$\pm$0.25 & -1.48$\pm$0.25 & 0.1\\
SN 2006gd & 0.155 & II P      & 11.02$\pm$0.06 & 101.9$\pm$7.8 & -0.33$\pm$1.00 & -3.10$\pm$1.00 & 0.6\\
SN 2006iw & 0.031 & II        & 9.73$\pm$0.15 & 34.2$\pm$2.0 & -0.52$\pm$0.36 & -2.15$\pm$0.36 & 0.0\\
SN 2006ix & 0.076 & II        & 9.57$\pm$0.08 & 48.3$\pm$2.5 & -1.01$\pm$0.15 & -2.81$\pm$0.15 & 0.1\\
SN 2006kh & 0.060 & II        & 9.47$\pm$0.10 & 65.5$\pm$1.0 & -0.70$\pm$0.10 & -1.18$\pm$0.10 & 0.3\\
SN 2006kn & 0.120 & II        & 10.08$\pm$0.16 & 51.6$\pm$3.4 & -0.30$\pm$0.39 & -2.52$\pm$0.39 & 0.0\\
SN 2006ns & 0.120 & II        & 9.88$\pm$0.06 & 43.4$\pm$1.8 & 0.33$\pm$0.26 & -1.71$\pm$0.26 & 0.2\\
SN 2006qk & 0.058 & Ic-BL & 9.55$\pm$0.11 & 57.4$\pm$1.4 & -0.60$\pm$0.13 & -1.30$\pm$0.13 & 0.1\\
SN 2007I & 0.022 & Ic-BL & 8.82$\pm$0.20 & 42.7$\pm$2.9 & -1.07$\pm$0.29 & -2.20$\pm$0.29 & 0.0\\
SN 2007bo & 0.044 & II        & 9.11$\pm$0.12 & 41.6$\pm$2.5 & -0.44$\pm$0.35 & -2.29$\pm$0.35 & 0.1\\
SN 2007bp & 0.028 & II        & 10.60$\pm$0.04 & 46.0$\pm$4.3 & -1.61$\pm$1.11 & -4.36$\pm$1.11 & 0.4\\
SN 2007dp & 0.033 & II        & 9.19$\pm$0.14 & 43.8$\pm$1.6 & -0.62$\pm$0.27 & -1.82$\pm$0.27 & 0.0\\
SN 2007fe & 0.033 & II        & 9.59$\pm$0.17 & 53.1$\pm$1.2 & -0.33$\pm$0.23 & -1.92$\pm$0.23 & 0.3\\
SN 2007fg & 0.026 & II        & 8.84$\pm$0.19 & 37.7$\pm$0.8 & -0.60$\pm$0.21 & -2.22$\pm$0.21 & 0.0\\
\hline
\end{tabular}
\end{table*}
\begin{table*}[htp!]
\centering
\scriptsize\begin{tabular}{lccccccc}
\hline
Name & {\it z} & Type & Mass & $\sigma_{\rm vel}$ & SFR (spec) & $\Sigma_{\rm SFR}$ & Fraction deV.\\
 &  &  & (log M$_{\odot}$) & (km s$^{-1}$) & (log M$_{\odot}$ yr$^{-1}$) & (log M$_{\odot}$ yr$^{-1}$ kpc$^{-2}$) & \\
\hline
SN 2007fy & 0.045 & II        & 10.03$\pm$0.11 & 59.2$\pm$0.8 & 0.27$\pm$0.25 & -1.61$\pm$0.25 & 0.1\\
SN 2007ib & 0.034 & II        & 10.00$\pm$0.14 & 57.1$\pm$0.9 & 0.12$\pm$0.25 & -1.84$\pm$0.25 & 0.1\\
SN 2007jf & 0.070 & II P      & 9.56$\pm$0.10 & 44.6$\pm$2.2 & -1.07$\pm$0.12 & -2.94$\pm$0.12 & 0.4\\
SN 2007jm & 0.091 & II n      & 9.70$\pm$0.10 & 56.9$\pm$1.8 & -0.55$\pm$0.13 & -2.00$\pm$0.13 & 0.3\\
SN 2007ky & 0.074 & II        & 11.03$\pm$0.07 & 79.4$\pm$4.4 & 0.47$\pm$0.45 & -2.41$\pm$0.45 & 0.6\\
SN 2007lx & 0.058 & II        & 10.72$\pm$0.10 & 69.2$\pm$2.7 & 0.79$\pm$0.33 & -1.46$\pm$0.33 & 0.8\\
SN 2007nw & 0.057 & II P      & 10.26$\pm$0.10 & 74.9$\pm$1.8 & 0.09$\pm$0.33 & -1.93$\pm$0.33 & 0.6\\
SN 2007qb & 0.079 & II        & 10.27$\pm$0.05 & 70.2$\pm$1.3 & 0.67$\pm$0.26 & -2.06$\pm$0.26 & 0.6\\
SN 2007qw & 0.151 & Ic-BL & 9.36$\pm$0.08 & 49.1$\pm$1.0 & -0.09$\pm$0.08 & -1.18$\pm$0.08 & 0.7\\
SN 2008bj & 0.019 & II        & 8.49$\pm$0.32 & 27.3$\pm$2.5 & -0.80$\pm$0.23 & -2.42$\pm$0.23 & 0.0\\
SN 2008fo & 0.030 & Ic        & 9.49$\pm$0.20 & 52.2$\pm$0.8 & 0.03$\pm$0.22 & -1.76$\pm$0.22 & 0.0\\
SN 2008gd & 0.059 & II        & 9.89$\pm$0.09 & 41.3$\pm$3.4 & -0.01$\pm$0.34 & -2.48$\pm$0.34 & 0.2\\
SN 2009bk & 0.039 & II        & 9.80$\pm$0.15 & 44.8$\pm$1.4 & 0.16$\pm$0.31 & -1.74$\pm$0.31 & 0.0\\
SN 2009bl & 0.040 & II        & 9.85$\pm$0.04 & 54.4$\pm$2.2 & 0.14$\pm$0.27 & -1.58$\pm$0.27 & 0.0\\
SN 2009ct & 0.057 & II        & 10.58$\pm$0.08 & 97.4$\pm$3.1 & 0.42$\pm$0.34 & -1.85$\pm$0.34 & 0.6\\
SN 2010ay & 0.067 & Ic-BL & 8.55$\pm$0.09 & 61.3$\pm$0.6 & -0.01$\pm$0.12 & 0.32$\pm$0.12 & 1.0\\
SN 2011bm & 0.022 & Ic        & 9.85$\pm$0.04 & 57.1$\pm$0.8 & -0.04$\pm$0.23 & -1.63$\pm$0.23 & 0.3\\
SN 2011cq & 0.017 & II pec    & 9.97$\pm$0.33 & 67.9$\pm$2.4 & -0.65$\pm$0.42 & -2.94$\pm$0.42 & 0.4\\
SN 2011en & 0.018 & II P      & 8.51$\pm$0.32 & 12.7$\pm$7.5 & -1.07$\pm$0.24 & -2.37$\pm$0.24 & 0.1\\
SN 2011hn & 0.014 & II P      & 9.95$\pm$0.04 & 12.6$\pm$2.8 & -0.52$\pm$0.41 & -2.35$\pm$0.41 & 0.2\\
SN 2011jm & 0.003 & Ic        & 8.82$\pm$0.06 & 31.9$\pm$0.3 & -2.51$\pm$0.15 & -3.23$\pm$0.15 & 0.4\\
SN 2012D & 0.026 & II P      & 9.65$\pm$0.03 & 47.4$\pm$0.5 & 0.14$\pm$0.21 & -1.34$\pm$0.21 & 0.0\\
SN 2012al & 0.038 & IIn       & 9.55$\pm$0.17 & 34.8$\pm$3.9 & -1.49$\pm$1.01 & -3.59$\pm$1.01 & 0.0\\
SN 2012dp & 0.036 & Ib        & 10.55$\pm$0.03 & 85.4$\pm$0.8 & 0.56$\pm$0.17 & -2.02$\pm$0.17 & 0.9\\
SN 2012ex & 0.023 & Ib        & 8.77$\pm$0.37 & 21.2$\pm$0.8 & -1.49$\pm$0.11 & -3.58$\pm$0.11 & 0.3\\
SN 2012hw & 0.038 & II P      & 10.06$\pm$0.13 & 39.4$\pm$2.4 & 0.05$\pm$0.31 & -2.36$\pm$0.31 & 0.0\\
SN 2013aw & 0.027 & II        & 9.37$\pm$0.21 & 40.3$\pm$1.2 & -0.57$\pm$0.26 & -2.21$\pm$0.26 & 0.1\\
\hline
\end{tabular}
\tablecomments{Column ``$z$ shows the SN spectroscopic redshift, while ``Type" lists the spectroscopic classification. The ``Mass" column shows the galaxy stellar mass we estimate by fitting broadband photometry with PEGASE2 \citep{fi99} stellar population synthesis models. The $\sigma_{\rm vel}$ estimate is of the gas velocity dispersion measured from the H$\alpha$ emission-line profile by the Portsmouth group \citep{thomassteele13} using the Penalized PiXel Fitting \citep{cappellariemsellem04} (pPXF) and the Gas and Absorption Line Fitting \citep{sarzifalconbarroso06} (GANDALF v1.5) codes. The column ``SFR (spec)" is a hybrid SFR estimate computed by the MPA-JHU group that is the sum of the SFR inside the SDSS fiber aperture inferred from the spectrum, and the SFR outside of the aperture from modeling {\it ugriz} photometry. Column $\Sigma_{\rm SFR}$ is the projected SFR surface density \mbox{$\Sigma_{\rm SFR} = {\rm log}_{10}({\rm SFR} \mathbin{/} 2 \mathbin{/} \pi A B)$} where SFR is the hybrid value in the adjacent column, while $A$ and $B$ are the semimajor and semiminor axes (in kpc) of the isophotal ellipse that contains half of the galaxy $r$-band flux. The column ``Fraction deV." shows the fraction of the total galaxy light attributed to the de Vaucouleurs $r^{1/4}$ component from a simultaneous fit by the SDSS {\tt photo} pipeline of a de Vaucouleurs and an exponential profile to the galaxy light distribution.}
\end{table*}

\end{document}